\documentclass[prepint,sort&compress]{article}

\usepackage{xcolor, colortbl}

\usepackage{lineno,hyperref}
\usepackage{textcomp}
\usepackage{xcolor}
\usepackage{algorithm}
\usepackage[noend]{algpseudocode}
\usepackage{placeins}
\usepackage{graphicx}
\usepackage{collectbox}
\DeclareGraphicsExtensions{.png,.jpg}
\usepackage{hyperref}
\hypersetup{
  colorlinks=true,
  linkcolor=blue,
  filecolor=magenta,   
  urlcolor=cyan,
}
\usepackage{caption,subcaption} 
\usepackage{multirow}
\usepackage{mwe} 
\usepackage[noend]{algpseudocode}
\usepackage{babel}[english]
\usepackage{geometry}
\usepackage{pdflscape}
\usepackage{multirow}
\usepackage[export]{adjustbox}

\usepackage{ulem} 

\makeatletter
\def\algbackskip{\hskip-\ALG@thistlm}
\makeatother

\begin{document}

\author{Alicia Beneyto-Rodriguez, Gregorio I. Sainz-Palmero \footnote{Corresponding author: gregorioismael.sainz@uva.es} \\Marta Galende-Hern\'{a}ndez, Mar\'{i}a J. Fuente \\
Department of System Engineering and Automatic Control\\ School of Industrial Engineering, Universidad de Valladolid, Spain \footnote{Emails:\{alicia.beneyto,marta.galende,mariajesus.fuente\}@uva.es}}
\title{Linguistic Ordered Weighted Averaging based deep learning pooling for fault diagnosis in a wastewater treatment plant \footnote{This work was supported by the Spanish Government through the Ministerio de Ciencia, Innovaci\'on y Universidades/ Agencia Estatal de Investigaci\'on (MICIU/AEI/10.13039/501100011033) under Grant PID2019-105434RB-C32.}}

\date{}

\maketitle

\begin{abstract}
{\color{black}
Nowadays, water reuse is a serious challenge to help address water shortages. Here, the wastewater treatment plants (WWTP) play a key role, and its proper operation is mandatory. So, fault diagnosis is a key activity for these plants. Their high complexity and large-scale require of smart methodologies for that fault diagnosis and safety operation. All these large-scale and complex industrial processes are  monitored, allowing the data collection about the plant operation, so data driven approaches for fault diagnosis can be applied. A popular approach to fault diagnosis is deep learning-based methodologies. Here, a fault diagnosis methodology is proposed  for a WWTP  using  a new  linguistic Ordered Weighted Averaging (OWA) pooling based Deep Convolutional Neural Network (DCNN) and  a sliding and overlapping time window. This  window slides over input data  based on the  monitoring  sampling time, then the diagnosis is carried out by the linguistic OWA pooling based DCNN. This alternative linguistic pooling uses  well-known linguistic OWA quantifiers, which permit terms such as \textsl{Most, AtLeast, etc.}, supplying new intuitive options for the pooling tasks. This sliding time window and the OWA pooling based network  permit  a better  and  earlier fault diagnosis, at each sampling time, using a few monitoring samples  and a fewer  learning iterations  than DCNN standard pooling. Several linguistic OWA operators have been checked with a benchmark for WWTPs. A set of 5 fault types has been used, taking into account 140 variables sampled at 15 minutes time intervals. The performance has been over $91\%$ for $Accuracy$, $Recall$ or $F1-Score$, and better than other competitive methodologies. Moreover, these linguistic OWA operators for DCNN pooling have shown a better performance than the standard \textsl{Max} and \textsl{Average} options.

}
\end{abstract}

\begin{center}
\textsl{Keywords}: Fault diagnosis, Deep learning,  Linguistic Ordered Weighted Averaging, Wastewater treatment plant, Convolutional neural network.
\end{center}

\section{Introduction} 
\label{sec:intro}

Water is a strategic and fundamental resource for human beings. Human activities in  industry, agriculture, and so on depend directly on access to water resources and any increase in these activities releases more polluted effluents into the fresh water systems, so accessibility to clean water is limited and a challenge for many people \cite{Fan-18}. Although around three-quarters of the overall surface of the Earth is covered by water, most of it is in the seas and oceans (97\%). There is only $3\%$ of fresh water, but more than two-thirds is frozen in glaciers and polar ice \cite{Salles-23}, so water is a strategic resource and must be managed adequately. Moreover, considering climate change,  water scarcity problems are expected to become more severe in several  European regions, such as Spain; so the reuse of water is an adaptation measure which can reduce the pressure on water resources and increase water security and resilience\footnote{https://climate-adapt.eea.europa.eu/en/metadata/adaptation-options/water-recycling/\#websites}. In this context, treated wastewater can help conserve groundwater if  used for irrigation and other uses. However, all this implies that the polluted water needs to be treated before use in appropriate places, such as in wastewater treatment plants (WWTP). These are facilities which  play an essential role in reducing environmental pollution and increasing water quality and its reuse. 

WWTPs are complex systems that must maintain a high performance, despite disturbances such as  water scarcity,  but this performance must be sustainable along time, with a high quality and low cost. To safely and optimally operate a WWTP it is necessary to monitor the treatment process on-line. Part of this process monitoring is the fault detection and diagnosis task (FDI), i.e., to detect abnormal events or anomalies as soon as possible so the necessary actions can be taken at the right time to prevent more serious consequences in the future. Fault diagnosis for any type of plant or process, mainly when these are large-scale, has been  a relevant target  for years  due to its high impact in the operation and performance of the plants, as well as for its economic or social/reputational impact, including in many cases  environmental effects and challenges,  as in the case of   wastewater treatment plants or other  similar facilities.

Abnormal or faulty events are those that occur when the process indicates a deviation from its normal behavior. In WWTPs, abnormal events include changes in the influent quality (e.g., rainfall, industrial discharge), modifications of microorganisms that impact treatment quality, irregularities or damage to treatment units, mechanical failures (e.g., pumps, air blowers) and sensor failures. All these faults can degrade the process performance and have to be detected and diagnosed as soon as they occur to maintain the effluent quality and to assess the environmental regulations. 

In recent years, sensor technology has been continuously developing, while distributed control systems have become more readily available for obtaining and storing data. This has created new opportunities for the development of data-driven monitoring techniques. There are principally two main approaches for this, the multivariate statistical process methods and the machine learning methodologies, including shallow, {\color{black}reinforcement} and deep learning. The former includes principal component analysis (PCA) \cite{Sanchez-15}\cite{Cheng-19} \cite{Li-22}, independent component analysis (ICA) \cite{Xu-21} \cite{Palla-23}, partial least-square (PLS) \cite{Liu-21}\cite{Sha-23}, canonical correlation analysis (CCA), and canonical variate (CVA) analysis \cite{Cheng-21} \cite{Fuente-23}; together with their kernel versions for non-linear systems (KPCA, KPLS, KICA and KCCA) that have successfully been applied in the field of process monitoring, where data dimensions are reduced to extract key information about the process data. A good review of the state of the art of these techniques applied to the WWTPs can be studied in \cite{Khurshid-23}.

The second approach considered is based on machine learning, and in general, artificial and computational  intelligence techniques. These techniques are successfully applied to this type of plant  due to its complex physical or chemical processes, as well as operating condition transformations and the nonlinearity of the wastewater treatment process. Support vector machine, multilayer perception, random forest, and gradient boosting methods are some of the most studied methods in this category \cite{Bellamoli-23} \cite{Khurshid-23}. {\color{black}Reinforcement learning algorithms have also gained much attention in recent years due to their capability for dealing with unknown environment \cite{Fan2022} \cite{Yang2023}}.
 
Deep learning strategies have also become increasingly popular in the face of complex nonlinearity for their power to extract knowledge from large and complex datasets, and they can be used for the modelling, control or management of WWTPs, as can be seen in \cite{Newhart-19} \cite{Alvi-23} \cite{Ismail-23}. However, very few studies have addressed fault detection problems in full WWTPs.

The most popular deep learning approaches for process monitoring of industrial plants are: Deep Belief networks (DBN) \cite{Harrou2018} \cite{Yu-19}, Autoencoders (AE) and their variants, such as Stacked Autoencoders (SAE), Denoising Autoencoders (DAE) or Variational Autoencoders (VAE), \cite{Cheng-19} \cite{Yu-21} \cite{Ba-Alawi-21} \cite{Ba-Alawi-22} \cite{Liu-23}.
{\color{black}A review of these approaches for  fault diagnosis in industrial processes is available in \cite{Qian2022}. Other popular approaches are } long short-term memory (LSTM) based models \cite{Mamandipoor-20}, and convolutional neural networks (CNN) \cite{Wu2018} \cite{Song-22}, or a combination of various methods, such as \cite{Chen-22}, which uses optimized CNN-LSTM, or \cite{Salles-23}, which uses Convolutional-AE and LSTM-AE models to detect failures in sensors in a WWTP. 

Focusing on  deep-learning methods for  fault diagnosis in wastewater treatment plants: \cite{Harrou2018} joins the deep belief networks (DBN) model and a one-class support vector machine (OCSVM)  to separate normal from abnormal features, probing its  effective for fault detection; while \cite{Mamandipoor-20} uses LSTM networks to identify collective faults in WWTP sensors,  comparing  its results  with regard to the autoregressive integrated moving average (ARIMA), principal component analysis (PCA) and support vector machines (SVM) models. Recurrent neural networks (RNN) to capture temporal auto-correlation features and Restricted Boltzmann Machines to delineate complex distributions (RNN-RBM) is in \cite {Dairi-19}. Autoencoders are frequently used in the WWTPs: \cite{Ba-Alawi-21} uses a stacked denoising autoencoder (SDAE) model to detect, identify, and reconcile faulty sensor data from BSM1 and real  influent WWTP data; \cite{Ba-Alawi-22} uses deep variational autoencoders (VAE) with  a good performance in detecting and reconstructing faulty sensors,  but the main drawback of this structure (VAE-ResNet) is its very high  computational requirements. Another approach based on AE is \cite{Salles-23}, where convolutional and Long short-term memory (LSTM) autoencoders (AEs) are used to identify failures for the  dissolved oxygen sensor used in WWTPs. Transformer-based models have also emerged as a powerful approach for fault detection in wastewater treatment processes, as in \cite{Peng-22}, where a transformer-based model incorporating a multihead attention mechanism and residual connections is used to detect faults in the BSM1 model of a WWTP.  

{\color{black} So, our work involves a new approach based on Deep Convolutional Neural Networks (DCNN)  incorporating linguistic pooling layers for a better performance and a fewer learning iterations,  not only to detect anomalies on the WWTP but also to diagnosis the different faults in the key elements of the plant. The benchmark Simulation Model No. 2 (BSM2), which was developed by the International Water Association (IWA) and is standard in the WWTP field, is used considering  faults in the oxygen sensor (DO) and the rest of key parts of the full process, in different fault magnitudes and times.}

On the other hand, CNN is a stack of different types of layers, such as convolutional and pooling layers,  to be tuned by a learning stage. These pooling layers are usually based  on two standard operations: \textsl{max} and \textsl{average}, but it is possible to find other approaches in several domains that provide other alternative  pooling operators: \cite{Zhao2024} proposes \textsl{T-Max-Avg} pooling for computer vision that incorporates  a threshold parameter T to control the output  based on the maximum values or weighted averages. Fuzzy logic is applied in \cite{Diamantis2021} by implementing a pooling layer  using a fuzzy aggregation and defuzzification of the fuzzified input features.  Ordered Weighted Averaging (OWA) operators \cite{Yager1988} have  been used to aggregate data within CNN pooling layers, training their weights in image recognition \cite{Forcen2020}. In \cite{DominguezCatena2021} and \cite{Catena2020}, an additional  OWA layer is considered, modifying  the usual CNN architecture, to be tuned in image classification. \cite{Hussain2022a} and \cite{Hussain2022} use  OWA operators in a prediction layer for QoS cloud data. \cite{Ghosal2021} proposed OWA operators  and neural networks to detect  various shades of sentiments. {\color{black}Here, a pooling based on linguistic OWA RIM operators is proposed, which permits a softer and fuzzy downsampling using linguistic terms which are commonly used by the users.

Moreover, the use of CNN's approaches involves to deal with  the challenge of their layout of layers: there are different success and popular CNN models, most of them  devoted  on computer vision,  but, in general, they imply high complexity and high computational resources as computation time as data, i.e., AlexNet \cite{Alex2017}, InceptionV3 \cite{Szegedy_2016_CVPR}, etc.  In this work, the CNN models are based on simple, lightweight  and straightforward layer layouts, with moderate computational requirements and fitting for faults diagnosis.

In modern plants, the monitoring systems supply an \textsl{'image'} shot of the plant condition at every sampling time, so a plant diagnosis should be available for a safe and valuable management of the operations. That is not usual in the  bibliographic references,  which are rarely focused on the fault diagnosis in a complete WWTP, most of them are devoted on fault detection only  considering  few fault types in some WWTP elements and,  in some cases,  managing few valuable data for complex approaches.
}

 {\color{black}
In accordance with all this, the current proposal addresses fault diagnosis in a complete WWTP using a sliding  and overlapping  time window to manage the data shots, which are provided by the monitoring system about the plant condition at every sampling time,  and  a linguistic OWA pooling based convolutional network to diagnosis. All this working together is able to provide a high fault diagnosis performance using only a few monitoring data, which can be supplied at every monitoring sampling time. So, the main contributions can be summarized in the following points:

\begin{enumerate}
		\item A new methodology for WWTP fault diagnosis, which is made up of a sliding  and overlapping   time  window for managing  input data and a DCNN implementing linguistic OWA based pooling layers to make the fault diagnosis.  This proposal has provided a better fault diagnosis than other approaches in similar problems,  as well as using fewer data and achieving the diagnostic at every sampling time.
		
		\item The linguistic OWA quantifiers proposed as pooling operators have shown  a better  fault diagnosis performance and a faster learning than the standard \textsl{Max} and \textsl{Average} pooling operators. 
		
				\item Two new  parametrized issues of  linguistic OWA quantifiers are introduced: \textsl{AtMiddle$_{\alpha}$($\alpha=0.2$)} and \textsl{AtLeast$_{\alpha}$ ($\alpha=0.75$)} for  a fuzzier and linguistic pooling.
		
		\item The linguistic, and fuzzy,  nature of these pooling operators permits an easier and more intuitive use for these pooling operations in DCNNs without any extra  network layers. 

\end{enumerate}
}

The remainder of this work is as follows: this current section, Section \ref{sec:intro},  has been  focused on a general introduction to the issue and the alternative approaches considered in many fields; the following section,  Section \ref{sec:stateofart}, is a brief and preliminary introduction to the theoretical foundations used in this work. Next, in Section  \ref{sec:proposal}, the key points of the proposal of this work are developed. The application, results and analysis  obtained when this approach has been applied on the BSM2 benchmark is described in detail in Section \ref{sec:casestudysetup}. Finally,  Section \ref{sec:conclusion} contains a brief summary of the main conclusions obtained from this work.

\section{Background and Related works}
\label{sec:stateofart}

\subsection{Deep Learning and convolutional neural networks}
\label{sec:dcnn}

\begin{figure}[th!]
  \begin{center}
		\includegraphics[width=1\textwidth]{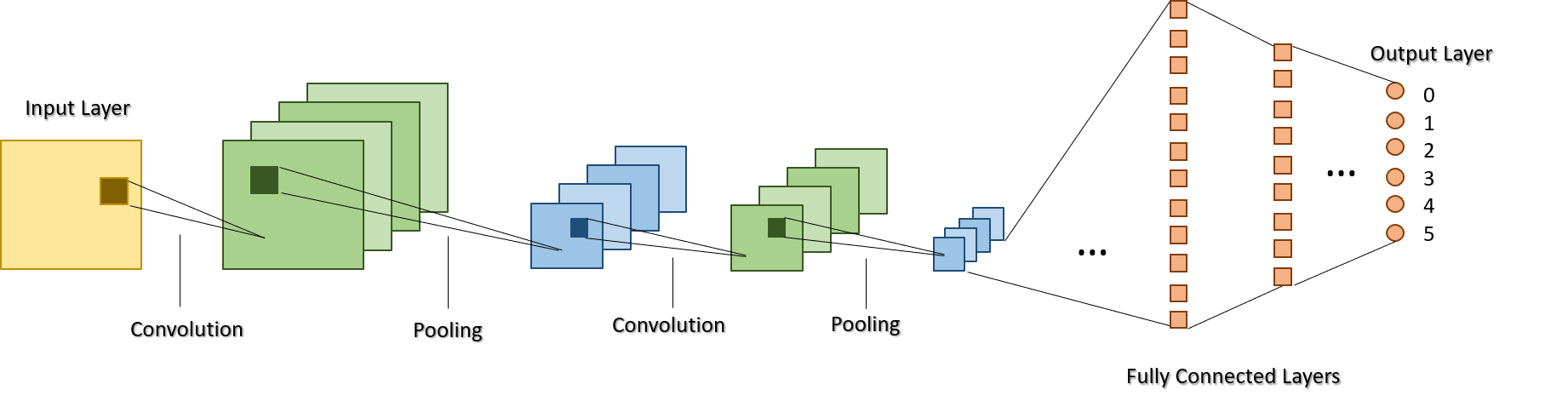}
		\caption{Schematic structure of a DCNN. }
		\label{fig:dcnn}
	\end{center}
\end{figure}

Deep Learning and convolutional networks  have been very popular in many fields for long time, especially in  computer vision, speech recognition and others  in which  this type of approach has been very successful, giving rise to a large bibliographic production as new neural network models as applications on a very wide range of domains \cite{Lecun2015} \cite{Samira2018}  \cite{Ausif2019}. Deep learning  permits  computational models  containing   multiple processing layers  (see Fig. \ref{fig:dcnn}) which are  able to learn with multiple levels of abstraction  from large amounts of data,  progressively extracting   higher-level features from data. All this is carried out by  supervised learning  during its training stage. The general structure of this type of networks  is very well known \cite{Ausif2019} \cite{Jun2022}, and is  based on feature extractor  and classifier, so a very brief description follows:  
 
\begin{itemize}
		\item \textsl{Extraction layers}, their goal  is to extract a high level  representation from data: map of features. The number of these layers  is arbitrary for each multilayer architecture
		or model.
				\begin{itemize}
						\item \textsl{Convolutional layers},  one or more filters are applied over an input to generate one or more feature maps,  summarizing the presence of detected features in the input.
						\item \textsl{Pooling layers}, or sub-sampling layers, follow  a convolutional layer producing a down-sampled version of the feature map, merging similar local features. 
						The pooling standard layers are  \textsl{max}  and \textsl{average},  computing the maximum and the average of  a local patch in a feature map,  respectively. Pooling layers reduce 
						the feature representation which implies a shorter time and computational  costs.
						In this proposal a new set of operators for pooling is proposed based on  \textsl{linguistic OWA} quantifiers, which allow the use of intuitive terms from natural language, such as 
						\textsl{Most}, \textsl{AtLeast}, ... into this type of layers (see Fig. \ref{fig:dcnnowa}).
						\item \textsl{Dropout},  aims to avoid overfitting, not considering  some feature randomly in each step  along the training stage. 
				\end{itemize}
		
		\item \textsl{Classification layers}, its  goal is to classify the features extracted from the data.
			\begin{itemize}
				\item \textsl{Full connected layer}, this is usually a backpropagation neural network with one-dimensional input.
			\end{itemize}
\end{itemize}

In this proposal, when Deep Learning networks  are mentioned, this means  Deep Convolutional Neural Networks. On the other hand, this type of networks implies defining their layouts  as a sequence of the said  layers which is a serious challenge. In the specialized bibliography,  many networks have been defined for different purposes, some of them being very complex and large neural models that implies another serious difficulty: there is a need for a sufficient amount of data for a fair, correct training in tuning all the parameters of the network layers. In this work, some layouts have been used that have shown a good performance in similar contexts, or  even in other fields,  but in any case maintaining  the goal of \textsl{lightweight} layouts.

\subsection{Linguistic OWA operators}
\label{sec:owa}

\begin{table}[t]
\centering	
\resizebox{15cm}{!} {	
\begin{tabular}{m{4cm}    m{6.5cm}  m{3.5cm}}
        \hline
        \textbf{Linguistic terms}& 							\textbf{Membership functions}&													\textbf{\color{black}Orness}\\
        \hline \hline
        There exists (Max)    							&\multicolumn{1}{m{9.6cm}}{\begin{displaymath}
																Q(x) = \left\{ \begin{array}{ll}
																0 & \mbox{if $ x=0$} \\
																1 & \mbox{if $ x\neq 0$}
																\end{array}
																\right.
															\end{displaymath}}											&$1$
       \\ 
       Average	    									&\multicolumn{1}{m{6.6cm}}{
																Q(x) = x($ 0 \leq x\leq 1$) 										
															}													&$0.5$
       \\ 
        Most    										&\multicolumn{1}{m{7cm}}{\begin{displaymath}
																Q(x) = \left\{ \begin{array}{ll}
																0 & \mbox{if $0\leq x\leq 0.3$} \\
																2(x-0.3) & \mbox{if $0.3 < x\leq 0.8$} \\
																1 & \mbox{if $0.8 < x\leq 1$}
																\end{array}
																\right.
															\end{displaymath}}											&$0.45$
      
       \\ 
        AtLeastHalf									&\multicolumn{1}{m{8.5cm}}{\begin{displaymath}
																Q(x) = \left\{ \begin{array}{ll}
																2x & \mbox{if $0\leq x\leq 0.5$} \\
																1 & \mbox{if $0.5 < x\leq 1$}
																\end{array}
																\right.
															\end{displaymath}}											&$0.75$
															
       \\ 
        AtMiddle$_{\alpha}$									&\multicolumn{1}{m{7cm}}{\begin{displaymath}
																Q(x) = \left\{ \begin{array}{ll}
																0 & \mbox{if $0\leq x\leq \alpha$} \\
																2(x-\alpha) & \mbox{if $\alpha < x\leq (1-\alpha)$} \\
																1 & \mbox{if $(1-\alpha) < x\leq 1$}
																\end{array}
																\right.
																\end{displaymath}} 									&$1+2\alpha(\alpha-1)$
       \\ 
        AtLeast$_{\alpha}$									&\multicolumn{1}{m{8.5cm}}{\begin{displaymath}
																Q(x) = \left\{ \begin{array}{ll}
																x/\alpha & \mbox{if $0\leq x\leq \alpha$} \\
																1 & \mbox{if $\alpha < x\leq 1$}
																\end{array}
																\right.
															\end{displaymath}}											&$1-\frac{\alpha}{2}$
       \\ \hline   \hline
	\end{tabular}

	}
	\caption{RIM quantifier linguistic terms (based on \cite{Liu2008}). }
	\label{tab:operaowa}
\end{table}

{\color{black}

The Ordered Weighted Averaging (OWA) operators \cite{Yager1988}, and their variants \cite{Yager2007},  are a popular type of parametrized aggregation operators in the field of multicriteria  decision-making and uncertain modelling, data mining, fuzzy systems, etc., for longtime \cite{flores2024} \cite{Shang2021} \cite{Serrano2020}. Regarding other aggregation operators, OWA does not assume that all elements carry equal weight, but rather  a specific weight is provided for each element,  reflecting its relative importance.

The idea  of \textsl{orness} is a central concept for these operators:  the \textsl{orness} gives  the \textsl{andlike} or \textsl{orlike} aggregation outcome of any OWA operator,  determining the  influence level  of each element in the overall aggregated result. By adjusting this parameter, decision-makers can tailor the OWA operator to align with their preferences and needs  in the problem at hand.

Mathematically, an OWA operator of dimension $n$ is a function $F: R^n \rightarrow R$  with an associated vector of $n$ weights 	$W=(w_1, \ldots , w_n)^T$ such that $w_i \in [0,1]$,  $1 \leq i \leq n$,   and  
\begin{equation}
	\sum_{i=1}^{n} w_i=w_1+ \ldots + w_n=1
\end{equation} 

where 

\begin{equation}
	F(a_1,\ldots, a_n)=\sum_{j=1}^{n} (w_j b_j)
\end{equation} 

$b_j$ being the jth largest element of the $a_n$. 

The \textsl{orness} associated to this operator is:

\begin{equation}
		Orness(W) = \sum_{j=1}^{n}\frac{n-j}{n-1} w_j  \hspace{2cm}  	Andness(W) = 1- Orness(W)
		\label{ec:orness}
\end{equation}

A must of the OWA operators is the re-ordering stage: a weight, $w_j$,  is associated with a particular ordered position, $j$,  of the aggregate of the elements  instead of a single element $a_j$.

On the other hand, there are different approaches to obtain the weights for the OWA operator \cite{Yager1993}, one popular  method is through Regular Increasing Monotone (RIM)  quantifiers \cite{Yager1996}. In this work,  this has been  done by fuzzy linguistic quantifiers through RIM linguistic quantifiers \cite{Yager1996}, which permit a semantic meaning in terms of natural language  such as \textsl{most, half, a few, etc.}, allowing a soft and fuzzy multicriteria decision-making.

This type of  linguistic quantifiers is defined as a function $Q: [0,1] \rightarrow [0,1]$ where $Q(0) = 0$, $Q(1) = 1$ and $Q(x) \geq Q(y)$ for $x > y$. For a given value $x \in [0,1]$, $Q(x)$ is the degree to which $x$ satisfies the fuzzy concept (\textsl{most, a few...})  represented by the quantifier.

Based on this function $Q$, the OWA weight vector is determined by the Eq. \ref{ec:rim}.

\begin{equation}
    w_i=Q(\frac{i}{n})-Q(\frac{i-1}{n})
    \label{ec:rim}
\end{equation}

These weights allow the importance of the different aggregation components to be increased or decreased, according to the semantics associated with the operator through $Q$. Thus, the quantifier determines the construction strategy of the weight vector, $W$, and its linked measures:  \textit{orness} and \textit{andness} (Eq. \ref{ec:orness}) .


In \cite{Liu2008} some RIM quantifier linguistic terms have been compiled  (see Table \ref{tab:operaowa}), with alternative formulation and linguistic meaning, which have been used in this work to implement the linguistic OWA operator  in the DCNN pooling layers of the proposal.  Moreover,  two new modified linguistic quantifiers have been added in this collection: 
\begin{itemize}
	\item AtMiddle$_{\alpha}$ for taking into account the \textsl{'elements around the middle except the  $\alpha$-top and $\alpha$-bottom elements'}.
	\item AtLeast$_{\alpha}$, which has been tuned for taking into account \textsl{'at least the  $\alpha$-top elements'}.
\end{itemize}

Moreover, \textsl{Max} and \textsl{Average} quantifiers  are  the operators  ordinarily used for pooling in DCNNs. Here,  other quantifiers are proposed as competitive alternatives for this task and goal.

}

\section {Fault diagnosis in a wastewater treatment plant}
\label{sec:proposal}

\begin{figure}[t!]
  \begin{center}
		\includegraphics[width=1.1\textwidth]{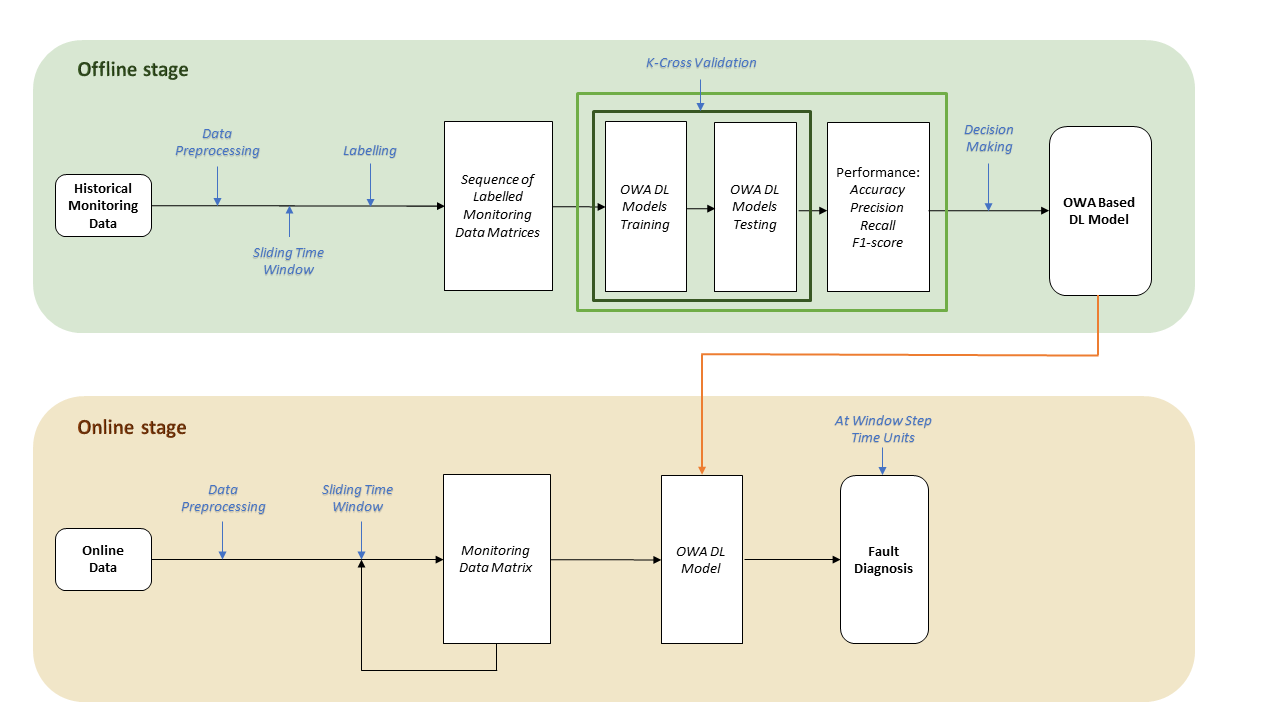}
		\caption{Window Sliding and linguist OWA pooling based Fault Diagnosis Scheme. }
		\label{fig:propuesta}
	\end{center}
\end{figure}

\begin{figure}[t!]
  \begin{center}
		\includegraphics[width=0.6\textwidth]{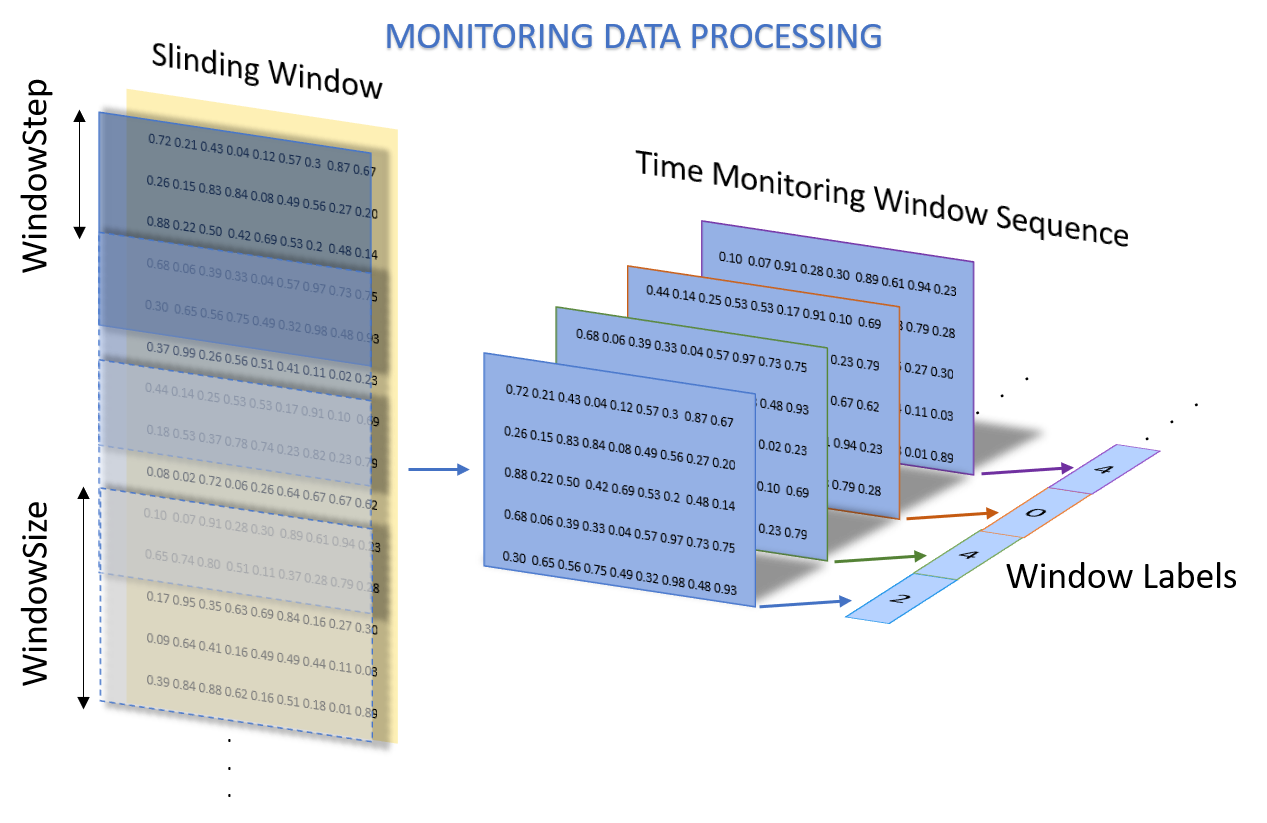}
		\caption{ Sliding window and labelling mechanisms. }
		\label{fig:slidingwindow}
	\end{center}
\end{figure}

\begin{figure}[t!]
  \begin{center}
		\includegraphics[width=1\textwidth]{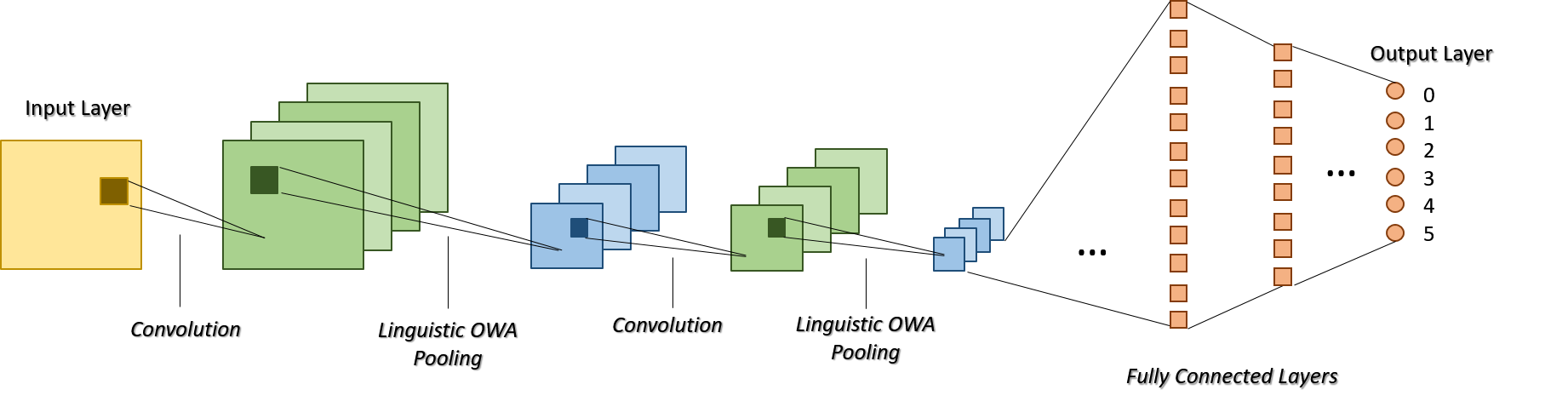}
		\caption{Schematic structure of a linguistic OWA pooling based DCNN. }
		\label{fig:dcnnowa}
	\end{center}
\end{figure}

This work deals with  the fault diagnosis of a  wastewater treatment plant. The proposal (see Fig. \ref{fig:propuesta})  is based on  a sliding time window  which captures the $m$ last monitoring samples of the plant, {\color{black} as an image shot about the plant condition}. This window slides along the time with a tunable sliding step and size (see Fig. \ref{fig:slidingwindow}). Then, this monitoring data window  is used as  input for a DCNN to diagnosis the condition type of the plant. This DCNN implements the pooling layers based on  (OWA) RIM linguistic quantifiers (see Fig. \ref{fig:dcnnowa}). The goal is to obtain a good performance with regard to the plant diagnosis using fewer monitoring data. On the other hand, this diagnosis must be provided as short a time as possible for its applicability in real environments where the  fault diagnosis is necessary as soon as possible to avoid serious damage and failures.

This DCNN based  fault diagnosis approach implies the two main and usual  stages in its setup (see Fig. \ref{fig:propuesta}):
\begin{enumerate}
		\item \textsl{Offline stage}, the linguistic OWA based DL models are  tuned based on historical monitoring data and using the sliding time window approach to manage the network inputs:
					\begin{enumerate}
							\item  Historical monitoring data are preprocessed: outliers, missing values, standarization, etc..
							\item  A sliding time window (see Fig. \ref{fig:slidingwindow}),  featuring its $WindowSize$ (number of monitoring samples to be included) and $WindowStep$ 
							(step of sliding, in time or monitoring data samples), 
							is applied along  this processed historical data,   resulting in a sequence of input matrices   of size: $[VariableNumber, $WindowSize$]$. 
							\item Each of these inputs is labelled  as one of the fault types to be considered (including \textsl{normal} mode).
							\item All these labelled inputs are organized by a \textsl{k-cross validation} scheme.
							\item Every alternative  DL model based on linguistic OWA pooling (see Tables \ref{tab:operaowa} and \ref{tab:modelos})  is trained and tested using \textsl{k-cross validation}  (details in Algorithm \ref{Alg:CV}).
							\item The performance of every DL model is checked by  quality indexes: \textsl{Accuracy, Recall, Precision, F1-score}.
							\item The best DL model is selected through decision making based on user preferences about \textsl{Accuracy, Recall, Precision...}.					
	
					\end{enumerate}
		
		\item \textsl{Online stage}, here the best DL model for fault diagnosis is used with data collected online coming from the wastewater treatment plant. An overview is shown in Fig. \ref{fig:propuesta}:
			\begin{enumerate}
						\item Monitoring data samples  coming from the plant are collected to generate an input with a size of  $[VariableNumber, WindowSize]$. Every  $WindowStep*SamplingTime$ time unit, 
						a new input matrix is generated using  $WindowSize-WindowStep$ monitoring data samples  of the previous input and newly collected $WindowStep$ monitoring samples (sliding window mechanism).
						\item Then each new generated  matrix input  or plant condition shot is diagnosed by the DL fault detection model.
						\item The diagnosis of the treatment plant  is provided at every  $WindowStep*SamplingTime$ time unit.						
			\end{enumerate}
\end{enumerate}

{\color{black} 
This two-stage methodology is possible due to  the monitoring systems  of the modern plants, which  periodically collect data about the plant operation.  So,  historical data  is available to be used for this type of data-driven approaches.

On the other hand, a  difficult and serious challenge must be addressed when DL networks are involved: to define some hyperparameters such as the  layout of the DL network. In this work, the CNN models are based on simple, lightweight  and straightforward layer layouts,   with moderate computational requirements and a good performance in similar challenges such as the layouts in \cite{Wu2018}, or even in other fields such as LeNet-5 \cite{Li2022} \cite{Lecun1998},  a popular model for image classification due to  its simple and straightforward architecture, which has been used  as foundation in other more complex models such as AlexNet \cite{Alex2017}. The layouts taken into account  (Table \ref{tab:modelos}) cannot be considered as large or complex layouts, which was a goal of this work. High computational requirements due to complex networks  can invalidate its applicability  in this domain.

Additionally, the \textsl{Offline stage} of this  approach can be used to update the DL model when  new conditions, or faults,  must  be  considered:  pre-trained or transfer learning.
	
}

\section{Case study}
\label{sec:casestudysetup}

\begin{figure}[h!]
		\includegraphics[width=0.9\textwidth]{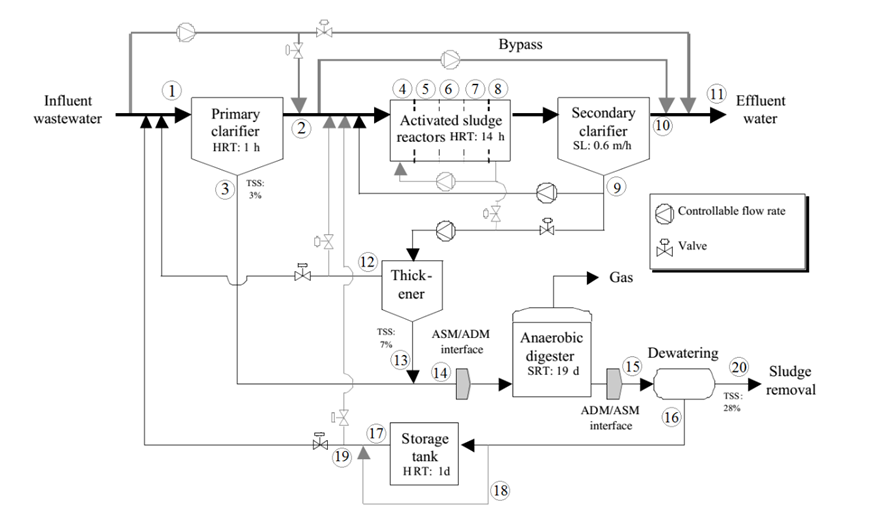}
		\caption{BSM2 Layout (http://iwa-mia.org/benchmarking/) \cite{Vrecko2006}. }
		\label{fig:BSM2}
\end{figure}

In order to check the performance of the proposal  described in previous sections,  a well-known  wastewater treatment plant benchmark  has been used:  BSM2 \cite{Vrecko2006}  \footnote{http://iwa-mia.org/benchmarking/\#BSM2} \footnote{https://github.com/wwtmodels/Benchmark-Simulation-Models}. {\color{black} It  has been developed by International Water Association (IWA)  as a model of a complete WWTP, according to the existing real plants. But in any case, this proposal can involve any WWTP  with available data.}  A general overview of this model is shown in Fig. \ref{fig:BSM2}, in which the available variables used for fault diagnosis were:

\begin{enumerate}
		\item $DQO$, chemical oxygen demand.
		\item $S_O$, dissolved oxygen.
		\item $S_{ALK}$, alkalinity.
		\item $N$, Nitrogen. 
		\item $T_{SS}$, total suspended solids.
		\item $Flow rate$.
		\item $Temperature$.
\end{enumerate}

These variables are measured {\color{black}  at the input/output of the main WWTP components (20 points, as shown in Fig. \ref{fig:BSM2}),}   so 140 variables are monitored and considered for fault diagnosis. The plant usually works in healthy conditions, so the plant has been running for {\color{black}'5 years'} on changing environment conditions, being these data used as training/test data for the \textsl{NonFault} mode. {\color{black}This time period allows to collect everything  about the standard operation of the plant, but, in any case, if this \textsl{standard} operating mode changed drastically, the network could be  retrained by new data about the new condition.}

On the other hand, a set of 5 {\color{black}remarkable} faults  were checked in the plant. These are:

\begin{enumerate}
	\item Biomass in reactor 1, $BR_1$.
	\item $Q_r$ and $Q_w$ flow rates.
	\item Flow rate from primary clarifier to anaerobic digestor, $Q_{CaD}$.
	\item Oxygen transfer coefficient (actuator), $Kla$.
	\item Fault in the oxygen sensor, $O_2$.
\end{enumerate}

Every fault dataset  collects the monitoring of the plant over a three-day period, {\color{black} that fits with the process time and the diagnosis requirements}, the fault happens  at $t_i$ during this time period with magnitude, $M_i$, concerning from soft/incipient faults to severe faults,   and from  early to recent faults. Ten combinations of $t_i - M_i$, $i \in [1,...10]$,  for every $Fault_j$ are taking into account and,  for each of them,  10 runs were carried out to capture the possible randomness of the plant,  as well as to deal with the issue regarding  imbalanced fault/non-fault data. 

\textsl{NonFault} data about the plant was also collected,  as previously described:  five 1-year periods of non-fault plant working were compiled. The sampling time for every case was 15 minutes. Finally, the data compilation is made up of $505$ available  data sets about the plant behaviour in fault/non-fault cases. All these data sets  were  organized  according to  a \textsl{5-cross validation} scheme, as shown in Algorithm \ref{Alg:CV}.

\subsection{Data preparation and experiment setup}

\begin{algorithm}[t!]
		\caption{\textsl{5-Cross Validation} Data Partition}\label{Alg:CV}
    \hspace*{\algorithmicindent} \textbf{Input: }{NonFault  and Fault Data sets} \\
    \hspace*{\algorithmicindent} \textbf{Output: }{Homogeneous \textsl{5-Cross Validation} partition} 
    \begin{algorithmic}[1]
			\For{CV=[1, 2, 3, 4, 5]}    
					\State{CV $\Rightarrow$ $DataSetAppend(NonFaultDataSet(CV))$}
					\For{FaultType=[1, 2, 3, 4, 5]}  						\Comment{Types of fault: 5}
							\For{$j=1$ to  $j_{FaultType}=[1,...,10]$}		\Comment{Fault Time  \& Magnitude}									
								\State{NRuns=NumberOfRuns/Size(CV)}		
								\State{}   \Comment{$FaultType$, $t_j$,$M_j$ data sets }
								\State{}                \Comment{set to include in each CV}
								\For{Run=$(CV-1)*NRuns$ to $Run=[$(CV-1)*NRuns$,...,NRuns]$}									
									  
										\State{CV$\Rightarrow$$AppendFile(FaultTypeDataSet_Run, t_j,M_j)$}
								\EndFor
						\EndFor
					\EndFor
			\EndFor
    \end{algorithmic}
\end{algorithm}

\begin{algorithm}
\label{alg:window}
  \caption{Generation of the DL Network Inputs: Sliding Window and Labelling \label{alg:slidewin}}
  \hspace*{\algorithmicindent} \textbf{Input:}{ Input Data Set} \\
  \hspace*{\algorithmicindent} \textbf{Output:}{ Input Windows  in $Size=NumberOfVariables*TimeWindowSize$  } 
  \begin{algorithmic}[1]
			\For{WindowSize=[4, 8, 12]}
				\For{WindowStep=[1, 2, 3, 4]}
					\For{$DataSet$ in CV=[1, 2, 3, 4, 5]}    			    
						\While{$DataSet$} 
							\State{InputWindowGenerate($FDataSet$, NumberOfVariables, WindowSize)}
							\State{LabelForInputWindow  $\leftarrow$ ($TimePosition$ in $DataSetP$ $>$,$<$,$FaultTime$?)}
							\State{$Sequence Of Input Time Windows$= $Append(InputWindow, LabelForInputWindow)$ }
							\State{Advance/Slide $WindowStep$  in $DataSet$}							
					\EndWhile
			\EndFor
		\EndFor
	\EndFor
	\end{algorithmic}
\end{algorithm}	

%

Monitoring data were  standardized and, in accordance with the methodology described in Section \ref{sec:proposal},  a sliding window mechanism (see Fig. \ref{fig:slidingwindow}) was applied over the monitoring data to create a  window time sequence of monitoring data as input to the DCNN. Each and every one of these time windows is a data matrix with dimension $VariableNumber*WindowSize$,  where $VariableNumber$  is the number of monitoring variables and $WindowSize$ is the number of monitoring samples of all these variables  taken into account for making up the window. Every monitoring time,  the former window  is overlapped with the following in $WindowSize-WindowStep$ monitoring samples, where $WindowStep$ is the number of samples for sliding in  this procedure. Each of these monitoring time windows is labelled  as $Fault_j$ if they contain, at least,  one  monitoring sample of $Fault_j$, $j \in FaultTypes$  or, otherwise,  as $NonFault$.  This procedure is synthesized in Algorithm \ref{alg:slidewin}.

\label{sec:prepdata}
	\begin{algorithm}[t!]
    \caption{Fault Diagnosis Experimental Methodology}
		\label{Alg:FDI}
    \hspace*{\algorithmicindent} \textbf{Input: }{Labeled Sequence of Time Monitoring Windows } \\
    \hspace*{\algorithmicindent} \textbf{Output: }{DL Network Performace for Fault Diagnosis} 
    \begin{algorithmic}[1]
		 \For{$DLModel$ in $Models$=\{in Table \ref{tab:modelos}\}}
				\For{{\color{black}LearningRate=[0.1, 0.01, 0.001]}}
					\For{{\color{black}BatchSize=[32, 50, 64, 128, 256]}}
						\For{Epochs=[200, 500, 700]}
							\For{CV=[1, 2, 3, 4, 5]}				\Comment{5-Cross Validation scheme}
									\State{Training(DLModel, CV)}
									\State{Test(DLMoldel, CV)}
								\EndFor			
							\EndFor
						\EndFor	
				\EndFor
			\State{$Performance(Model)$: $Accuracy$, $Precision$, $Recall$, $F1-score$}
		\EndFor
		\State{DLModelPerformance for Fault Diagnosis}
   \end{algorithmic}
  \end{algorithm}

\begin{table}[t!]
\begin{center}
	\begin{tabular}{ |l|c |c |c| c| c| c|} \hline
					\textbf{Fault Types}						&$\mathit{NonFault}$	&$\mathit{BR_1}$  &$\mathit{Q_r  Q_w}$ &$\mathit{Q_{CaD}}$ &$\mathit{Kla}$ &$\mathit{O_2}$ \\ \hline \hline
					\textbf{Monitoring  Data  Windows} 	& 179540 		& 19620		&21550 		&20870		& 21740 & 23660 \\ \hline
	\end{tabular}
	\caption{Number of monitoring time windows for each $Fault_j$ or $NonFault$.}
	\label{tab:nwin}
	\end{center}
\end{table}

{\color{black}

On the other hand, the fault diagnosis approach and the DL network layouts considered in this work are shown in Table \ref{tab:modelos}, where the layer layout is  described in detail: the models are based on \cite{Wu2018} and \cite{Lecun1998}, lightweight models  including \textsl{linguistic OWA based pooling layers} (see Table \ref{tab:operaowa}). {\color{black}Model7 and Model3 \cite{Wu2018} are competitive models that have provided good results in the field of fault diagnosis, as well as having simple layer layouts. LeNet-5 \cite{Lecun1998} has also a relative simple layout and well performing in many fields  \cite{Elakkiya2024} \cite{Wang2024} \cite{Rai2020}.}

Several Linguist OWA cases such as \textsl{Most}, \textsl{AtLeastHalf}, \textsl{AtMiddle$_{\alpha}$}, \textsl{Average}, etc., based on  \cite{Liu2008},  are taken into account. The usual (\textsl{Max and Average}) pooling layers for Deep Learning approaches  are only two of these  linguistic OWA options. {\color{black}The parameter $\alpha$ can be tuned by  each user. In this case, this parameter has not been optimized, it has been tuned by a search grid and a  $cross-validation$ scheme following the linguistic meaning of the operators. \textsl{AtMiddle$_{\alpha}$} takes into account the \textsl{'most centered'}  of the features, discarding about $\alpha$(\%) of the lowest and $\alpha$(\%) of the highest, and \textsl{AtLeast$_{\alpha}$} takes into account  \textsl{''at least''} the $\alpha$-highest features.

{\color{black}
The diagnosis performance of each model is evaluated in terms of well-known performance indexes: $F1-score$, $Accuracy$, $Recall$ and $Precision$  \cite{powers2020}. Decisions are carried out based on $F1-score$, but any other can be taken into account if it is necessary. The hyperparameter tuning for every model was made by a grid search as shown in  Algorithm \ref{Alg:FDI} under $cross-validation$ scheme as shown in Table \ref{Alg:FDI}.}
}

\begin{table}[t!]
\centering	
\resizebox{17cm}{!} {
	\begin{tabular}{|l|l|}
		\hline
		\textbf{DL Model}& \textbf{Layer Layout}\\
		\hline \hline
		\textsl{\textbf{Model7-\cite{Wu2018}}}&Conv(64)-Conv(64)-MaxPool(2 × 2)-Conv(128)-MaxPool(2 × 1)-FC(300)-FC(6)\\
		\hline
		Model7-Pooling(Average)&Conv(64)-Conv(64)-AvgPool(2 × 2)-Conv(128)-AvgPool(2 × 1)-FC(300)-FC(6)\\
		\hline
		Model7-Pooling(Most)	&Conv(64)-Conv(64)-OWAPoolMost(2 × 2)-Conv(128)-OWAPoolMost(2 × 1)-FC(300)-FC(6)\\
		\hline
		Model7-Pooling(AtMiddle$_{\alpha}$ $\alpha=0.2$)&Conv(64)-Conv(64)-OWAPoolAtMiddle$_{\alpha}$02(2 × 2)-Conv(128)-OWAPoolAtMiddle$_{\alpha}$02(2 × 1)-FC(300)-FC(6)\\
		\hline
		Model7-Pooling(AtLeastHalf)&Conv(64)-Conv(64)-OWAPoolAtLeastHalf(2 × 2)-Conv(128)-OWAPoolAtLeastHalf(2 × 1)-FC(300)-FC(6)\\
		\hline
		Model7-Pooling(AtLeast$_{\alpha}$ $\alpha=0.75$)&Conv(64)-Conv(64)-OWAPoolAtLeast$_{\alpha}$075(2 × 2)-Conv(128)-OWAPoolAtLeast$_{\alpha}$075(2 × 1)-FC(300)-FC(6)\\
		\hline	\hline
		\textsl{\textbf{Model3-\cite{Wu2018}}}&Conv(128)-Conv(128)-Conv(128)-MaxPool(2 × 1)-FC(300)-FC(6)\\
		\hline
		Model3-Pooling(Average)&Conv(128)-Conv(128)-Conv(128)-AvgPool(2 × 1)-FC(300)-FC(6)\\
		\hline
		Model3-Pooling(Most)	&Conv(128)-Conv(128)-Conv(128)-OWAPoolMost(2 × 1)-FC(300)-FC(6)\\
		\hline
		Model3-Pooling(AtMiddle$_{\alpha}$$\alpha=0.2$)&Conv(128)-Conv(128)-Conv(128)-OWAPoolAtMiddle$_{\alpha}$02(2 × 1)-FC(300)-FC(6)\\
		\hline
		Model3-Pooling(AtLeastHalf)&Conv(128)-Conv(128)-Conv(128)-OWAPoolAtLeastHalf(2 × 1)-FC(300)-FC(6)\\
		\hline
		Model3-Pooling(AtLeast$_{\alpha}$ $\alpha=0.75$)&Conv(128)-Conv(128)-Conv(128)-OWAPoolAtLeast$_{\alpha}$075(2 × 1)-FC(300)-FC(6)\\
		\hline\hline
		\textsl{\textbf{LeNet-5 \cite{Lecun1998}}}&Conv(6)-MaxPool(2 × 2)-Conv(16)-MaxPool(2 × 2)-FC(120)-FC(84)-FC(6)\\
		\hline
		LeNet-5-Pooling(Average)&Conv(6)-AvgPool(2 × 2)-Conv(16)-AvgPool(2 × 2)-FC(120)-FC(84)-FC(6)\\
		\hline
		LeNet-5-Pooling(Most)	&Conv(6)-OWAPoolMost(2 × 2)-Conv(16)-OWAPoolMost(2 × 2)-FC(120)-FC(84)-FC(6)\\
		\hline
		LeNet-5-Pooling(AtMiddle$_{\alpha}$$\alpha=0.2$)&Conv(6)-OWAPoolAtMiddle$_{\alpha}$02(2 × 2)-Conv(16)-OWAPoolAtMiddle$_{\alpha}$02(2 × 2)-FC(120)-FC(84)-FC(6)\\
		\hline
		LeNet-5-Pooling(AtLeastHalf)&Conv(6)-OWAPoolAtLeastHalf(2 × 2)-Conv(16)-OWAPoolAtLeastHalf(2 × 2)-FC(120)-FC(84)-FC(6)\\
		\hline
		LeNet-5-Pooling(AtLeast$_{\alpha}$ $\alpha=0.75$)&Conv(6)-OWAPoolAtLeast$_{\alpha}$075(2 × 2)-Conv(16)-OWAPoolAtLeast$_{\alpha}$075(2 × 2)-FC(120)-FC(84)-FC(6)\\
		\hline	\hline
		
	\end{tabular}
}
\caption{Models summary. Baseline models: \textsl{Model7-\cite{Wu2018}},  \textsl{Model3-\cite{Wu2018}}  and \textsl{LeNet-5 \cite{Lecun1998}.}}
\label{tab:modelos}
\end{table}

A quantitative summary of the entire procedure applied to the available data  is shown in Table \ref{tab:nwin}, when $WindowSize=4$ and $WindowStep=1$. This table shows that the addressed fault diagnosis is strongly imbalanced, but any real-world plant is usually working in a {\color{black}$NonFault$} way.

}


\subsection{Experiments and results}
\label{sec_results}

This section  shows a brief summary of the fault diagnosis results  obtained by the proposed sliding window and the linguistic OWA pooling layers on the waste water treatment plant. First of all,  this fault diagnosis proposal has been compared  regarding \cite{Wu2018}. Next, using the proposed  sliding window mechanism,    the linguistic OWA operators are analyzed  for their use  in pooling layers for fault diagnosis. 
Here, three baselines have been considered: the two best models (Model 7 and Model 3) for fault diagnosis in  \cite{Wu2018} and the standard LeNet-5 model \cite{Lecun1998}. All these baseline models use \textsl{Max Pooling} layers, and they have also been implemented by linguistic OWA operators in their pooling layers, including the well-known \textsl{Average} operator/layer for deep learning. A brief summary of all these models is shown in Table \ref{tab:modelos}.

\begin{table}[t!]
	\hspace{-2cm}	
	\resizebox{19cm}{!} {
			\begin{tabular}{l||c|c|c||c|c|c||c|c|c||c|c|c||}		\cline{2-13}		\cline{2-13}
																								&\multicolumn{12}{|c||}{Fault Diagnosis}\\ \cline{2-13}
												&\multicolumn{3}{|c||}{Accuracy}				&	\multicolumn{3}{|c||}{Precision} 			  &\multicolumn{3}{|c||}{Recall}					 &\multicolumn{3}{|c||}{F1-score}  \\ \cline{2-13} \cline{2-13}

	\textbf{Approach:}		 							&Acc.	&$\triangle(\%)$ &Epochs					& Pre.  & $\triangle(\%)$ & Epochs 			  &Rcll.& $\triangle(\%)$	&Epochs				&F1	&$\triangle(\%)$ & Epochs 
	\\\hline \hline		
	 \textsl{Hao Wu} in Model7 \cite{Wu2018}					&$0.88$  &$-$ &$500$						&$0.82$  &$-$  &$700$					&$0.78$ &$-$ &$500$						&$0.77$ &$-$ &$500$	
	\\ \hline
	Model7- Sliding Window +Pooling(AtMiddle$_{\alpha}$$\alpha=0.2$)			&$\textbf{0.94}$ & $\textbf{6.82}$ &$700$			&$0.94$ &$14.63$ &$700$				&$\textbf{0.91}$ &$\textbf{16.67}$ &$700$			&$\textbf{0.92}$ &$\textbf{19.48}$ &$700$			
	\\ \hline
	Model3- Sliding Window +Pooling(Most)						&$0.94$ & $6.82$ &$700$					&$\textbf{0.94}$ &$\textbf{14.63}$ &$700$		&$0.91$ &$16.67$ &$700$					&$0.92$ &$19.48$ &$700$			
	\\ \hline
	\end{tabular}
	}
	\caption{Fault diagnosis: Sliding Window ($WindowSize=4$, $WindowStep=1$) and Linguistic OWA pooling proposal regarding \cite{Wu2018}.}
		\label{tab:compara}
\end{table}

\begin{table}[t!]
	\hspace{-2cm}	
	\resizebox{19cm}{!} {
			\begin{tabular}{l||c|c|c|c|c|c||}		\cline{2-7}		\cline{2-7}
																								&\multicolumn{6}{|c||}{F1-score}\\ \cline{2-7}

	\textbf{Approach:}		 							&\textsl{NonFault}		&$BR_1$ 		&$Q_r$$Q_w$		& $Q_{CaD}$ 		 & $Kla$ 		& $O_2$ 				
	\\\hline \hline		
	 \textsl{Hao Wu} in Model7 \cite{Wu2018}					&$0.96$  			&$0.97$ 		&$1.00$			&$0.87$  			&$0.34$  		&$0.50$					
	\\ \hline
	Model7- Sliding Window +Pooling(AtMiddle$_{\alpha}$$\alpha=0.2$)			&$\textbf{0.96}$  		&$0.98$ 		&$1.00$			&$0.99$  			&$0.73$  		&$\textbf{0.84}$						
	\\ \hline
	Model3- Sliding Window +Pooling(Most)						&$0.96$  			&$\textbf{0.98}$ 	&$\textbf{1.00}$		&$\textbf{1.00}$  		&$\textbf{0.74}$  	&$0.82$				
	\\ \hline
	\end{tabular}
	}
	\caption{F1-score: Sliding Window ($WindowSize=4$, $WindowStep=1$) and Linguistic OWA pooling proposal regarding \cite{Wu2018}.}
	\label{tab:comparaf1}
\end{table}

\begin{table}[t!]
	\hspace{-2cm}	
	\resizebox{19cm}{!} {
			\begin{tabular}{l||c|c|c|c|c|c||}		\cline{2-7}		\cline{2-7}
																								&\multicolumn{6}{|c||}{Precision}\\ \cline{2-7}

	\textbf{Approach:}		 							&\textsl{NonFault}		&$BR_1$ 		&$Q_r$$Q_w$		& $Q_{CaD}$ 		 & $Kla$ 		& $O_2$ 				
	\\\hline \hline		
	 \textsl{Hao Wu} in Model7 \cite{Wu2018}					&$0.95$  			&$1.00$ 		&$1.00$			&$1.00$  			&$0.39$  		&$0.56$					
	\\ \hline
	Model7- Sliding Window +Pooling(AtMiddle$_{\alpha}$$\alpha=0.2$)			&$\textbf{0.95}$  		&$1.00$ 		&$1.00$			&$1.00$  			&$\textbf{0.78}$  	&$0.90$						
	\\ \hline
	Model3- Sliding Window +Pooling(Most)						&$0.95$  			&$1.00$ 		&$1.00$			&$\textbf{1.00}$  		&$0.71$  		&$\textbf{0.99}$				
	\\ \hline
	\end{tabular}
	}
	\caption{Precision: Sliding Window ($WindowSize=4$, $WindowStep=1$) and Linguistic OWA pooling proposal regarding \cite{Wu2018}.}
 \label{tab:comparaprecision}
\end{table}

\begin{table}[t!]
	\hspace{-2cm}	
	\resizebox{19cm}{!} {
			\begin{tabular}{l||c|c|c|c|c|c||}		\cline{2-7}		\cline{2-7}
																								&\multicolumn{6}{|c||}{Recall}\\ \cline{2-7}

	\textbf{Approach:}		 							&\textsl{NonFault}		&$BR_1$ 		&$Q_r$$Q_w$		& $Q_{CaD}$ 		 & $Kla$ 		& $O_2$ 				
	\\\hline \hline		
	 \textsl{Hao Wu} in Model7 \cite{Wu2018}					&$0.97$  			&$0.94$ 		&$\textbf{1.00}$		&$0.77$  			&$0.40$  		&$0.63$					
	\\ \hline
	Model7- Sliding Window +Pooling(AtMiddle$_{\alpha}$$\alpha=0.2$)			&$0.97$  			&$0.96$ 		&$1.00$			&$0.99$  			&$0.70$  		&$\textbf{0.83}$						
	\\ \hline
	Model3- Sliding Window +Pooling(Most)						&$\textbf{0.97}$  		&$\textbf{0.96}$ 	&$1.00$			&$\textbf{0.99}$  		&$\textbf{0.78}$  	&$0.73$				
	\\ \hline
	\end{tabular}
	}
	\caption{Recall: Sliding Window ($WindowSize=4$, $WindowStep=1$) and Linguistic OWA pooling proposal regarding \cite{Wu2018}.}
  \label{tab:compararecall}
\end{table}

Table \ref{tab:compara} contains the fault diagnosis results,    showing  the proposal detailed in Section \ref{sec:proposal}    regarding  \cite{Wu2018},  taking into account its Model7 and Model3 DL  layouts, but  implemented by linguistic OWA operators in pooling layers,  as well as the sliding time window with variable size and step. This table shows the best result for the  performance indexes ($Accuracy, Precision,  Recall, F1-score$). In fact, the best tuned hyper-parameters were: $WindowSize=4$, $WindowStep=1$ and AtMiddle$_{\alpha}(\alpha=0.2)$ for Model7   and  $Most$ for Model3 as  linguistic OWA pooling operators, both cases  improving all the baseline performance indexes,  from around $6\%$ up to almost $20\%$ for \textsl{F1-Score}.
Table \ref{tab:comparaf1} shows the diagnosis for every fault type  based on \textsl{F1-score}: linguistic OWA pooling  offers better results in all cases,  the Model7  layout is once more the best one, in particular its higher performance  in $Kla$ and  $O_2$ faults. Quite similar remarks can be observed  in Tables \ref{tab:comparaprecision} and  \ref{tab:compararecall}, taking into account the \textsl{Precision} and the \textsl{Recall} indexes.

When the diagnosis is focused on the most difficult faults, such as $Kla$ and $O_2$ in this case, this proposal also clearly outperforms  the approach in \cite{Wu2018}, as shown in Tables \ref{tab:compalka} and \ref{tab:compalco2}. In Fig. \ref{fig:compara}, all these global results are graphically displayed.

\begin{table}[t!]
	\hspace{-2cm}	
	\resizebox{19cm}{!} {
	\begin{tabular}{l||c|c|c||c|c|c||c|c|c||}		\cline{2-10}		\cline{2-10}
																								&\multicolumn{9}{|c||}{Fault Diagnosis}\\ \cline{2-10}		
											&\multicolumn{3}{|c||}{Precision} 			 				 &\multicolumn{3}{|c||}{Recall}									 &\multicolumn{3}{|c||}{F1-score}  \\ \cline{2-10} \cline{2-10}
	\textbf{Models:}		 						& Pre. 	 &$\triangle(\%)$	& Epochs 						&Rcll.		&$\triangle(\%)$	&Epochs							&F1		&$\triangle(\%)$	& Epochs
	\\\hline \hline	
	\textsl{Hao Wu} in Model7 \cite{Wu2018}				&$0.39$ 		&$-$		& $700$ 						&$0.40$ 		&$-$		&$500$ 							&$0.34$ 		&$-$		& $500$
	\\ \hline
	Model7- Sliding Window +Pooling(AtMiddle$_{\alpha}$$\alpha=0.2$)		&$\textbf{0.78}$	&$\textbf{100.00}$	&$700$				&$0.70$		&$75.00$	&$700$							&$0.73$		&$114.71$	&$700$	
	\\ \hline
	Model3- Sliding Window +Pooling(Most)					&$0.71$	&$82.05$	&$700$							&$\textbf{0.78}$	&$\textbf{95.00}$	&$700$						&$\textbf{0.74}$	&$\textbf{117.65}$	&$700$	
	\\ \hline
	\end{tabular}
	}
	\caption{$Kla$ Fault diagnosis: Sliding Window ($WindowSize=4$, $WindowStep=1$) and Linguistic OWA pooling proposal regarding \cite{Wu2018}.}
	\label{tab:compalka}
\end{table}

\begin{table}[t!]
	\hspace{-2cm}	
	\resizebox{19cm}{!} {
	\begin{tabular}{l||c|c|c||c|c|c||c|c|c||}		\cline{2-10}		\cline{2-10}
																								&\multicolumn{9}{|c||}{Fault Diagnosis}\\ \cline{2-10}		
											&\multicolumn{3}{|c||}{Precision} 			 				 &\multicolumn{3}{|c||}{Recall}									 &\multicolumn{3}{|c||}{F1-score}  \\ \cline{2-10} \cline{2-10}
	\textbf{Models:}		 						& Pre. 	 &$\triangle(\%)$	& Epochs 						&Rcll.		&$\triangle(\%)$	&Epochs							&F1		&$\triangle(\%)$	& Epochs	
	\\\hline \hline	
	\textsl{Hao Wu} in Model7 \cite{Wu2018}				&$0.56$ 		&$-$		& $700$ 						& $0.67$ 		&$-$		& $700$							 & $0.52$   		&$-$		& $700$
	\\ \hline
	Model7- Sliding Window +Pooling(AtMiddle$_{\alpha}$$\alpha=0.2$)		&$0.90$ 	&$60.71$	&$700$					&$0.86$ 	&$28.36$	&$500$								&0.84$$ 		&$61.54$	&$700$			
	\\ \hline
	{\color{black}Model3- Sliding Window +Pooling(AtLeast$_{\alpha}$$\alpha=0.75$)}		&{\color{black}$\textbf{0.91}$} &{\color{black}$\textbf{62.50}$}&$700$		&{\color{black}$\textbf{0.86}$} &{\color{black}$\textbf{28.36}$}	&{\color{black}$500$}			&{\color{black}\textbf{$\textbf{0.87}$}}&{\color{black}\textbf{$\textbf{67.31}$}}&$700$			
	\\ \hline
	\end{tabular}
	}
	\caption{$O_2$ Fault diagnosis: Sliding Window ($WindowSize=4$, $WindowStep=1$) and Linguistic OWA pooling proposal regarding \cite{Wu2018}.}
	\label{tab:compalco2}
\end{table}

In fact,  the sliding time window has allowed the plant to be diagnosed at every $WindowStep$, 15 minutes when $WindowStep=1$ and $T_s$=15 minutes,  with as little monitoring data  as  the last $WindowSize$ ($WindowSize=4 \rightarrow$ 4 monitoring data samples), achieving a  better performance than \cite{Wu2018}, where  a fault detection approach is applied to the  TE process. Other different values for $WindowStep$  were also checked  by a grid search  without finding a better performance and even, in some cases, with  much worse results,  except  for the  computation time  for higher  $WindowSize$ or  $WindowStep$.  

Moreover, here, the number of monitorized process variables is much larger (140) and the number of samples for the window is  shorter  than in \cite{Wu2018}, obtaining   a diagnosis at every  $Ts*WindowStep$, in this best case every 15 minutes, which can allow incipient faults to be addressed in the short term,  with a high level of diagnosis on difficult $Kla$ and $O_2$ faults.

\begingroup

\setlength{\tabcolsep}{1pt} 
\renewcommand{\arraystretch}{0.3} 

\setlength{\tabcolsep}{0pt}
	\begin{figure}[t!]
	\hspace{-3.2cm}
	\begin{tabular}{lll}
		\includegraphics[width=0.35\textheight]{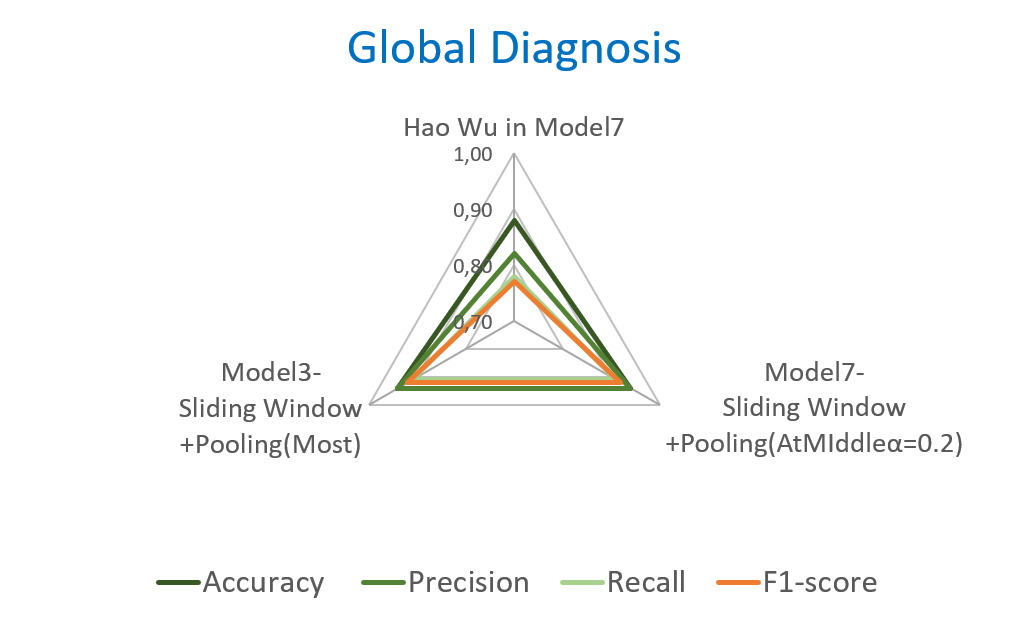} &  \includegraphics[width=0.35\textheight]{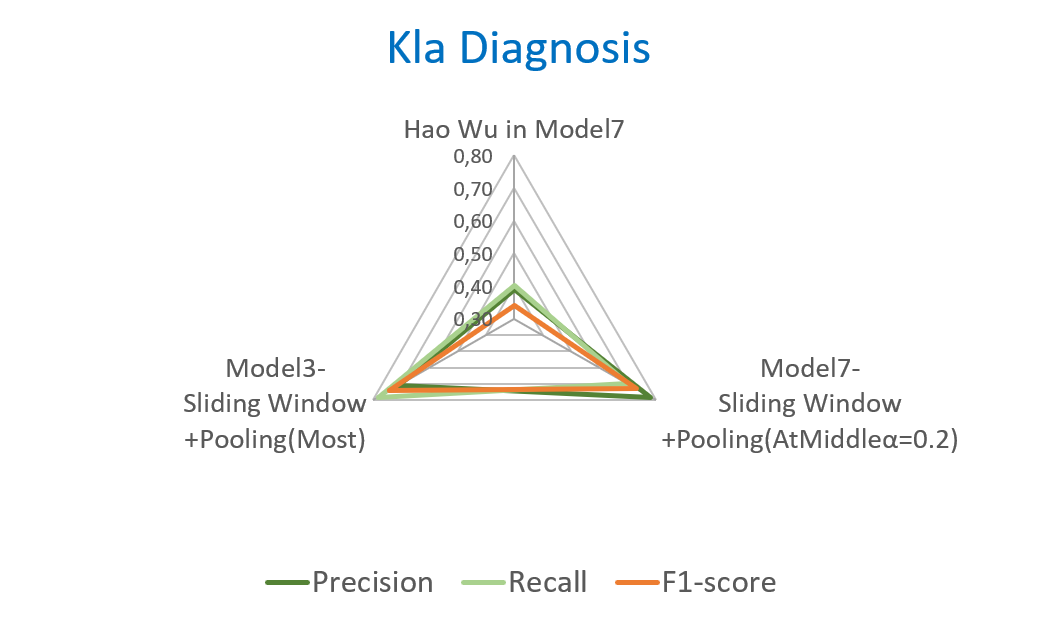} &\includegraphics[width=0.35\textheight]{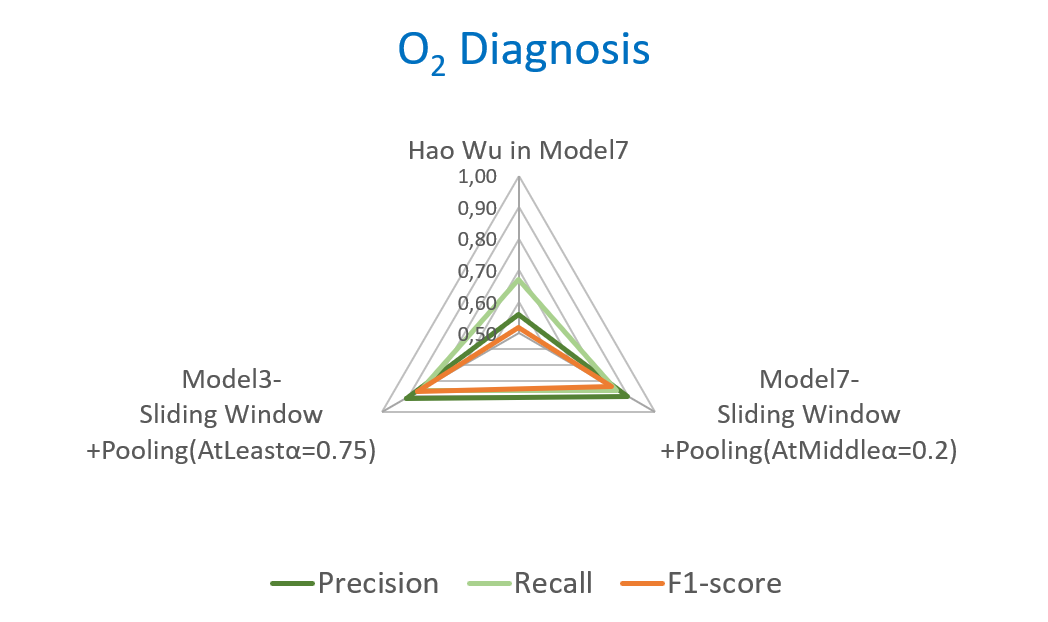}  \\
	\end{tabular}
		\caption{Sliding Time Window \& Linguistic OWA based fault diagnosis.}
		\label{fig:compara}
	\end{figure}

\endgroup

\subsection{Comparison of standard pooling operators vs linguistic OWA operators in fault diagnosis}

Another goal of this work was to check  out  the linguistic OWA operators for pooling in fault diagnosis, as  previously described in Section \ref{sec:owa}, regarding the \textsl{Max} and \textsl{Average} based standard pooling layers. 

Here the proposal, as  described in Section \ref{sec:proposal} (with $WindowStep=1$  and $WindowSize=4$), and the  layouts referenced by  Model7 and Model 3 in \cite{Wu2018}  and LeNet-5 \cite{Lecun1998}, are used for  checking  out the performance of the linguistic OWA operators with regard to  the \textsl{Max} and \textsl{Average} standard pooling layers in  fault diagnosis. 

\subsubsection{Model7}
Considering the layout of Model7, the different linguistic OWA operators have been implemented in corresponding pooling  layers, and compared with regard to the Model7 versions containing the \textsl{max} and \textsl{average} standard pooling layers.

\begin{table}[t!]
	\hspace{-2cm}	
	\resizebox{19cm}{!} {
	\begin{tabular}{l||c|c|c||c|c|c||c|c|c||c|c|c||}		\cline{2-13}		\cline{2-13}
																								&\multicolumn{12}{|c||}{Fault Diagnosis}\\ \cline{2-13}
												&\multicolumn{3}{|c||}{Accuracy}				&	\multicolumn{3}{|c||}{Precision} 			  &\multicolumn{3}{|c||}{Recall}					 &\multicolumn{3}{|c||}{F1-score}  \\ \cline{2-13} \cline{2-13}
	\textbf{Models:}		 	&Acc.	&$\triangle(\%)$ &Epochs					& Pre.  & $\triangle(\%)$ & Epochs 			  &Rcll.& $\triangle(\%)$	&Epochs				&F1	&$\triangle(\%)$ & Epochs  \\\hline \hline	
	\textit{Model7-MaxPooling}\cite{Wu2018} &$\mathit{0.91}$ &$-$&$\mathit{700}$				&$\mathit{0.84}$&$-$&$\mathit{700}$			&$\mathit{0.84}$&$-$&$\mathit{700}$				&$\mathit{0.83}$&$-$&$\mathit{700}$ 	\\ \hline
	
	Model7-Pooling(Average)	&$\mathbf{0.91}$&$\mathbf{0.00}$&$	500$			&$\mathbf{0.91}$&$\mathbf{8.33}$&$700$		&$\mathbf{0.85}$&$\mathbf{1.19}$&$	700$			   &$\mathbf{0.85}$&$\mathbf{2.41}$&$	700$	 	\\ \hline
	
	Model7-Pooling(Most)	&$\mathbf{0.93}$&$\mathbf{2.20}$&$\mathbf{200}$	&$\mathbf{0.91}$&$\mathbf{8.33}$&$	500$	&$\mathbf{0.88}$&$\mathbf{4.76}$&$\mathbf{200}$	 &$\mathbf{0.89}$&$\mathbf{7.23}$&$\mathbf{200}$  	\\ \hline
	
	Model7-Pooling(AtMiddle$_{\alpha}$$\alpha=0.2$)	&\textcolor{teal}{$\mathbf{0.94}$}&\textcolor{teal}{$\mathbf{3.30}$}&$700$			&\textcolor{teal}{$\mathbf{0.94}$}&\textcolor{teal}{$\mathbf{11.90}$}&$700$				&\textcolor{teal}{$\mathbf{0.91}$}&\textcolor{teal}{$\mathbf{8.33}$}&$700$				&\textcolor{teal}{$\mathbf{0.92}$}&\textcolor{teal}{$\mathbf{10.84}$}&$700$	
	\\ \hline
	
	Model7-Pooling(AtLeastHalf)							&$\mathbf{0.92}$&$\mathbf{1.10}$&$	500$			&$\mathbf{0.90}$&$\mathbf{7.14}$&$	700$		&$\mathbf{0.86}$&$\mathbf{2.38}$&$	500$			&$\mathbf{0.86}$&$\mathbf{3.61}$&$	500$	
	\\ \hline
	
	Model7-Pooling(AtLeast$_{\alpha}$ $\alpha=0.75$)					&$\mathbf{0.93}$&$\mathbf{2.20}$&$	500$			&$\mathbf{0.90}$&$\mathbf{7.14}$&$	700$		&$\mathbf{0.87}$&$\mathbf{3.57}$&$	500$			&$\mathbf{0.88}$&$\mathbf{6.02}$&$	500$	
	\\ \hline
	
		\end{tabular}
	}
	\caption{Fault diagnosis ($WindowSize=4$, $WindowStep=1$): Linguistic OWA pooling vs. Max pooling in Model7.}	
	\label{tab:maxcomparaM7}
\end{table}

\begin{table}[t!]
\hspace{-2cm}
\resizebox{19cm}{!} {
	\begin{tabular}{l||c|c|c||c|c|c||c|c|c||}		\cline{2-10}		\cline{2-10}
																								&\multicolumn{9}{|c||}{Fault Diagnosis}\\ \cline{2-10}		
											&\multicolumn{3}{|c||}{Precision} 			 				 &\multicolumn{3}{|c||}{Recall}									 &\multicolumn{3}{|c||}{F1-score}  \\ \cline{2-10} \cline{2-10}
	\textbf{Models:}		 						& Pre.  & $\triangle(\%)$ & Epochs 							  &Rcll.& $\triangle(\%)$	&Epochs								&F1	&$\triangle(\%)$ & Epochs 
	\\\hline \hline	
	 \textit{Model7-MaxPooling}\cite{Wu2018}				&$\mathit{0.72}$ & $- $& $\mathit{700} $						&\textcolor{teal}{\textbf{ $\mathit{0.86}$}} &$ -$ & $\mathit{500}$			 &$\mathit{ 0.65}$ &$ -$ & $\mathit{500}$
	\\ \hline
	Model7-Pooling(Average)							&$\mathbf{ 0.93}$ & $\mathbf{29.17}$ & $700$ 					&  $0.66 $ &  $-23.26 $ &  $500 $ 								&  $0.62 $ &  $-4.62 $ &  $500 $
	\\ \hline
	Model7-Pooling(Most)							&\textcolor{teal}{$\mathbf{0.96}$} &\textcolor{teal}{$\mathbf{ 33.33}$} &$ 500$ 	     &  $0.72 $ &  $-16.28 $ &  $700 $							 &$\mathbf{0.76}$ &$\mathbf{ 16.92}$ & $\mathbf{200}$ 
	\\ \hline
	Model7-Pooling(AtMiddle$_{\alpha}$$\alpha=0.2$)				&$\mathbf{0.90} $&$\mathbf{25.00} $& 700						 &  $0.86 $ &  $0.00 $ &  $500 $ 								& \textcolor{teal}{$\mathbf{0.84}$} &\textcolor{teal}{$\mathbf{29.23}$} & $700$
	\\ \hline
	Model7-Pooling(AtLeastHalf)						&$\mathbf{ 0.85}$ &$\mathbf{18.06 }$& $700$ 					& $ 0.75 $ &  $-12.79 $ &  $200 $								&$\mathbf{0.68 }$& $\mathbf{4.62}$& $\mathbf{200}$
	\\ \hline
	Model7-Pooling(AtLeast$_{\alpha}$ $\alpha=0.75$)				&$\mathbf{ 0.88} $&$\mathbf{ 22.22} $&$ 700$ 					& $0.82 $ &  $-4.65 $ &  $500 $ 									& $\mathbf{ 0.79 }$&$\mathbf{ 21.54} $& $500$
	\\ \hline
	\end{tabular}
	}
	\caption{$O_2$ Fault diagnosis ($WindowSize=4$, $WindowStep=1$): Linguistic OWA pooling vs. Max pooling in Model7. }
	\label{tab:maxco27}
\end{table}

\begin{table}[t!]
\hspace{-2cm}
\resizebox{19cm}{!} {
	\begin{tabular}{l||c|c|c||c|c|c||c|c|c||}		\cline{2-10}		\cline{2-10}
																								&\multicolumn{9}{|c||}{Fault Diagnosis}\\ \cline{2-10}		
											&\multicolumn{3}{|c||}{Precision} 			 				 &\multicolumn{3}{|c||}{Recall}									 &\multicolumn{3}{|c||}{F1-score}  \\ \cline{2-10} \cline{2-10}
	\textbf{Models:}		 						& Pre.  & $\triangle(\%)$ & Epochs 							  &Rcll.& $\triangle(\%)$	&Epochs								&F1	&$\triangle(\%)$ & Epochs 
	\\\hline \hline	
	 \textit{Model7-MaxPooling}\cite{Wu2018}				&$\mathit{0.47}$ & $-$ &$\mathit{ 500}$ 						&$\mathit{0.46 }$& $-$ &$\mathit{200}$ 							&$\mathit{0.41}$ & $- $&$\mathit{ 200}$
	\\ \hline
	Model7-Pooling(Average)							&$\mathbf{0.57}$ & $\mathbf{21.28}$ & $500$ 					&$\mathbf{0.79}$ & $\mathbf{71.74}$ & $700$ 						&$\mathbf{ 0.65}$ & $\mathbf{58.54}$ & $700$ 
	\\ \hline
	Model7-Pooling(Most)							&$\mathbf{0.65} $& $\mathbf{38.30 }$& $\mathbf{200}$				& $\mathbf{0.76}$ & $\mathbf{65.22}$ & $\mathbf{200}$ 					& $\mathbf{0.70}$ & $\mathbf{70.73}$ & $\mathbf{200}$
	\\ \hline
	Model7-Pooling(AtMiddle$_{\alpha}$$\alpha=0.2$)				&\textcolor{teal}{$\mathbf{0.78}$} &\textcolor{teal}{$\mathbf{ 65.96}$} & $700$ 	& $\mathbf{0.70}$ &$\mathbf{52.17}$ & $700$ 					& \textcolor{teal}{$\mathbf{0.73}$} &\textcolor{teal}{$\mathbf{78.05 }$}& $700$
	\\ \hline
	Model7-Pooling(AtLeastHalf)						&$\mathbf{0.73}$ & $\mathbf{55.32}$& $\mathbf{200}$ 				& \textcolor{teal}{$\mathbf{0.80 }$}& \textcolor{teal}{$\mathbf{73.91}$} & $700$   & $\mathbf{0.66}$ &$\mathbf{60.98}$ &$500$  
	\\ \hline
	Model7-Pooling(AtLeast$_{\alpha}$ $\alpha=0.75$)				&$\mathbf{0.61}$ &$\mathbf{29.79}$ & $500$ 					& $\mathbf{0.70}$ & $\mathbf{52.17}$ & $700$ 						&$\mathbf{0.64 }$&$\mathbf{56.10}$ & $700$
	\\ \hline
	\end{tabular}
	}
	\caption{$Kla$ Fault diagnosis ($WindowSize=4$, $WindowStep=1$): Linguistic OWA pooling vs. Max pooling in Model7. }
	\label{tab:maxkla7}
\end{table}
\begin{itemize}

	\item \textbf{with regard to the \textsl{Max} pooling layer:}

				Table \ref{tab:maxcomparaM7} shows the results obtained when the different linguistic OWA operators were implemented in pooling layers of the Model7 layout. In this case, the diagnosis is improved by all the linguistic OWA operators for all the performance indexes,  even with a lower training,  such as for the \textsl{Most} operator which  outperforms the \textsl{Max} based  standard pooling in 200 epochs. The best linguistic operator was \textsl{AtMiddle$_{\alpha}$($\alpha=0.2$)}, improving most of the indexes from $3.30\%$ up to $10.84\%$.
	
			There are two of the faults with a more difficult diagnosis:  $O_2$ and $Kla$.  The diagnosis focused on these two faults is shown in Tables \ref{tab:maxco27} and \ref{tab:maxkla7}. Here, the $O_2$ fault has the better \textsl{Recall} for the \textsl{Max} pooling, but \textsl{F1-score} is better for the alternative OWA based pooling operators than for the \textsl{Max} case, \textsl{AtMiddle$_{\alpha}$($\alpha=0.2$)} once more being the best. It is similar for  \textsl{Precision}, where \textsl{Max} is the worst. Focusing on the $Kla$ fault, the alternative linguistic OWA based operators outperform the original \textsl{Max}  pooling. \textsl{AtMiddle$_{\alpha}$($\alpha=0.2$}) is once more the one that gives the best \textsl{F1-score} and \textsl{Precision} performance.

\begin{table}[t!]
\hspace{-2cm}
\resizebox{19cm}{!} {
	\begin{tabular}{l||c|c|c||c|c|c||c|c|c||c|c|c||}		\cline{2-13}		\cline{2-13}
																								&\multicolumn{12}{|c||}{Fault Diagnosis}\\ \cline{2-13}
												&\multicolumn{3}{|c||}{Accuracy}				&	\multicolumn{3}{|c||}{Precision} 			  &\multicolumn{3}{|c||}{Recall}					 &\multicolumn{3}{|c||}{F1-score}  \\ \cline{2-13} \cline{2-13}
	\textbf{Models:}		 							&Acc.	&$\triangle(\%)$ &Epochs					& Pre.  & $\triangle(\%)$ & Epochs 			  &Rcll.& $\triangle(\%)$	&Epochs				&F1	&$\triangle(\%)$ & Epochs 
	\\\hline \hline	
	Model7-MaxPooling\cite{Wu2018}						&$0.91$ & $0.00 $& $700$ 					& $0.84$ & $-7.69$ & $700$ 				  & $0.84$ & $-1.18$ & $700$ 					& $0.83$ & $-2.35$ & $700$
	\\ \hline
	\textit{Model7-AveragePooling}							&$\mathit{0.91}$ & $-$ &$\mathit{500}$ 			&$\mathit{0.91}$ & $-$ &$\mathit{700}$		  &$\mathit{0.85}$ &$ -$ & $\mathit{700}$ 			&$\mathit{0.85}$ &$ - $& $\mathit{700}$
	\\ \hline
	Model7-Pooling(Most)								&$\mathbf{0.93}$ & $\mathbf{2.20 }$& $\mathbf{200}$	& $\mathbf{0.91}$ &$\mathbf{ 0.00}$ & 500 		  &$\mathbf{0.88} $& $\mathbf{3.53}$ & $\mathbf{200}$ 	& $\mathbf{0.89} $&$\mathbf{4.71}$ & $\mathbf{200}$
	\\ \hline
	Model7-Pooling(AtMiddle$_{\alpha}$$\alpha=0.2$)					&\textcolor{teal}{$\mathbf{0.94}$} & \textcolor{teal}{$\mathbf{3.30}$} & $700$ 		&\textcolor{teal}{$\mathbf{0.94}$} &\textcolor{teal}{$\mathbf{3.30}$ }& $700$ 			& \textcolor{teal}{$\mathbf{0.91}$} & \textcolor{teal}{$\mathbf{7.06} $}& $700 $		& \textcolor{teal}{$\mathbf{0.92}$} &\textcolor{teal}{$\mathbf{8.24}$} & $700$
	\\ \hline
	Model7-Pooling(AtLeastHalf)							&$\mathbf{0.92}$ & $\mathbf{1.10}$ & $500$		 & $0.90$ & $-1.10$ & $700$ 				  & $\mathbf{0.86}$ &$\mathbf{1.18}$ & $500$ 		  & $\mathbf{0.86}$ &$\mathbf{1.18}$ & $500$
	\\ \hline
	Model7-Pooling(AtLeast$_{\alpha}$ $\alpha=0.75$)					&$\mathbf{0.93 }$& $\mathbf{2.20}$ & $500$ 		& $0.90$ & $-1.10$ & $700 $				  & $\mathbf{0.87}$ &$\mathbf{2.35}$ &$ 500 $		  &$\mathbf{0.88}$ &$\mathbf{3.53}$ & $500$
	\\ \hline
	\end{tabular}
	}
	\caption{Fault diagnosis ($WindowSize=4$, $WindowStep=1$): Linguistic OWA pooling vs. Average pooling in Model7.}	
	\label{tab:averco7}
\end{table}

\begin{table}[t!]
\hspace{-2cm}
\resizebox{19cm}{!} {
	\begin{tabular}{l||c|c|c||c|c|c||c|c|c||}		\cline{2-10}		\cline{2-10}
																								&\multicolumn{9}{|c||}{Fault Diagnosis}\\ \cline{2-10}		
											&\multicolumn{3}{|c||}{Precision} 			 				 &\multicolumn{3}{|c||}{Recall}									 &\multicolumn{3}{|c||}{F1-score}  \\ \cline{2-10} \cline{2-10}
	\textbf{Models:}		 						& Pre.  & $\triangle(\%)$ & Epochs 							  &Rcll.& $\triangle(\%)$	&Epochs								&F1	&$\triangle(\%)$ & Epochs 
	\\\hline \hline	
	Model7-MaxPooling\cite{Wu2018}					&$0.72$ & $-22.58$ & $700$ 								&\textcolor{teal}{$\mathbf{ 0.86 }$}& \textcolor{teal}{$\mathbf{30.30}$} & $500$	 & $\mathbf{0.65}$ &$\mathbf{4.84 }$& $500$
	\\ \hline
	 \textit{Model7-AveragePooling}						&$\mathit{0.93}$ &$ -$ & $\mathit{700}$ 						& $\mathit{0.66}$ & $-$ & $\mathit{500}$							 & $\mathit{0.62}$ & $-$ & $\mathit{500}$
	\\ \hline
	Model7-Pooling(Most)							&\textcolor{teal}{$\mathbf{0.96}$} &\textcolor{teal}{$\mathbf{ 3.23}$} & $500$ 		&$\mathbf{ 0.72}$ & $\mathbf{9.09}$ & $700$ 				& $\mathbf{ 0.76}$&$\mathbf{22.58 }$ &$\mathbf{200 }$
	\\ \hline
	Model7-Pooling(AtMiddle$_{\alpha}$$\alpha=0.2$)				&$0.90$ & $-3.23$ & $700$								& $\mathbf{ 0.86}$ & $\mathbf{30.30}$ &$500$						 &\textcolor{teal}{$\mathbf{0.84 }$}& \textcolor{teal}{$\mathbf{35.48}$} & $700$
	\\ \hline
	Model7-Pooling(AtLeastHalf)						&$0.85$ & $-8.60$ & $700$								&$\mathbf{ 0.75} $& $\mathbf{13.64} $& $\mathbf{200}$ 					&$\mathbf{ 0.68}$ &$\mathbf{ 9.68}$ & $\mathbf{200}$
	\\ \hline
	Model7-Pooling(AtLeast$_{\alpha}$ $\alpha=0.75$)				&$0.88$ & $-5.38$ & $700$ 								&$\mathbf{ 0.82} $& $\mathbf{24.24 }$& $500$ 						&$\mathbf{0.79 }$&$\mathbf{ 27.42}$ & $500$
	\\ \hline
	\end{tabular}
	}
	\caption{$O_2$ Fault diagnosis: Linguistic OWA pooling vs. Average pooling in Model7. }
	\label{tab:averagelco2}
\end{table}

\begin{table}[t!]
\hspace{-2cm}
\resizebox{19cm}{!} {
	\begin{tabular}{l||c|c|c||c|c|c||c|c|c||}		\cline{2-10}		\cline{2-10}
																								&\multicolumn{9}{|c||}{Fault Diagnosis}\\ \cline{2-10}		
											&\multicolumn{3}{|c||}{Precision} 			 				 &\multicolumn{3}{|c||}{Recall}									 &\multicolumn{3}{|c||}{F1-score}  \\ \cline{2-10} \cline{2-10}
	\textbf{Models:}		 						& Pre.  & $\triangle(\%)$ & Epochs 							  &Rcll.& $\triangle(\%)$	&Epochs								&F1	&$\triangle(\%)$ & Epochs 
	\\\hline \hline	
	Model7-MaxPooling\cite{Wu2018}					&$0.47$ & $-17.54$ & $500$ 								& $0.46$ & $-41.77$ & $\mathbf{200}$ 							& $0.41$ & $-36.92$ & $\mathbf{200}$
	\\ \hline
	\textit{Model7-AveragePooling}						&$\mathit{0.57}$ &$ $- & $\mathit{500}$ 						&$\mathit{0.79}$ & $-$ &$\mathit{700}$ 							&$\mathit{0.65}$ & $-$ & $\mathit{700}$
	\\ \hline
	Model7-Pooling(Most)							&$\mathbf{0.65}$ & $\mathbf{14.04}$ & $\mathbf{200}$				 & $0.76$ & $-3.80$ & $\mathbf{200}$ 							&$\mathbf{0.70}$ & $\mathbf{7.69}$ & $\mathbf{200}$
	\\ \hline
	Model7-Pooling(AtMiddle$_{\alpha}$$\alpha=0.2$)				&\textcolor{teal}{$\mathbf{0.78}$} & \textcolor{teal}{$\mathbf{36.84}$} & $700$ 		&$ 0.70$ & $-11.39$ & $700$ 							&\textcolor{teal}{$\mathbf{0.73}$} &\textcolor{teal}{$\mathbf{ 12.31}$} & $700$
	\\ \hline
	Model7-Pooling(AtLeastHalf)						&$\mathbf{0.73} $&$\mathbf{28.07}$ & $\mathbf{200}$ 				&\textcolor{teal}{$\mathbf{ 0.80}$} & \textcolor{teal}{$\mathbf{1.27}$} & $700$ 	& $\mathbf{0.66}$ & $\mathbf{1.54}$ & $500$
	\\ \hline
	Model7-Pooling(AtLeast$_{\alpha}$ $\alpha=0.75$)				&$\mathbf{0.61}$ &$\mathbf{7.02}$ & $500$					 & $0.70$ & $-11.39$ & $700$ 									& $0.64$ & $-1.54$ & $700$
	\\ \hline
	\end{tabular}
	}
	\caption{$Kla$ Fault diagnosis: Linguistic OWA pooling vs. Average pooling in Model7. }
	\label{tab:averagekla}
\end{table}

	\item \textbf{with regard to the \textsl{Average} pooling layer:}

			Now, a similar analysis is carried out taking into account the \textsl{Average} standard pooling and the Model7 layout, and  applying the diagnosis methodology proposed in this work. Table \ref{tab:averco7} shows a summary of this analysis. The \textsl{AtMiddle$_{\alpha}$($\alpha=0.2$)} operator is the best, obtaining the best diagnosis performance, achieving  an improvement of up to $3.30\%$ and  $7.06\%$  for $Precision$ and $Recall$, and $8.24\%$ for $F1-score$. The remaining linguistic OWA operators also obtain better results than the standard pooling options,  \textsl{Max} and \textsl{Average}. Focusing on the $O_2$ and $Kla$ faults,  the results are quite similar to the previous case, where \textsl{AtMiddle$_{\alpha}$($\alpha=0.2$)}  gives the better global result and $Precision$ and \textsl{AtLeastHalf} the best \textsl{Recall}, as shown in Tables \ref{tab:averagelco2} and \ref{tab:averagekla}.

\end{itemize}

Fig. \ref{fig:compara7MaxAver} displays a  graphical summary of all these results  for a faster and intuitive view.

\begingroup

\setlength{\tabcolsep}{1pt} 
\renewcommand{\arraystretch}{0.3} 

\setlength{\tabcolsep}{0pt}
	\begin{figure}[t!]
	\hspace{-3.2cm}
	\begin{tabular}{lll}
		\includegraphics[width=0.35\textheight]{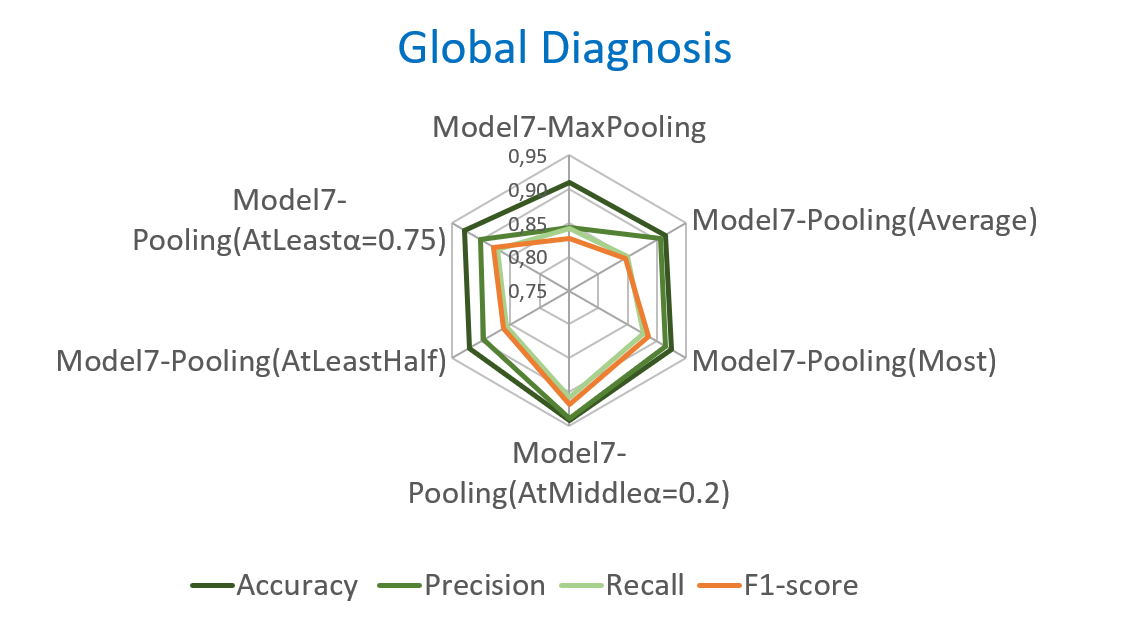} &  \includegraphics[width=0.35\textheight]{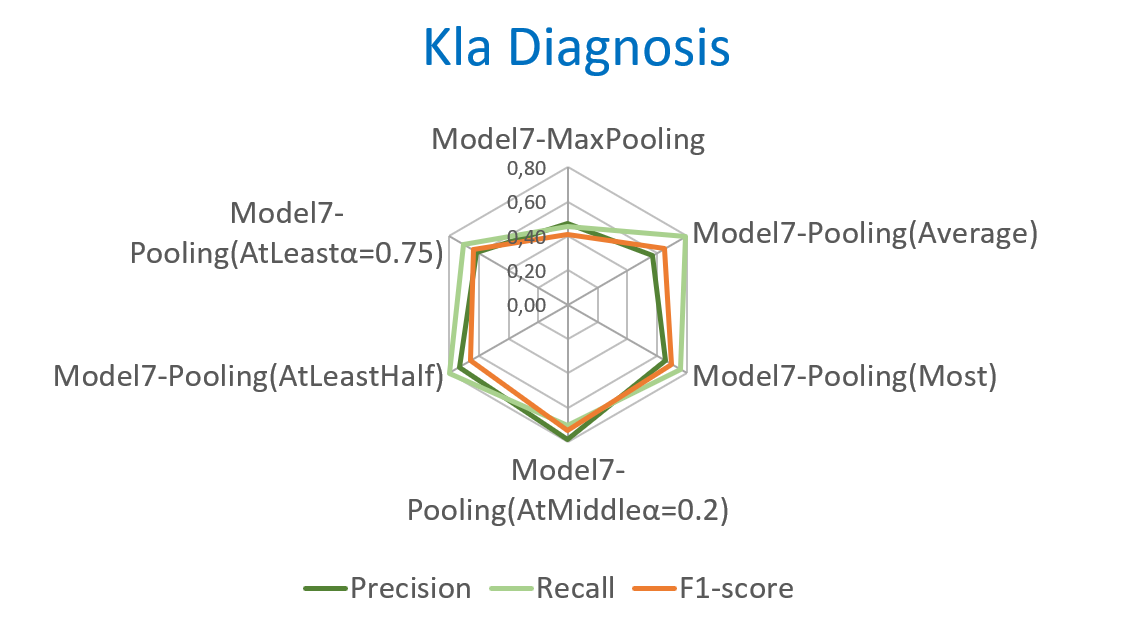} & \includegraphics[width=0.35\textheight]{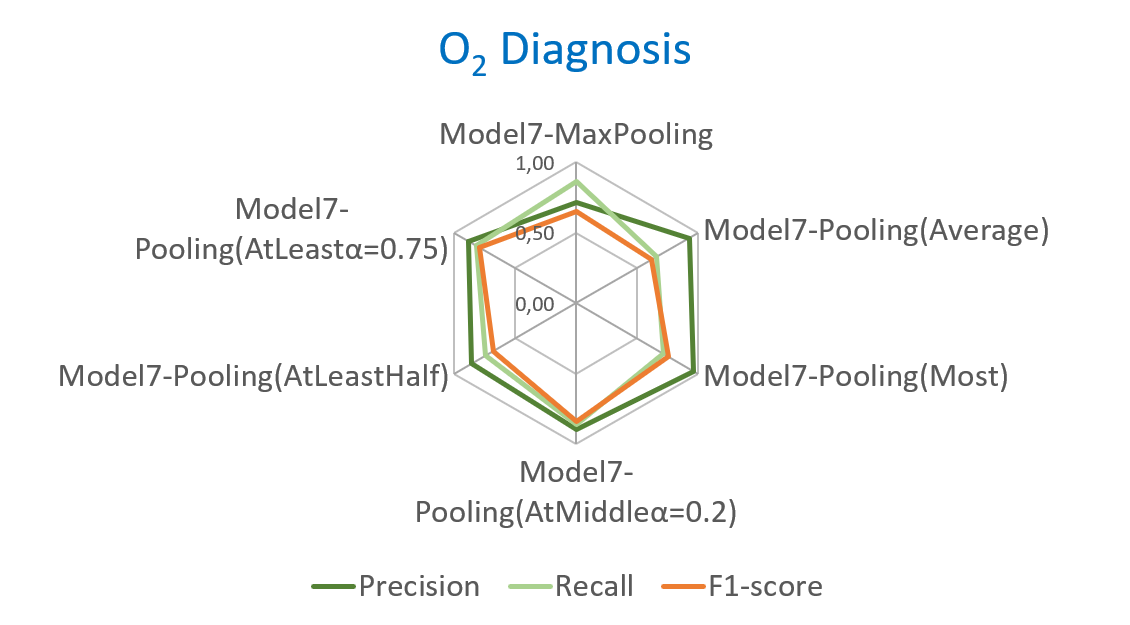} \\
	\end{tabular}
		\caption{Sliding Time Window \& Linguistic OWA based fault diagnosis in Model7.}
		\label{fig:compara7MaxAver}
\end{figure}
\endgroup


\subsubsection{Model3}
\begin{itemize}

	\item \textbf{with regard to the \textsl{Max} pooling layer:}
	
	In a similar way, the Model3 layout  has been used in our diagnosis methodology implemented by different pooling options.  Tables  \ref{tab:compamax3},  \ref{tab:maxO23} and \ref{tab:maxKla3} display a summary of the analysis  with regard to the baseline  based on \textsl{Max} pooling. Here, the  best global results, which can be seen in Table  \ref{tab:compamax3},  are given by the \textsl{Most} linguistic OWA operator for all the performances indexes, followed by {\color{black}\textsl{AtLeastHalf}}, with the reaming operators give more irregular adjusted  results regarding \textsl{Max} pooling. 
{\color{black}The diagnosis  for $O_2$ shows that the linguistic OWA operators offer a better \textsl{Precision}   than the \textsl{Max} standard pooling.  $F1-score$ and $Recall$ performance is higher with \textsl{AtLeast$_{\alpha}$ ($\alpha=0.75$)}.  In the case of the $Kla$ fault, the \textsl{Most} pooling clearly outperforms the rest of the options, except for $Precision$ where \textsl{AtLeastHalf} is the best option achieving an improvement of up to  $30.65\%$.}

\begin{table}[t!]  	
	\hspace{-2cm}	
	\resizebox{19cm}{!} {
	\begin{tabular}{l||c|c|c||c|c|c||c|c|c||c|c|c||}		\cline{2-13}		\cline{2-13}
																								&\multicolumn{12}{|c||}{Fault Diagnosis}\\ \cline{2-13}
												&\multicolumn{3}{|c||}{Accuracy}				&	\multicolumn{3}{|c||}{Precision} 			  &\multicolumn{3}{|c||}{Recall}					 &\multicolumn{3}{|c||}{F1-score}  \\ \cline{2-13} \cline{2-13}
	\textbf{Models:}		 							&Acc.	&$\triangle(\%)$ &Epochs					& Pre.  & $\triangle(\%)$ & Epochs 			  &Rcll.& $\triangle(\%)$	&Epochs				&F1	&$\triangle(\%)$ & Epochs 
	\\\hline \hline	
	\textit{Model3-MaxPooling}\cite{Wu2018}               				&$\mathit{0.94}$&$-$&$\mathit{700}$				&$\mathit{0.90}$&$-$&$\mathit{700}$			&$\mathit{0.90}$&$-$&$\mathit{700}$				&$\mathit{0.90}$&$-$&$\mathit{700}$
	\\ \hline
	Model3-Pooling(Average)								&$0.92$&$-2.13$&$500$						&$\mathbf{0.91}$&$\mathbf{1.11}$&$700$		&$0.84$&$-6.67$&$700$						&$0.84$&$-6.67$&$700$	
	\\ \hline
	Model3-Pooling(Most)								&\textcolor{teal}{$\mathbf{0.94}$}&\textcolor{teal}{$\mathbf{0.00}$}&$700$	&\textcolor{teal}{$\mathbf{0.95}$}&\textcolor{teal}{$\mathbf{4.44}$}&$700$		&\textcolor{teal}{$\mathbf{0.91}$}&\textcolor{teal}{$\mathbf{1.11}$}&$700$	&\textcolor{teal}{$\mathbf{0.92}$}&\textcolor{teal}{$\mathbf{2.22}$}&$700$
	\\ \hline
	Model3-Pooling(AtMiddle$_{\alpha}$$\alpha=0.2$)					&$\mathbf{0.94}$&$\mathbf{0.00}$&$500$			&$\mathbf{0.93}$&$\mathbf{3.33}$&$500$		&$0.89$&$-1.11$&$500$						&$\mathbf{0.90}$&$\mathbf{0.00}$&$500$	
	\\ \hline
	Model3-Pooling(AtLeastHalf)							&$\mathbf{0.94}$&$\mathbf{0.00}$&$500$			&$\mathbf{0.93}$&$\mathbf{3.33}$&$500$		&$\mathbf{0.90}$&$\mathbf{0.00}$&$500$			&$\mathbf{0.91}$&$\mathbf{1.11}$&$500$	
	\\ \hline
	Model3-Pooling(AtLeast$_{\alpha}$ $\alpha=0.75$)					&$0.94$&$0.00$&$700$						&$\mathbf{0.93}$&$\mathbf{3.33}$&$700$		&$\mathbf{0.91}$&$\mathbf{1.11}$&$700$			&$\mathbf{0.91}$&$\mathbf{1.11}$&$700$	
	\\ \hline
		\end{tabular}
	}
	\caption{Fault diagnosis ($WindowSize=4$, $WindowStep=1$): Linguistic OWA pooling vs. Max pooling in Model3.}	
	\label{tab:compamax3}
\end{table}

\begin{table}[t!]	
\hspace{-2cm}
\resizebox{19cm}{!} {
	\begin{tabular}{l||c|c|c||c|c|c||c|c|c||}		\cline{2-10}		\cline{2-10}
																										&\multicolumn{9}{|c||}{Fault Diagnosis}\\ \cline{2-10}		
											&\multicolumn{3}{|c||}{Precision} 			 						 &\multicolumn{3}{|c||}{Recall}									 &\multicolumn{3}{|c||}{F1-score}  \\ \cline{2-10} \cline{2-10}
	\textbf{Models:}		 						& Pre.  & $\triangle(\%)$ & Epochs 							 		 &Rcll.& $\triangle(\%)$	&Epochs								&F1	&$\triangle(\%)$ & Epochs 
	\\\hline \hline	
	 \textit{Model3-MaxPooling}\cite{Wu2018}				&$\mathit{0.84}$ & $- $& $\mathit{700} $								&$\mathit{0.86}$&$ -$ & $\mathit{700}$			 				&$\mathit{0.83}$ &$ -$ & $\mathit{700}$
	\\ \hline
	Model3-Pooling(Average)							&$\mathbf{0.89}$ & $\mathbf{5.95}$ & $700$ 							&  $0.80 $ &  $-6.98 $ &  $200 $ 									&  $0.65 $ &  $-21.69$ &  $200 $
	\\ \hline
	Model3-Pooling(Most)							&\textcolor{teal}{$\mathbf{0.99}$} &\textcolor{teal}{$\mathbf{17.86}$} &$ 700$ 	    	 &  $0.73 $ &  $-15.12$ &  $700 $								 &$0.82$ &$-1.20$ & $700$ 
	\\ \hline
	Model3-Pooling(AtMiddle$_{\alpha}$$\alpha=0.2$)				&$\mathbf{0.94} $&$\mathbf{11.90} $& $700$							 &  $0.81 $ &  $-5.81$ &  $500 $ 									& $\mathbf{0.83}$&$\mathbf{0.00} $& $500$
	\\ \hline
	Model3-Pooling(AtLeastHalf)						&$\mathbf{0.84}$ &$\mathbf{0.00}$& $700$ 							& $ 0.83 $ &  $-3.49 $ &  $500 $									&$0.77$& $-7.23$& $700$
	\\ \hline
	Model3-Pooling(AtLeast$_{\alpha}$ $\alpha=0.75$)				&$\mathbf{0.91} $&$\mathbf{8.33}$&$ 700$ 							&\textcolor{teal}{$\mathbf{0.86}$} &\textcolor{teal}{$\mathbf{0.00}$} &$ 500$  	& \textcolor{teal}{$\mathbf{0.87}$}&\textcolor{teal}{$\mathbf{4.82}$} & $700$
	\\ \hline
	\end{tabular}
	}
	\caption{$O_2$ Fault diagnosis: Linguistic OWA pooling vs. Max pooling in Model3. }
	\label{tab:maxO23}
\end{table}

\begin{table}[t!]	
\hspace{-2cm}
\resizebox{19cm}{!} {
	\begin{tabular}{l||c|c|c||c|c|c||c|c|c||}		\cline{2-10}		\cline{2-10}
																									&\multicolumn{9}{|c||}{Fault Diagnosis}\\ \cline{2-10}		
											&\multicolumn{3}{|c||}{Precision} 			 					 &\multicolumn{3}{|c||}{Recall}									 &\multicolumn{3}{|c||}{F1-score}  \\ \cline{2-10} \cline{2-10}
	\textbf{Models:}		 						& Pre.  & $\triangle(\%)$ & Epochs 							 	 &Rcll.& $\triangle(\%)$	&Epochs								&F1	&$\triangle(\%)$ & Epochs 
	\\\hline \hline	
	 \textit{Model3-MaxPooling}\cite{Wu2018}				&$\mathit{0.62}$ & $-$ &$\mathit{700}$ 							&$\mathit{0.71}$& $-$ &$\mathit{500}$ 							&$\mathit{0.61}$ & $- $&$\mathit{700}$
	\\ \hline
	Model3-Pooling(Average)							&$0.59$ & $-4.84$ & $700$ 									&$\mathbf{0.72}$ & $\mathbf{1.41}$ & $700$ 						&$0.60$ & $-1.64$ & $700$ 
	\\ \hline
	Model3-Pooling(Most)							&$\mathbf{0.71} $&$\mathbf{14.52 }$& $700$						& \textcolor{teal}{$\mathbf{0.78}$} & \textcolor{teal}{$\mathbf{9.86}$} & $700$ 	& \textcolor{teal}{$\mathbf{0.74}$} & \textcolor{teal}{$\mathbf{21.31}$} & $700$
	\\ \hline
	Model3-Pooling(AtMiddle$_{\alpha}$$\alpha=0.2$)				&$\mathbf{0.75}$ &$\mathbf{20.97}$ & $500$ 						& $\mathbf{0.76}$ &$\mathbf{7.04}$ & $700$ 						& $\mathbf{0.73}$ &$\mathbf{19.67 }$& $700$
	\\ \hline
	Model3-Pooling(AtLeastHalf)						&\textcolor{teal}{$\mathbf{0.81}$} & \textcolor{teal}{$\mathbf{30.65}$}& $500$ 	&$0.66$& $-7.04$ & $700$   									& $\mathbf{0.71}$ &$\mathbf{16.39}$ &$500$  
	\\ \hline
	Model3-Pooling(AtLeast$_{\alpha}$ $\alpha=0.75$)				&$\mathbf{0.71}$ &$\mathbf{14.52}$ & $500$ 						& $\mathbf{0.71}$ & $\mathbf{0.00}$ & $200$ 						&$\mathbf{0.70}$&$\mathbf{14.75}$ & $700$
	\\ \hline
	\end{tabular}
	}
	\caption{$Kla$ Fault diagnosis: Linguistic OWA pooling vs. Max pooling in Model3. }
	\label{tab:maxKla3}
\end{table}

	\item \textbf{with regard to the \textsl{Average} pooling layer:}
	
	Now the Model3 layout is used to compare the diagnosis scheme proposed with regard to   \textsl{Average} pooling. Tables \ref{tab:compa3}, 	\ref{tab:AveO23} and  \ref{tab:AveKla3} display the main results.  \textsl{Most} pooling is the best for all the performance indexes, improving the baseline performance from $3.30\%$ up to  $9.52\%$ for all the indexes. All the linguistic pooling operators almost outperform  \textsl{Average} pooling, as  Table \ref{tab:compa3} shows.
	
When the $O_2$ and $Kla$ faults diagnoses are focused, the {\color{black} \textsl{AtLeast$_{\alpha}$ ($\alpha=0.75$)}} and, mainly, the  \textsl{Most} options give the best results. In general, the linguistic OWA  operators offer a better performance in $F1-score$ and $Precision$ than the \textsl{Average} standard pooling, except in {\color{black}$O_2$} \textsl{Recall}, where \textsl{AtLeastHalf} and \textsl{AtLeast$_{\alpha}$ ($\alpha=0.75$)} have a worse performance.
	
\begin{table}[t!]	
\hspace{-2cm}
\resizebox{19cm}{!} {
	\begin{tabular}{l||c|c|c||c|c|c||c|c|c||c|c|c||}		\cline{2-13}		\cline{2-13}
																								&\multicolumn{12}{|c||}{Fault Diagnosis}\\ \cline{2-13}
												&\multicolumn{3}{|c||}{Accuracy}				&	\multicolumn{3}{|c||}{Precision} 			  &\multicolumn{3}{|c||}{Recall}					 &\multicolumn{3}{|c||}{F1-score}  \\ \cline{2-13} \cline{2-13}
	\textbf{Models:}		 							&Acc.	&$\triangle(\%)$ &Epochs					& Pre.  & $\triangle(\%)$ & Epochs 			  &Rcll.& $\triangle(\%)$	&Epochs				&F1	&$\triangle(\%)$ & Epochs 
	\\\hline \hline	
	Model3-MaxPooling\cite{Wu2018}						&$\mathbf{0.94}$&$\mathbf{2.17}$&$700$			&$0.90$&$-1.10$&$700$					&$\mathbf{0.90}$&$\mathbf{7.14}$&$700$			&$\mathbf{0.90}$&$\mathbf{7.14}$&$700$
	\\ \hline
	\textit{Model3-AveragePooling}							&$\mathit{0.92}$&$-$&$\mathit{500}$				&$\mathit{0.91}$&$-$&$\mathit{700}$			&$\mathit{0.84}$&$-$&$\mathit{700}$				&$\mathit{0.84}$&$-$&$\mathit{700}$
	\\ \hline
	Model3-Pooling(Most)								&\textcolor{teal}{$\mathbf{0.94}$}&\textcolor{teal}{$\mathbf{2.17}$}&$700$	&\textcolor{teal}{$\mathbf{0.94}$}&\textcolor{teal}{$\mathbf{3.30}$}&$700$		&\textcolor{teal}{$\mathbf{0.91}$}&\textcolor{teal}{$\mathbf{8.33}$}&$700$	&\textcolor{teal}{$\mathbf{0.92}$}&\textcolor{teal}{$\mathbf{9.52}$}&$700$
	\\ \hline
	Model3-Pooling(AtMiddle$_{\alpha}$$\alpha=0.2$)					&$\mathbf{0.94}$&$\mathbf{2.17}$&$500$			&$\mathbf{0.93}$&$\mathbf{2.20}$&$500$		&$\mathbf{0.89}$&$\mathbf{5.95}$&$500$			&$\mathbf{0.90}$&$\mathbf{7.14}$&$500$
	\\ \hline
	Model3-Pooling(AtLeastHalf)							&$\mathbf{0.94}$&$\mathbf{2.17}$&$500$			&$\mathbf{0.93}$&$\mathbf{2.20}$&$500$		&$\mathbf{0.90}$&$\mathbf{7.14}$&$500$			&$\mathbf{0.91}$&$\mathbf{8.33}$&$500$
	\\ \hline
	Model3-Pooling(AtLeast$_{\alpha}$ $\alpha=0.75$)					&$\mathbf{0.94}$&$\mathbf{2.17}$&$700$			&$\mathbf{0.93}$&$\mathbf{2.20}$&$700$		&$\mathbf{0.91}$&$\mathbf{8.33}$&$700$			&$\mathbf{0.91}$&$\mathbf{8.33}$&$700$	
	\\ \hline
	\end{tabular}
	}
	\caption{Fault diagnosis ($WindowSize=4$, $WindowStep=1$): Linguistic OWA pooling vs. Average pooling in Model3.}	
	\label{tab:compa3}	
\end{table}

\begin{table}[t!]	
\hspace{-2cm}
\resizebox{19cm}{!} {
	\begin{tabular}{l||c|c|c||c|c|c||c|c|c||}		\cline{2-10}		\cline{2-10}
																								&\multicolumn{9}{|c||}{Fault Diagnosis}\\ \cline{2-10}		
											&\multicolumn{3}{|c||}{Precision} 			 				 &\multicolumn{3}{|c||}{Recall}									 &\multicolumn{3}{|c||}{F1-score}  \\ \cline{2-10} \cline{2-10}
	\textbf{Models:}		 						& Pre.  & $\triangle(\%)$ & Epochs 							  &Rcll.& $\triangle(\%)$	&Epochs								&F1	&$\triangle(\%)$ & Epochs 
	\\\hline \hline	
	Model3-MaxPooling\cite{Wu2018}					&$0.84$ & $-5.62$& $700 $								&$\mathbf{0.86} $ &  $\mathbf{7.50}$ &  $700 $						&$\mathbf{0.83}$ &$ \mathbf{27.69}$ & $700$
	\\ \hline
	\textit{Model3-AveragePooling}						&$\mathit{0.69}$ & $-$ & $\mathit{700}$ 						&  $\mathit{0.80} $ &  $- $ &  $\mathit{200} $ 						& $\mathit{0.65} $ &  $-$ &  $\mathit{200} $
	\\ \hline
	Model3-Pooling(Most)							&\textcolor{teal}{$\mathbf{0.99}$} &\textcolor{teal}{$\mathbf{11.24}$} &$ 700$ 	     &  $0.73 $ &  $-8.75$ &  $700 $							 &$\mathbf{0.82}$ & $\mathbf{ 26.15}$& $700$
	\\ \hline
	Model3-Pooling(AtMiddle$_{\alpha}$$\alpha=0.2$)				&$\mathbf{0.94} $&$\mathbf{5.62} $& $700$					& $\mathbf{0.81} $ &  $\mathbf{1.25}$ &  $500 $ 						& $\mathbf{0.83}$& $\mathbf{27.69}$& $500$
	\\ \hline
	Model3-Pooling(AtLeastHalf)						&$0.84$ &$-5.62$& $700$ 								& $\mathbf{0.83} $ &  $\mathbf{3.75}$ &  $500 $ 						&$\mathbf{0.77}$& $\mathbf{18.46}$& $700$
	\\ \hline
	Model3-Pooling(AtLeast$_{\alpha}$ $\alpha=0.75$)				&$\mathbf{0.91} $&$\mathbf{2.25} $&$ 700$ 					& \textcolor{teal}{$\mathbf{0.86}$}&\textcolor{teal}{$\mathbf{7.50}$} & $500$ 	& \textcolor{teal}{$\mathbf{0.87}$}&\textcolor{teal}{$\mathbf{33.85}$} & $700$
	\\ \hline
	\end{tabular}
	}
	\caption{$O_2$ Fault diagnosis: Linguistic OWA pooling vs. Average pooling in Model3.}
	\label{tab:AveO23}
\end{table}

\begin{table}[t!]	
\hspace{-2cm}
\resizebox{19cm}{!} {
	\begin{tabular}{l||c|c|c||c|c|c||c|c|c||}		\cline{2-10}		\cline{2-10}
																									&\multicolumn{9}{|c||}{Fault Diagnosis}\\ \cline{2-10}		
											&\multicolumn{3}{|c||}{Precision} 			 					 &\multicolumn{3}{|c||}{Recall}									 &\multicolumn{3}{|c||}{F1-score}  \\ \cline{2-10} \cline{2-10}
	\textbf{Models:}		 						& Pre.  & $\triangle(\%)$ & Epochs 								  &Rcll.& $\triangle(\%)$	&Epochs								&F1	&$\triangle(\%)$ & Epochs 
	\\\hline \hline	
	Model3-MaxPooling\cite{Wu2018}					&$\mathbf{0.62}$ &$\mathbf{5.08}$&$700$ 							&$0.71$& $-1.39$ &$500$ 									&$\mathbf{0.61}$ & $\mathbf{1.67}$&$700$
	\\ \hline
	\textit{Model3-AveragePooling}						&$\mathit{0.59}$ & $-$ & $\mathit{700}$ 							&$\mathit{0.72}$ & $-$ & $\mathit{700}$ 							&$\mathit{0.60}$ & $-$ & $\mathit{700}$
	\\ \hline
	Model3-Pooling(Most)							& $\mathbf{0.71}$&$\mathbf{20.34}$ & $700$						&\textcolor{teal}{ $\mathbf{0.78}$} & \textcolor{teal}{$\mathbf{8.33}$} & $700$ 	&  \textcolor{teal}{$\mathbf{0.74}$ }&  \textcolor{teal}{$\mathbf{23.33}$} & $700$
	\\ \hline
	Model3-Pooling(AtMiddle$_{\alpha}$$\alpha=0.2$)				&$\mathbf{0.75}$ &$\mathbf{27.12}$ & $500$ 						& $\mathbf{0.76}$&$\mathbf{5.56}$ & $700$ 						&$\mathbf{0.73}$ &$\mathbf{21.67}$& $700$
	\\ \hline
	Model3-Pooling(AtLeastHalf)						&\textcolor{teal}{$\mathbf{0.81} $}& \textcolor{teal}{$\mathbf{37.29}$}& $500$ 	&$0.66$& $-8.33$ & $700$   									& $\mathbf{0.71}$ &$\mathbf{18.33}$ &$500$
	\\ \hline
	Model3-Pooling(AtLeast$_{\alpha}$ $\alpha=0.75$)				&$\mathbf{0.71}$ &$\mathbf{20.34}$ & $500$ 						& $0.71$ & $-1.39$ & $200$ 									&$\mathbf{0.70}$&$\mathbf{16.67}$ & $700$
	\\ \hline
	\end{tabular}
	}
	\caption{$Kla$ Fault diagnosis: Linguistic OWA pooling vs. Average pooling in Model3. }
	\label{tab:AveKla3}
\end{table}

\end{itemize}

\begingroup

\setlength{\tabcolsep}{1pt} 
\renewcommand{\arraystretch}{0.3} 

\setlength{\tabcolsep}{0pt}
	\begin{figure}[t!]
	\hspace{-3.2cm}
	\begin{tabular}{lll}
		\includegraphics[width=0.35\textheight]{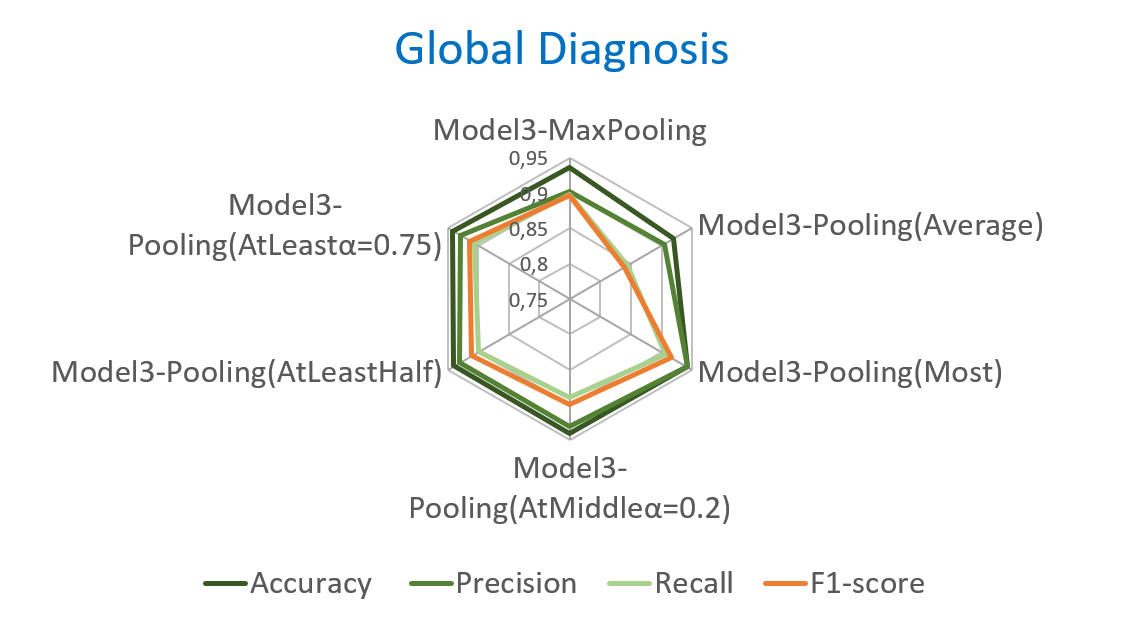} &  \includegraphics[width=0.35\textheight]{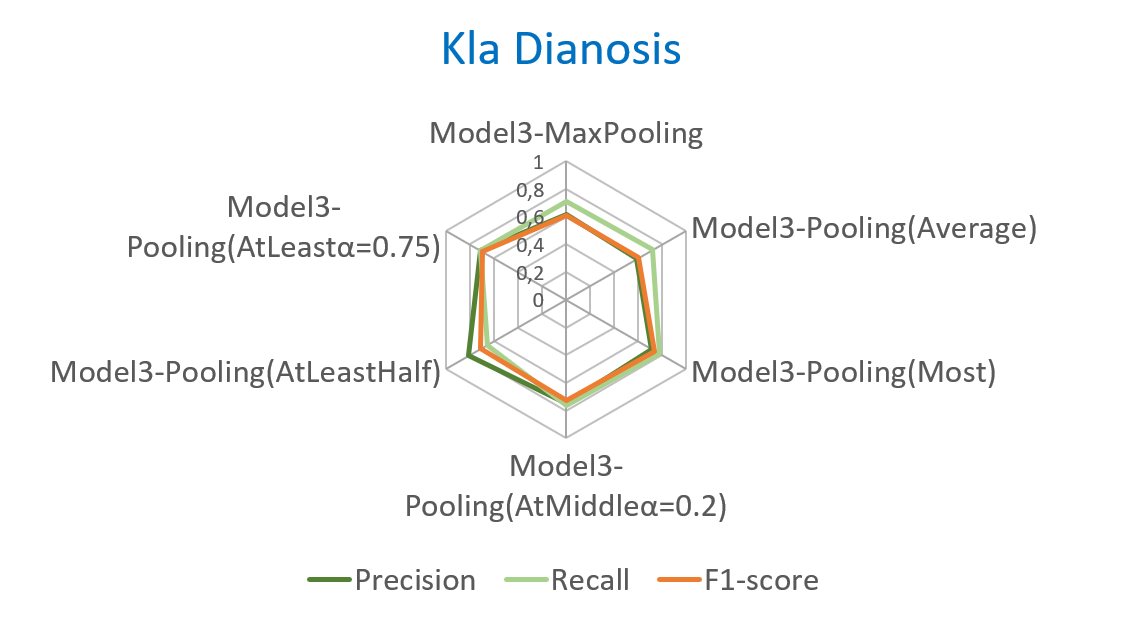} & \includegraphics[width=0.35\textheight]{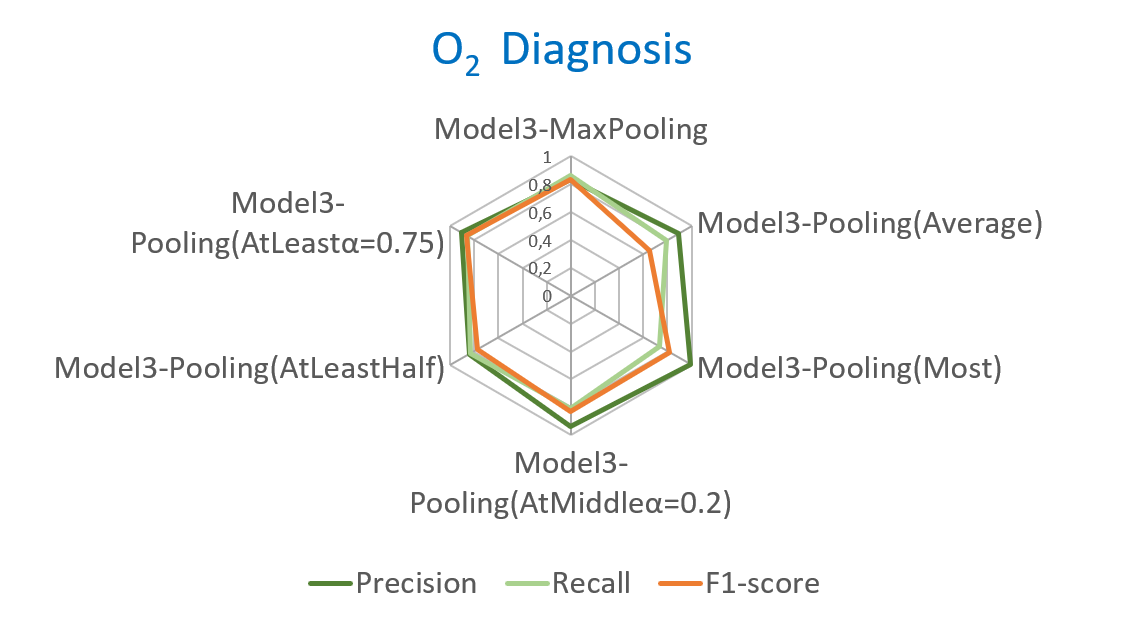} \\
	\end{tabular}
		\caption{Sliding Time Window \& Linguistic OWA based fault diagnosis in Model3.}
		\label{fig:compara3}
	\end{figure}

\endgroup

Fig. \ref{fig:compara3} displays  a summary of all these results in graphics for easier reading.


\subsubsection{LeNet-5}

Here, a well-known DL network is used to implement our fault diagnosis methodology and the linguistic OWA operators. The results obtained are as follows:

\begin{itemize}

	\item  \textbf{with regard to the \textsl{Max} pooling layer}:

	In Table \ref{tab:compaLeMax}, the global diagnosis  obtained is displayed. The \textsl{Most} pooling is the best, although the rest of the options also notably  improve the \textsl{Max} pooling by up to $51.85 \%$ in $F1-score$. Similar results, or even better,  are obtained for the $O_2$ and $Kla$ fault diagnosis,  except in the $Kla$ fault for a couple of cases in \textsl{Recall} (see Tables \ref{tab:O2LeMax} and \ref{tab:KlaLeMax}).


\begin{table}[t!]
\hspace{-2cm}
\resizebox{19cm}{!} {
	\begin{tabular}{l||c|c|c||c|c|c||c|c|c||c|c|c||}		\cline{2-13}		\cline{2-13}
																									&\multicolumn{12}{|c||}{Fault Diagnosis}\\ \cline{2-13}
												&\multicolumn{3}{|c||}{Accuracy}					&\multicolumn{3}{|c||}{Precision} 					  &\multicolumn{3}{|c||}{Recall}						 &\multicolumn{3}{|c||}{F1-score}  \\ \cline{2-13} \cline{2-13}
	\textbf{Models:}		 							&Acc.	&$\triangle(\%)$ &Epochs						& Pre.  & $\triangle(\%)$ & Epochs 					  &Rcll.& $\triangle(\%)$	&Epochs					&F1	&$\triangle(\%)$ & Epochs 
	\\\hline \hline	
	\textit{LeNet-5-MaxPooling}\cite{Lecun1998}					&$\mathit{0.79}$ &$ -$ &$\mathit{700}$ 				&$\mathit{0.57}$ &$ - $&$\mathit{700 }$				&$\mathit{0.56}$ &$ - $& $\mathit{700}$ 				& $\mathit{0.54 }$& $- $& $\mathit{700}$
	\\ \hline
	LeNet-5-Pooling(Average)								&$\mathbf{ 0.86} $&$\mathbf{  8.86 }$& $\mathbf{200}$ 		& $\mathbf{ 0.73}$&$\mathbf{28.07}$ & $\mathbf{200}$		 &$\mathbf{ 0.73} $&$\mathbf{30.36} $& $\mathbf{200}$ 		&$\mathbf{0.71 }$& $\mathbf{31.48}$ & $\mathbf{200}$
	\\ \hline
	LeNet-5-Pooling(Most)								&\textcolor{teal}{$\mathbf{0.90}$} &\textcolor{teal}{$\mathbf{13.92}$} & $500$ 			& \textcolor{teal}{$\mathbf{0.84}$} &\textcolor{teal}{$\mathbf{47.37}$} & $500$ 			&\textcolor{teal}{$\mathbf{0.83}$} & \textcolor{teal}{$\mathbf{48.21}$} & $500$ 			&\textcolor{teal}{$\mathbf{0.82}$} & \textcolor{teal}{$\mathbf{ 51.85}$} & $500$
	\\ \hline
	LeNet-5-Pooling(AtMiddle$_{\alpha}$$\alpha=0.2$)					& $\mathbf{0.88 }$& $\mathbf{11.39}$ & $\mathbf{200 }$  	&$\mathbf{ 0.79} $&$\mathbf{ 38.60 }$&$ \mathbf{200} $		&$\mathbf{ 0.78}$ &$ \mathbf{39.29}$ &$ \mathbf{200 }$		&$\mathbf{ 0.77}$ &$\mathbf{42.59 }$&$ \mathbf{200}$
	\\ \hline
	LeNet-5-Pooling(AtLeastHalf)							&$\mathbf{ 0.87}$& $\mathbf{10.13}$ & $\mathbf{200 }$		&$\mathbf{0.80 }$&$\mathbf{40.35 }$& $\mathbf{200}$		 &$\mathbf{0.75} $& $\mathbf{33.93} $& $\mathbf{200}$ 		&$\mathbf{0.74}$ &$\mathbf{37.04} $&$ \mathbf{200}$
	\\ \hline
	LeNet-5-Pooling(AtLeast$_{\alpha}$ $\alpha=0.75$)					&$\mathbf{0.89}$ & $\mathbf{12.66}$ & $\mathbf{200 }$		&$\mathbf{ 0.77} $&$\mathbf{35.09}$ &$\mathbf{ 200 }$		&$\mathbf{0.81}$ &$ \mathbf{44.64}$ & $\mathbf{200}$		 & $\mathbf{0.78} $&$\mathbf{44.44}$ &$ \mathbf{200 } $
	\\ \hline
	\end{tabular}
	}
	\caption{Fault diagnosis ($WindowSize=4$, $WindowStep=1$): Linguistic OWA pooling vs. Max pooling in LeNet-5.}
	\label{tab:compaLeMax}
\end{table}

\begin{table}[t!]
\hspace{-2cm}
\resizebox{19cm}{!} {
	\begin{tabular}{l||c|c|c||c|c|c||c|c|c||}		\cline{2-10}		\cline{2-10}
																								&\multicolumn{9}{|c||}{Fault Diagnosis}\\ \cline{2-10}		
											&\multicolumn{3}{|c||}{Precision} 			 				 &\multicolumn{3}{|c||}{Recall}									 &\multicolumn{3}{|c||}{F1-score}  \\ \cline{2-10} \cline{2-10}
	\textbf{Models:}		 						& Pre.  & $\triangle(\%)$ & Epochs 							  &Rcll.& $\triangle(\%)$	&Epochs								&F1	&$\triangle(\%)$ & Epochs 
	\\\hline \hline	
	\textit{LeNet-5-MaxPooling}\cite{Lecun1998}				&$\mathit{0.51}$ &$ -$ &$\mathit{700}$						&$\mathit{0.27}$ &$ - $& $\mathit{200} $							&$\mathit{0.31} $& $-$ &$\mathit{200}$
	\\ \hline
	LeNet-5-Pooling(Average)							&$\mathbf{0.53}$ & $\mathbf{3.92} $& $500$ 					& $\mathbf{0.61 }$& $\mathbf{125.93}$ & $500$ 						&$\mathbf{ 0.55 }$&$\mathbf{77.42}$ & $500$ 
	\\ \hline
	LeNet-5-Pooling(Most)							&$\mathbf{0.66}$ &$\mathbf{29.41}$ & $\mathbf{200}$				& \textcolor{teal}{$\mathbf{0.77 }$}& \textcolor{teal}{$\mathbf{185.19}$} & $500$ 		& \textcolor{teal}{$\mathbf{0.62}$} &\textcolor{teal}{$\mathbf{100.00}$} & $500$ 
	\\ \hline
	LeNet-5-Pooling(AtMiddle$_{\alpha}$$\alpha=0.2$)				& $\mathbf{0.60}$ &  $\mathbf{17.65}$& $700$ 					& $\mathbf{0.55 }$&  $\mathbf{103.70}$ & $700$ 						&  $\mathbf{0.57}$ &  $\mathbf{83.87}$ & $700$
	\\ \hline
	LeNet-5-Pooling(AtLeastHalf)						&\textcolor{teal}{$\mathbf{ 0.66}$} & \textcolor{teal}{$\mathbf{29.41}$} &$ 500$ 		&$\mathbf{0.41 }$&$\mathbf{51.85} $& $500$ 				&$\mathbf{0.40}$ &$\mathbf{29.03 }$& $500$
	\\ \hline
	LeNet-5-Pooling(AtLeast$_{\alpha}$ $\alpha=0.75$)				&$\mathbf{0.58}$&$\mathbf{13.73}$ & $700$ 					&$\mathbf{0.67}$ & $\mathbf{148.15}$ & $\mathbf{200}$					&$\mathbf{0.51}$ & $\mathbf{64.52}$ & $\mathbf{200}$
	\\ \hline
	\end{tabular}
	}
	\caption{$O_2$ Fault diagnosis: Linguistic OWA pooling vs. Max pooling LeNet-5.}
	\label{tab:O2LeMax}
\end{table}

\begin{table}[t!]
\hspace{-2cm}
\resizebox{19cm}{!} {
	\begin{tabular}{l||c|c|c||c|c|c||c|c|c||}		\cline{2-10}		\cline{2-10}
																								&\multicolumn{9}{|c||}{Fault Diagnosis}\\ \cline{2-10}		
											&\multicolumn{3}{|c||}{Precision} 			 				 &\multicolumn{3}{|c||}{Recall}									 &\multicolumn{3}{|c||}{F1-score}  \\ \cline{2-10} \cline{2-10}
	\textbf{Models:}		 						& Pre.  & $\triangle(\%)$ & Epochs 							  &Rcll.& $\triangle(\%)$	&Epochs								&F1	&$\triangle(\%)$ & Epochs 
	\\\hline \hline	
	\textit{LeNet-5-MaxPooling}\cite{Lecun1998}				&$\mathit{0.30}$ & $-$ & $\mathit{700}$ 						& $\mathit{0.47}$ & $-$ &$\mathit{700 }$							& $\mathit{0.37}$ &$ -$ & $\mathit{700}$
	\\ \hline
	LeNet-5-Pooling(Average)							&$\mathbf{0.37} $& $\mathbf{23.33 }$& $500$ 					& $\mathbf{0.47} $&$\mathbf{0.00}$ &$ 700 $						& $\mathbf{0.37}$ & $\mathbf{0.00}$ & $700$
	\\ \hline
	LeNet-5-Pooling(Most)							& \textcolor{teal}{$\mathbf{ 0.55}$} &\textcolor{teal}{$\mathbf{ 83.33}$} & $500$ 		&$ 0.45$ & $-4.26$ & $\mathbf{200}$ 						&$\mathbf{ 0.42}$ & $\mathbf{13.51}$ & $\mathbf{200}$ 
	\\ \hline
	LeNet-5-Pooling(AtMiddle$_{\alpha}$$\alpha=0.2$)				&$\mathbf{0.46}$ & $\mathbf{53.33}$ & $\mathbf{200}$ 				& \textcolor{teal}{$\mathbf{ 0.74}$} &\textcolor{teal}{$\mathbf{ 57.45}$} & $\mathbf{200}$ 		&\textcolor{teal}{$\mathbf{0.57}$} & \textcolor{teal}{$\mathbf{54.05}$} & $\mathbf{200}$
	\\ \hline
	LeNet-5-Pooling(AtLeastHalf)						&$\mathbf{0.48}$ & $\mathbf{50.00} $& $\mathbf{200}$ 				&$\mathbf{ 0.73} $&$\mathbf{55.32} $& $\mathbf{200} $					&$\mathbf{0.57}$ &$\mathbf{54.05}$ & $\mathbf{200}$
	\\ \hline
	LeNet-5-Pooling(AtLeast$_{\alpha}$ $\alpha=0.75$)				&$\mathbf{0.44}$ & $\mathbf{46.67}$ & $700$ 					& $0.43$ & $-8.51$ & $500$ 									& $\mathbf{0.39 }$&$\mathbf{5.41 }$& $500$
	\\ \hline
	\end{tabular}
	}
	\caption{$Kla$ Fault diagnosis: Linguistic OWA pooling vs. Max pooling LeNet-5. }
	\label{tab:KlaLeMax}
\end{table}

\item  \textbf{with regard to the \textsl{Average} pooling layer}:

    Now, using as baseline the \textsl{Average} pooling in LeNet5, Tables   \ref{tab:compaLeAvera}, \ref {tab:LeO2Avera} and \ref{tab:LeKlaAvera} summarize the main results. The  \textsl{Most} is the best pooling option for the global  and $O_2$ diagnoses. For the $Kla$ diagnosis, the best pooling option is \textsl{AtMiddle$_{\alpha}$($\alpha=0.2)$}.

\begin{table}[t!]
\hspace{-2cm}
\resizebox{19cm}{!} {
	\begin{tabular}{l||c|c|c||c|c|c||c|c|c||c|c|c||}		\cline{2-13}		\cline{2-13}
																									&\multicolumn{12}{|c||}{Fault Diagnosis}\\ \cline{2-13}
												&\multicolumn{3}{|c||}{Accuracy}					&\multicolumn{3}{|c||}{Precision} 					  &\multicolumn{3}{|c||}{Recall}						 &\multicolumn{3}{|c||}{F1-score}  \\ \cline{2-13} \cline{2-13}
	\textbf{Models:}		 							&Acc.	&$\triangle(\%)$ &Epochs						& Pre.  & $\triangle(\%)$ & Epochs 					  &Rcll.& $\triangle(\%)$	&Epochs					&F1	&$\triangle(\%)$ & Epochs 
	\\\hline \hline	
	LeNet-5-MaxPooling \cite{Lecun1998}						&$\mathit{0.79}$ &$ \mathit{-8.14}$ &$\mathit{700}$ 		&$\mathit{0.57}$ &$\mathit{ -21.92}$&$\mathit{700 }$		&$\mathit{0.56}$ &$\mathit{ - 23.29}$& $\mathit{700}$ 		& $\mathit{0.54 }$& $\mathit{-23.94} $& $\mathit{700}$
	\\ \hline
	\textit{LeNet-5-AveragePooling}							&$\mathbf{ 0.86} $&$-$& $\mathbf{200}$ 				& $\mathbf{ 0.73}$&$-$ & $\mathbf{200}$				 &$\mathbf{ 0.73} $&$- $& $\mathbf{200}$ 				&$\mathbf{0.71 }$& $-$ & $\mathbf{200}$
	\\ \hline
	LeNet-5-Pooling(Most)								&\textcolor{teal}{$\mathbf{0.90}$} &\textcolor{teal}{$\mathbf{4.65}$} & $500$ 			& \textcolor{teal}{$\mathbf{0.84}$} &\textcolor{teal}{$\mathbf{15.07}$} & $500$ 			&\textcolor{teal}{$\mathbf{0.83}$} & \textcolor{teal}{$\mathbf{13.70}$} & $500$ 			&\textcolor{teal}{$\mathbf{0.82}$} & \textcolor{teal}{$\mathbf{15.49}$} & $500$
	\\ \hline

	LeNet-5-Pooling(AtMiddle$_{\alpha}$$\alpha=0.2$)					& $\mathbf{0.88 }$& $\mathbf{2.33}$ & $\mathbf{200 }$ 	 	&$\mathbf{ 0.79} $&$\mathbf{ 8.22}$&$ \mathbf{200} $		&$\mathbf{ 0.78}$ &$ \mathbf{6.85}$ &$ \mathbf{200 }$		&$\mathbf{ 0.77}$ &$\mathbf{8.45 }$&$ \mathbf{200}$
	\\ \hline
	LeNet-5-Pooling(AtLeastHalf)							&$\mathbf{ 0.87}$& $\mathbf{1.16}$ & $\mathbf{200 }$		&$\mathbf{0.80 }$&$\mathbf{9.59}$& $\mathbf{200}$		 &$\mathbf{0.75} $& $\mathbf{2.74} $& $\mathbf{200}$ 		&$\mathbf{0.74}$ &$\mathbf{4.23} $&$ \mathbf{200}$
	\\ \hline
	LeNet-5-Pooling(AtLeast$_{\alpha}$ $\alpha=0.75$)					&$\mathbf{0.89}$ & $\mathbf{3.49}$ & $\mathbf{200 }$		&$\mathbf{ 0.77} $&$\mathbf{5.48}$ &$\mathbf{ 200 }$		&$\mathbf{0.81}$ &$ \mathbf{10.96}$ & $\mathbf{200}$		 & $\mathbf{0.78} $&$\mathbf{9.86}$ &$ \mathbf{200 } $
	\\ \hline
	\end{tabular}
}
	\caption{Fault diagnosis ($WindowSize=4$, $WindowStep=1$): Linguistic OWA pooling vs. Average pooling LeNet-5.}
	\label{tab:compaLeAvera}
\end{table}

\begin{table}[t!]
\hspace{-2cm}
\resizebox{19cm}{!} {
	\begin{tabular}{l||c|c|c||c|c|c||c|c|c||}		\cline{2-10}		\cline{2-10}
																								&\multicolumn{9}{|c||}{Fault Diagnosis}\\ \cline{2-10}		
											&\multicolumn{3}{|c||}{Precision} 			 				 &\multicolumn{3}{|c||}{Recall}									 &\multicolumn{3}{|c||}{F1-score}  \\ \cline{2-10} \cline{2-10}
	\textbf{Models:}		 						& Pre.  & $\triangle(\%)$ & Epochs 							  &Rcll.& $\triangle(\%)$	&Epochs								&F1	&$\triangle(\%)$ & Epochs 
	\\\hline \hline	
	LeNet-5-MaxPooling \cite{Lecun1998}					&$0.51$ & $-3.77$ & $700$ 								& $0.27$ & $-55.74$ & $200$ 									& $0.31$ & $-43.64$ & $200$
	\\ \hline
	\textit{LeNet-5-AveragePooling}						&$\mathit{0.53}$ & $-$ & $\mathit{500}$ 						&$\mathit{ 0.61}$ & $-$ &$\mathit{500}$ 							&$\mathit{0.55}$ & $-$ &$\mathit{500}$
	\\ \hline
	LeNet-5-Pooling(Most)							&$\mathbf{0.66 }$& $\mathbf{24.53 }$& $\mathbf{200}$				&\textcolor{teal}{$\mathbf{ 0.77}$} & \textcolor{teal}{$\mathbf{26.23}$} & $500$ 		& \textcolor{teal}{$\mathbf{ 0.62}$} & \textcolor{teal}{$\mathbf{12.73}$} & $500$
	\\ \hline
	LeNet-5-Pooling(AtMiddle$_{\alpha}$$\alpha=0.2$)				&$\mathbf{0.60}$ & $\mathbf{13.21 }$& $700$ 					& $0.55$ & $-9.84$ & $700$ 									&$\mathbf{0.57}$ &$\mathbf{3.64} $& $700$
	\\ \hline
	LeNet-5-Pooling(AtLeastHalf)						&\textcolor{teal}{$\mathbf{ 0.66}$}& \textcolor{teal}{$\mathbf{ 24.53}$} & $500$		& $0.41$ & $-32.79$ & $500$ 							& $0.40$ & $-27.27$ & $500$
	\\ \hline
	LeNet-5-Pooling(AtLeast$_{\alpha}$ $\alpha=0.75$)				&$\mathbf{0.58 }$& $\mathbf{9.43}$& $700$ 					&$\mathbf{ 0.67 }$& $\mathbf{9.84 }$&  $\mathbf{200}$					& $0.51$ & $-7.27$ & $\mathbf{200}$
	\\ \hline
	\end{tabular}
	}
	\caption{$O_2$ Fault diagnoisis: Linguistic OWA pooling vs. Average pooling LeNet-5.}
	\label{tab:LeO2Avera}
\end{table}

\begin{table}[t!]
\hspace{-2cm}
\resizebox{19cm}{!} {
	\begin{tabular}{l||c|c|c||c|c|c||c|c|c||}		\cline{2-10}		\cline{2-10}
																								&\multicolumn{9}{|c||}{Fault Diagnosis}\\ \cline{2-10}		
											&\multicolumn{3}{|c||}{Precision} 			 				 &\multicolumn{3}{|c||}{Recall}									 &\multicolumn{3}{|c||}{F1-score}  \\ \cline{2-10} \cline{2-10}
	\textbf{Models:}		 						& Pre.  & $\triangle(\%)$ & Epochs 							  &Rcll.& $\triangle(\%)$	&Epochs								&F1	&$\triangle(\%)$ & Epochs 
	\\\hline \hline	
	LeNet-5-MaxPooling \cite{Lecun1998}					& $0.30$ & $-18.92$ & $700$ 								& $0.47$ & $0.00$ & $700$ 									& $0.37$ & $0.00$ & $700$
	\\ \hline
	\textit{LeNet-5-AveragePooling}						&$\mathit{0.37 }$ & $ -$  & $\mathit{500 }$ 						& $\mathit{0.47}$  &$  -$  & $\mathit{700}$  							&$\mathit{0.37}$  & $ -$  & $\mathit{700}$ 
	\\ \hline
	LeNet-5-Pooling(Most)							&\textcolor{teal}{$\mathbf{0.55 }$}&\textcolor{teal}{$\mathbf{48.65}$} &$ 500$ 		& $0.45$ & $-4.26$ & $\mathbf{200} $						&$\mathbf{0.42}$ & $\mathbf{13.51}$ & $\mathbf{200}$ 
	\\ \hline
	LeNet-5-Pooling(AtMiddle$_{\alpha}$$\alpha=0.2$)				&$\mathbf{ 0.46 }$&$\mathbf{24.32}$ &$\mathbf{ 200}$				&\textcolor{teal}{$\mathbf{0.74}$} &\textcolor{teal}{$\mathbf{57.45}$} & $\mathbf{200}$ 		& \textcolor{teal}{$\mathbf{0.57 }$}& \textcolor{teal}{$\mathbf{ 54.05}$} & $\mathbf{200}$
	\\ \hline
	LeNet-5-Pooling(AtLeastHalf)						&$\mathbf{ 0.48} $&$\mathbf{ 29.73} $& $\mathbf{200}$ 				& $\mathbf{0.73}$ &$\mathbf{55.32}$ & $\mathbf{200} $					&$\mathbf{0.57}$ &$\mathbf{ 54.05}$ & $\mathbf{200}$
	\\ \hline
	LeNet-5-Pooling(AtLeast$_{\alpha}$ $\alpha=0.75$)				&$\mathbf{0.44}$ & $\mathbf{18.92}$ & $700$ 					& $0.43$ & $-8.51$ & $500$ 									& $\mathbf{0.39 }$&$\mathbf{5.41}$ &$ 500$
	\\ \hline
	\end{tabular}
	}
	\caption{$Kla$ Fault diagnoisis: Linguistic OWA pooling vs. Average pooling LeNet-5.}
	\label{tab:LeKlaAvera}
\end{table}

\end{itemize}

As in previous cases, Fig. \ref{fig:comparaLenet5} shows  a graphic view   with all these results.

\begingroup

\setlength{\tabcolsep}{1pt} 
\renewcommand{\arraystretch}{0.3} 

\setlength{\tabcolsep}{0pt}
	\begin{figure}[t!]
	\hspace{-3.2cm}
	\begin{tabular}{lll}
		\includegraphics[width=0.35\textheight]{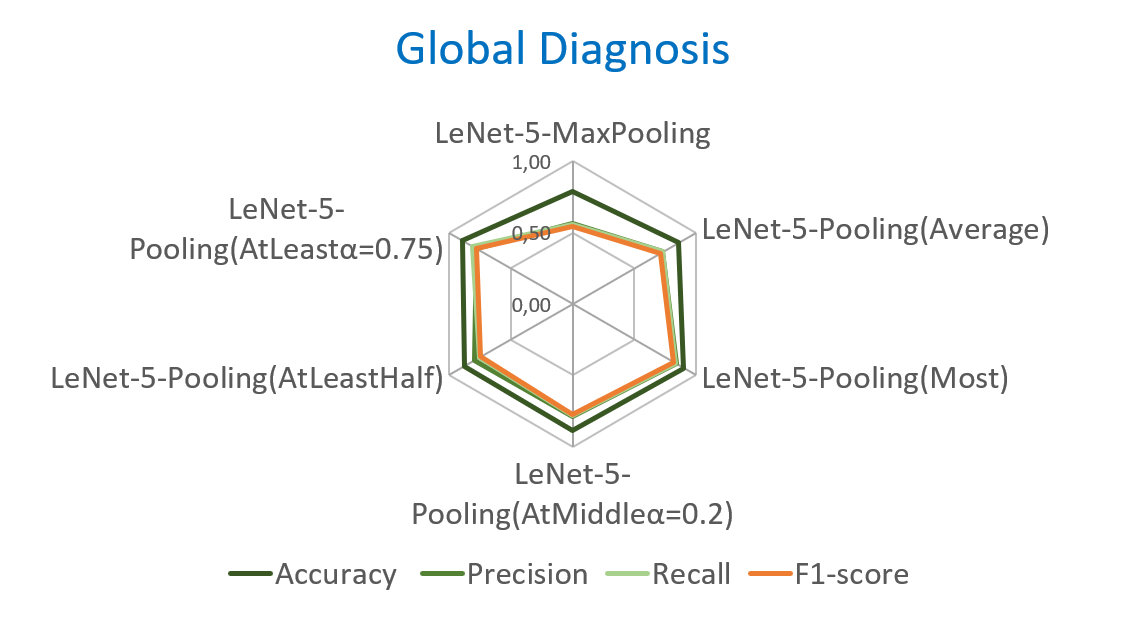} &  \includegraphics[width=0.35\textheight]{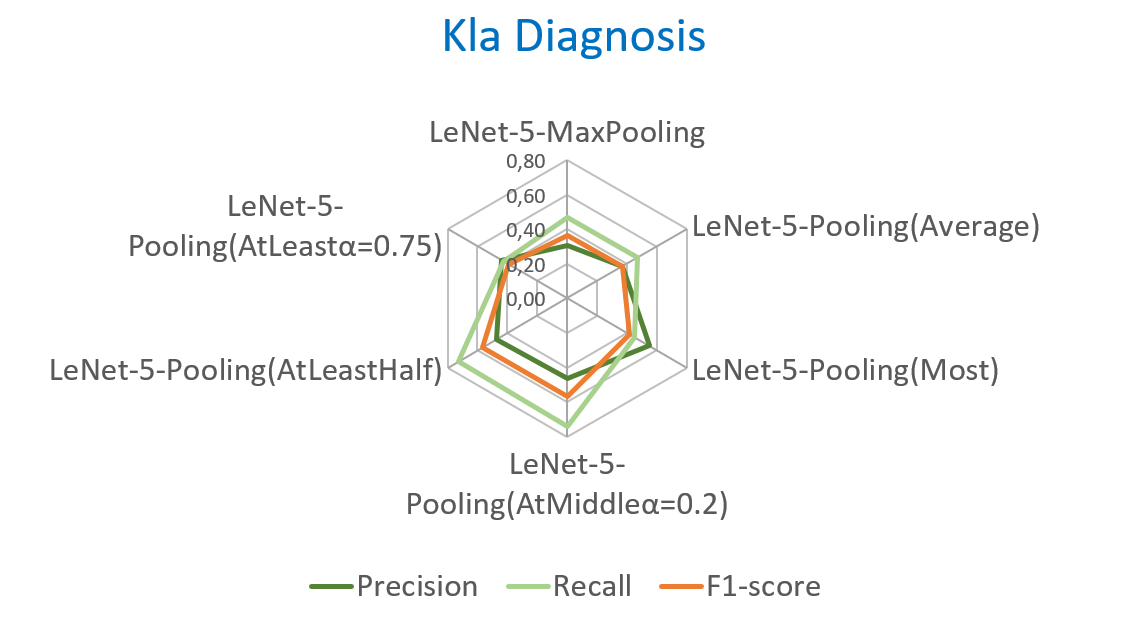} &\includegraphics[width=0.35\textheight]{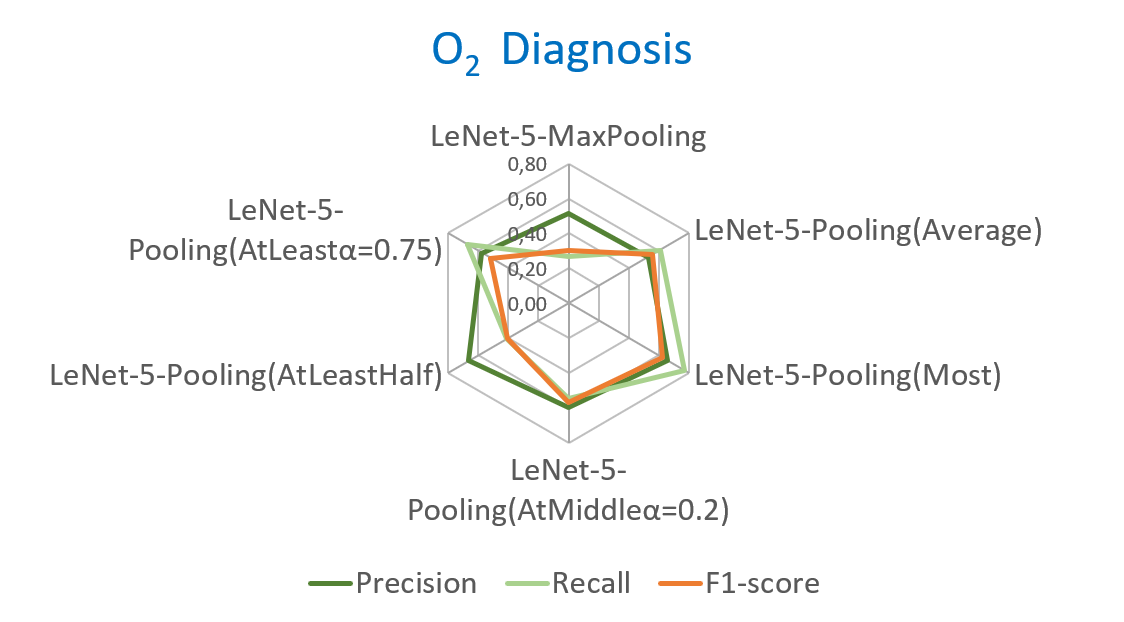}  \\
	\end{tabular}
		\caption{Sliding Time Window \& Linguistic OWA based fault diagnosis in LeNet5.}
		\label{fig:comparaLenet5}
	\end{figure}

\endgroup

{\color{black}
\subsection{Faster Learning}

This DCNN pooling layers based on well-known linguistic terms not only provide a better diagnosis performance, but also that permits to achieve a diagnosis performance with a fewer learning iterations than standard pooling options: $Max$ and $Average$. The OWA based DCNN pooling supplies better diagnosis performance for any of the checked learning periods.

Considering the layout of Model7, Fig. \ref{fig:FasterM7} summarizes the performance  obtained for 200, 500 and 700 learning iterations (epochs). $Most$ achieves  a better learning speed in  all performance indexes, supplying even a better performance  in 200 epochs than the standard pooling layers in 700 epochs. $Most$ is followed by \textsl{AtMiddle$_{\alpha}$($\alpha=0.2$)} and \textsl{AtLeastHalf} operators.

\begin{figure}[th]

  \begin{center}
		\includegraphics[width=1\textwidth]{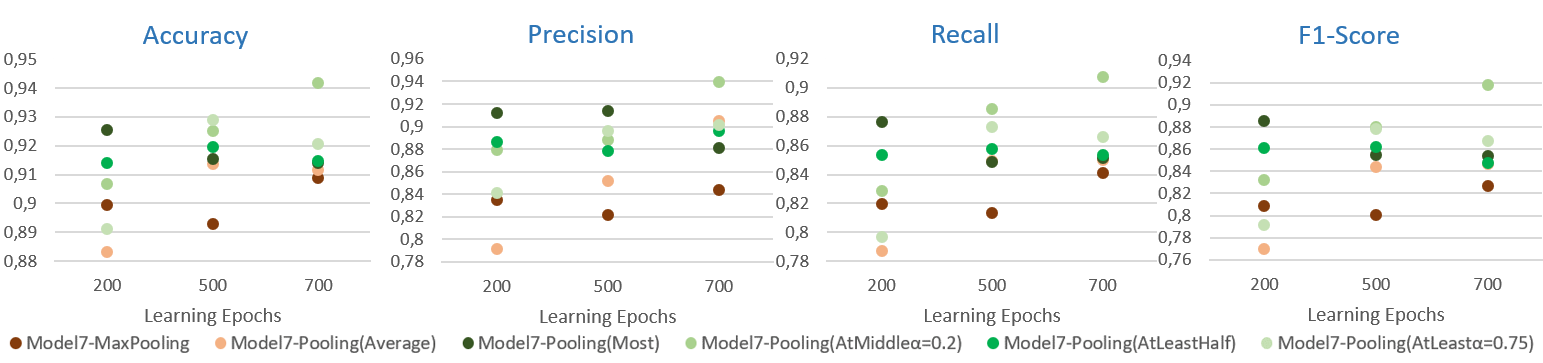}
		\caption{Model7: performance and  learning period. }
		\label{fig:FasterM7}
	\end{center}
\end{figure}

For Model3 (Fig. \ref{fig:FasterM3});  $Accuracy$, $Recall$ and $F1-score$ provided by \textsl{AtLeastHalf} in only 200 epochs surpasses the \textsl{Average} performance in 700 epochs. On the other hand, \textsl{AtLeastHalf} in 200 epochs  overcomes the \textsl{Max} $Precision$ performance in 700 epochs. The proposed linguistic OWA pooling operators provide a better diagnosis,  achieving better performance in any stage of learning.

\begin{figure}[th]
  \begin{center}
		\includegraphics[width=1\textwidth]{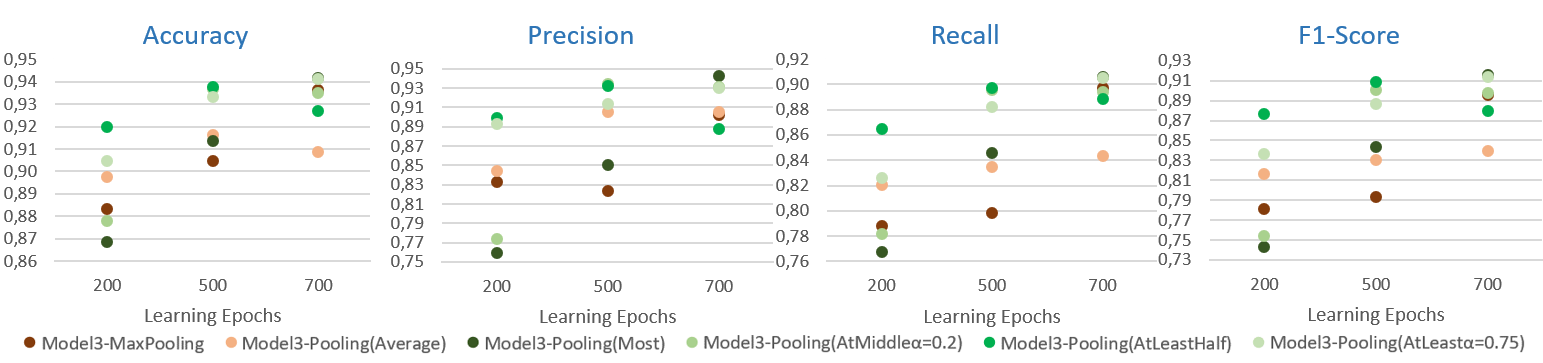}
		\caption{Model3: performance and  learning period. }
		\label{fig:FasterM3}
	\end{center}
\end{figure}

Finally, Fig. \ref{fig:FasterLN5} shows the performance of the LeNet-5 layout. The OWA  based pooling shows better performance  in all the leaning periods. Here, to highlight that \textsl{AtLeast$_{\alpha}$ ($\alpha=0.75$)}, \textsl{AtMiddle$_{\alpha}$($\alpha=0.2$)} and \textsl{AtLeastHalf} have a need  for a fewer learning epochs to provide the best performances in  all cases, except for $Precision$.

\begin{figure}[th]
  \begin{center}
		\includegraphics[width=1\textwidth]{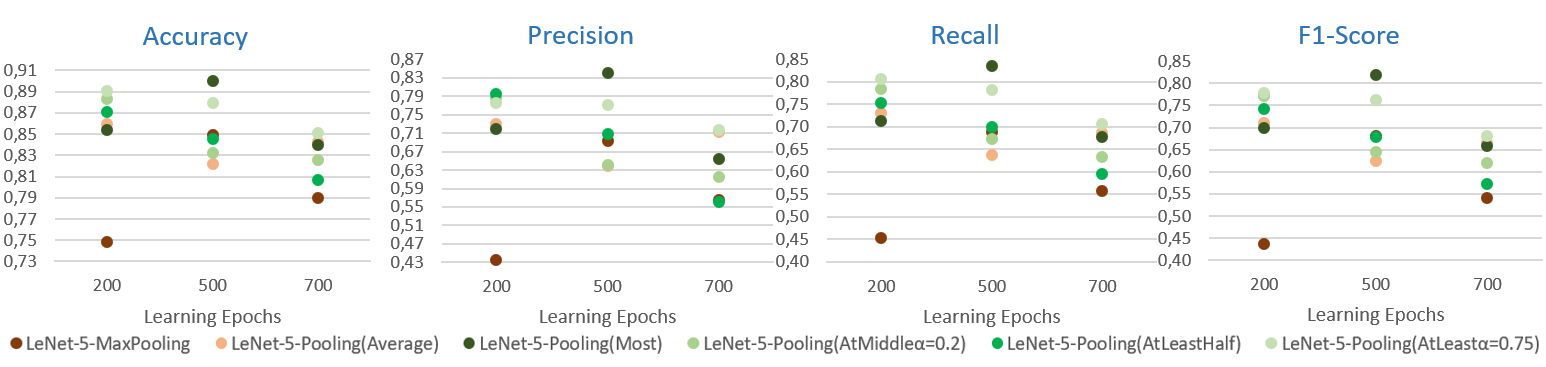}
		\caption{LeNet-5: performance and learning period. }
		\label{fig:FasterLN5}
	\end{center}
\end{figure}

To sum up,  the linguistic OWA pooling has shown a better fault diagnosis performance, as well needing fewer training epochs  than $Max$ and $Average$ to achieve a goal performance.

}


\subsection{Results Summary}

The sliding time window and the \textsl{Most} and {\color{black} \textsl{AtMiddle$_{\alpha}$} linguistic OWA pooling operators} have given the best performances  in the fault diagnosis of the WTTP, around $91-94\%$ for all the performance indexes, considering  a set of faults in different  magnitudes and time characteristics, improving the results with  regard to such methodologies  as  \cite{Wu2018}, which has been  taken as reference;  as well as managing a large number of monitoring variables (140),  a small number samples (4) for the diagnosis, and an earlier diagnosis every a $SamplingTime$, here 15 minutes.

{\color{black}
On the other hand,  in a simultaneous way, a set of alternative operators for pooling have been experimented on  three lightweight CNN layouts (used in fault diagnosis and other tasks) and compared  with regard to the standard pooling operation (\textsl{Max/Average}), being these  outperformed by linguistic OWA operators. }

In general,  these proposed linguistic OWA operators for pooling  give a better performance, in diagnosis,  than  implementing  the  standard pooling  for all the  baselines involved,   taking into account the performance indexes (\textsl{Accuracy, Precision, Recall, F1-score}). These pooling operators have shown better performance for any learning period.

\begin{figure}[t!]
	\hspace{-2cm}
	\begin{tabular}{ccc}
		\includegraphics[width=0.3\textheight]{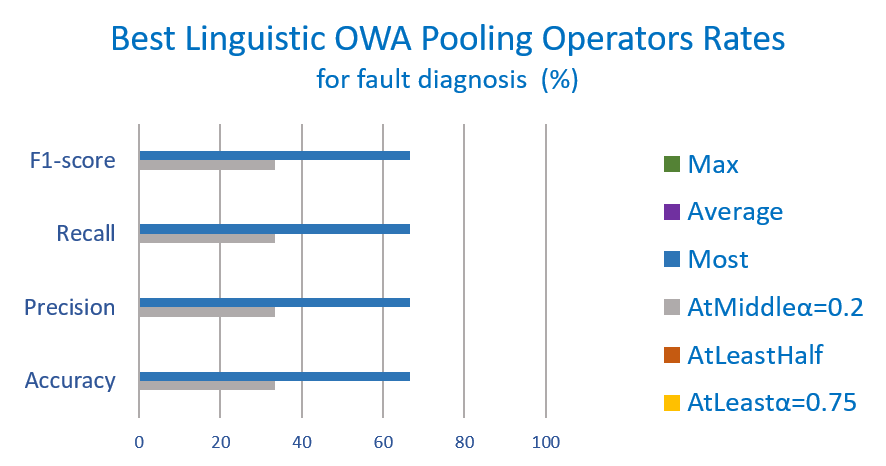} &  \includegraphics[width=0.3\textheight]{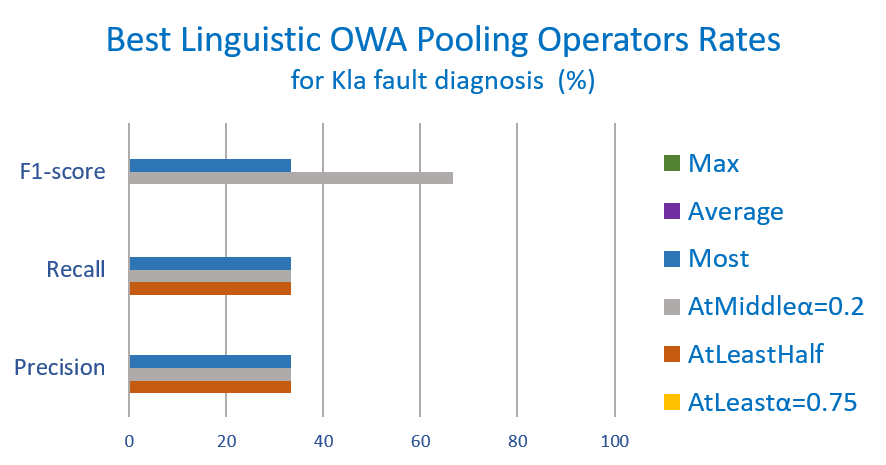} & \includegraphics[width=0.3\textheight]{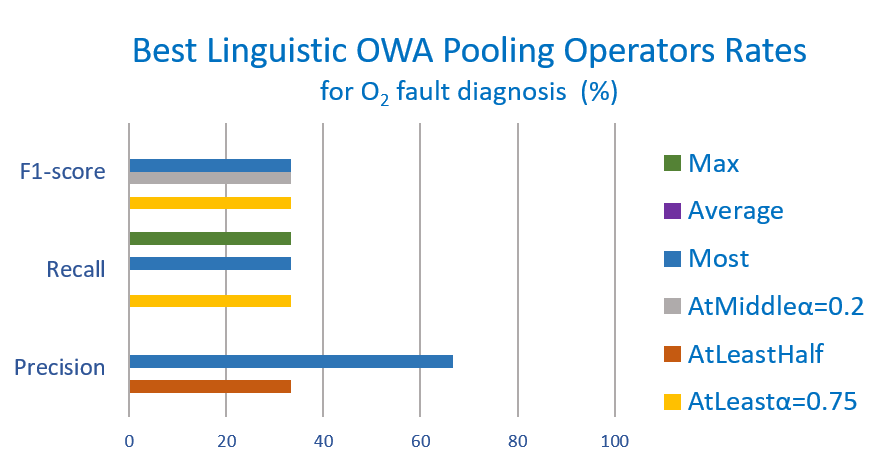} \\
	\end{tabular}
		\caption{Best Linguist OWA Operators in diagnosis.}
		\label{fig:bestOwaOperators}
\end{figure}

Fig. \ref {fig:bestOwaOperators} shows that the \textsl{Most} and \textsl{AtMiddle$_{\alpha}$($\alpha=0.2$)} pooling operators are the ones with the best scoring in global diagnosis, with around $66\%$ and $34\%$  of cases for each one.
{\color{black}Focusing on the $Kla$ fault, the scoring is quite similar, but $AtLeastHalf$ also provides the best scoring with around $34\%$  of cases for $Recall$ and $Precision$. In the case of the $O_2$ fault, \textsl{AtLeast$_{\alpha}$ ($\alpha=0.75$)} provides the best result in around $34\%$  of cases for $F1-Score$ and $Recall$.}
On the other hand, these alternative pooling operators give a  linguistic and intuitive meaning to the pooling, in the case of \textsl{Most}, 'most' of the features are taken into account for pooling, discarding about $20\%$ of the lowest  and $30\%$ of the highest, in the case \textsl{AtMiddle$_{\alpha}$($\alpha=0.2$)} only $20\%$ are discarded on both sides.

{\color{black}

Moreover, the networks incorporating the linguistic OWA based pooling show a faster learning in comparison with the standard pooling based models. A goal performance is achieved by OWA pooling based networks  in fewer training epochs than the networks using the standard pooling.
}

\section{Conclusions}
\label{sec:conclusion}

 First of all, the methodology based on a sliding  time  window  and alternative pooling based on  linguistic OWA operators in a DCNN  has permitted us to obtain a high level of fault diagnosis in the WWTP case study:
\textsl{Accuracy}: $0.94$,  \textsl{Precision}: $0.94$, \textsl{Recall}: $0.91$, \textsl{F1-score}: $0.92$,  considering faults in different magnitude conditions and time context, without any selection/extraction of the sampled variables (140). The fault diagnosis  was achieved in the short term, every 15 minutes or one sampling time, based on a data window concerning the last hour of WWTP working, that is $60/15$ samples of the data.  {\color{black} This approach can be applied in any other WWTP model if data about this plant are available.}

On the other hand, simultaneously, a set of alternative operators for pooling based on linguistic OWA have been experimented on, based on known DCNN  layouts in order to check their performance with regard to the \textsl{Max} and \textsl{Average} standard pooling layers. These standard poolings, which are particular cases of linguistic OWA  pooling,   have been outperformed by the alternative  linguistic OWA  pooling proposed in this work. Worth noting is the performance of {\color{black}\textsl{Most} and  \textsl{AtMiddle$_{\alpha}$($\alpha=0.2$)}}, which gives the best performance for most of the  fault diagnosis indexes: \textsl{Accuracy}, \textsl{Precision}, \textsl{Recall} and, in particular the \textsl{F1-score}, even in the most difficult faults, such as $Kla$ and $O_2$. {\color{black}All this follows and  reinforces the idea and performance of the called OWA-centered operators \cite{Yager2007}.}

Moreover, this type of operator for pooling has a linguistic meaning which can be useful for a more comprehensive decision making when feature pooling must be addressed. {\color{black} Additionally, this OWA pooling  permits a faster learning in comparison  standard pooling operators.}

{\color{black} The work in progress is focused on this approach for WWTP diagnosis when some faults could have roots in cyber attack to the plant. This is a  very serious  safety challenge to be addressed in order to maintain the process performance and the effluent quality and safety.}

\section*{Acknowledgement}
This work was supported by the Spanish Government through the Ministerio de Ciencia, Innovación y Universidades/ Agencia estatal de Investigación (MICIU/AEI/10.13039/501100011033) under Grant PID2019-105434RB-C32.\\
\FloatBarrier

\bibliographystyle{elsarticle-num} 
\bibliography{./OWApoolingT2}

\begin{thebibliography}{10}
\expandafter\ifx\csname url\endcsname\relax
  \def\url#1{\texttt{#1}}\fi
\expandafter\ifx\csname urlprefix\endcsname\relax\def\urlprefix{URL }\fi
\expandafter\ifx\csname href\endcsname\relax
  \def\href#1#2{#2} \def\path#1{#1}\fi

\bibitem{Fan-18}
M.~Fan, J.~Hu, R.~Cao, W.~Ruan, X.~Wei, A review on experimental design for
  pollutants removal in water treatment with the aid of artificial
  intelligence, Chemosphere 200 (2018) 330--343.
\newblock \href
  {https://doi.org/https://doi.org/10.1016/j.chemosphere.2018.02.111}
  {\path{doi:https://doi.org/10.1016/j.chemosphere.2018.02.111}}.

\bibitem{Salles-23}
R.~Salles, J.~Mendes, R.~P. Ribeiro, J.~Gama, Fault detection in wastewater
  treatment plants: Application of autoencoders models with streaming data,
  in: Machine Learning and Principles and Practice of Knowledge Discovery in
  Databases, Springer Nature Switzerland, 2023, pp. 55--70.
\newblock \href {https://doi.org/https://doi.org/10.1007/978-3-031-23618-1_4}
  {\path{doi:https://doi.org/10.1007/978-3-031-23618-1_4}}.

\bibitem{Sanchez-15}
A.~S{\'{a}}nchez-Fern{\'{a}}ndez, M.~J. Fuente, G.~I. Sainz-Palmero, {Fault
  detection in wastewater treatment plants using distributed PCA methods}, in:
  2015 IEEE 20th Conference on Emerging Technologies Factory Automation (ETFA),
  2015, pp. 1--7.
\newblock \href {https://doi.org/10.1109/ETFA.2015.7301504}
  {\path{doi:10.1109/ETFA.2015.7301504}}.

\bibitem{Cheng-19}
H.~Cheng, J.~Wu, Y.~Liu, D.~Huang, A novel fault identification and
  root-causality analysis of incipient faults with applications to wastewater
  treatment processes, Chemometrics and Intelligent Laboratory Systems 188
  (2019) 24--36.
\newblock \href
  {https://doi.org/https://doi.org/10.1016/j.chemolab.2019.03.004}
  {\path{doi:https://doi.org/10.1016/j.chemolab.2019.03.004}}.

\bibitem{Li-22}
Z.~Li, L.~Tian, Q.~Jiang, X.~Yan, Dynamic nonlinear process monitoring based on
  dynamic correlation variable selection and kernel principal component
  regression, Journal of the Franklin Institute 359~(9) (2022) 4513--4539.
\newblock \href
  {https://doi.org/https://doi.org/10.1016/j.jfranklin.2022.04.021}
  {\path{doi:https://doi.org/10.1016/j.jfranklin.2022.04.021}}.

\bibitem{Xu-21}
C.~Xu, D.~Huang, D.~Li, Y.~Liu, Novel process monitoring approach enhanced by a
  complex independent component analysis algorithm with applications for
  wastewater treatment, Industrial \& Engineering Chemistry Research 60~(38)
  (2021) 13914–13926.
\newblock \href {https://doi.org/https://doi.org/10.1021/acs.iecr.1c01990}
  {\path{doi:https://doi.org/10.1021/acs.iecr.1c01990}}.

\bibitem{Palla-23}
G.~{Lakshmi Priya Palla}, A.~{Kumar Pani}, Independent component analysis
  application for fault detection in process industries: Literature review and
  an application case study for fault detection in multiphase flow systems,
  Measurement 209 (2023) 112504.
\newblock \href
  {https://doi.org/https://doi.org/10.1016/j.measurement.2023.112504}
  {\path{doi:https://doi.org/10.1016/j.measurement.2023.112504}}.

\bibitem{Liu-21}
H.~Liu, J.~Yang, Y.~Zhang, C.~Yang,
  \href{https://www.sciencedirect.com/science/article/pii/S0957582020317511}{Monitoring
  of wastewater treatment processes using dynamic concurrent kernel partial
  least squares}, Process Safety and Environmental Protection 147 (2021)
  274--282.
\newblock \href {https://doi.org/https://doi.org/10.1016/j.psep.2020.09.034}
  {\path{doi:https://doi.org/10.1016/j.psep.2020.09.034}}.
\newline\urlprefix\url{https://www.sciencedirect.com/science/article/pii/S0957582020317511}

\bibitem{Sha-23}
X.~Sha, N.~Diao,
  \href{https://www.sciencedirect.com/science/article/pii/S0263224122014798}{Robust
  kernel principal component analysis and its application in blockage detection
  at the turn of conveyor belt}, Measurement 206 (2023) 112283.
\newblock \href
  {https://doi.org/https://doi.org/10.1016/j.measurement.2022.112283}
  {\path{doi:https://doi.org/10.1016/j.measurement.2022.112283}}.
\newline\urlprefix\url{https://www.sciencedirect.com/science/article/pii/S0263224122014798}

\bibitem{Cheng-21}
H.~Cheng, J.~Wu, D.~Huang, Y.~Liu, Q.~Wang,
  \href{https://www.sciencedirect.com/science/article/pii/S0019057821000495}{Robust
  adaptive boosted canonical correlation analysis for quality-relevant process
  monitoring of wastewater treatment}, ISA Transactions 117 (2021) 210--220.
\newblock \href {https://doi.org/https://doi.org/10.1016/j.isatra.2021.01.039}
  {\path{doi:https://doi.org/10.1016/j.isatra.2021.01.039}}.
\newline\urlprefix\url{https://www.sciencedirect.com/science/article/pii/S0019057821000495}

\bibitem{Fuente-23}
M.~J. De~La~Fuente, G.~I. Sainz-Palmero, M.~Galende-Hern\'andez, Dynamic
  decentralized monitoring for large-scale industrial processes using
  multiblock canonical variate analysis based regression, IEEE Access 11 (2023)
  26611--26623.
\newblock \href {https://doi.org/10.1109/ACCESS.2023.3256719}
  {\path{doi:10.1109/ACCESS.2023.3256719}}.

\bibitem{Khurshid-23}
A.~Khurshid, A.~K. {Pani}, Machine learning approaches for data-driven process
  monitoring of biological wastewater treatment plant: A review of research
  works on benchmark simulation model no. 1(bsm1), Environmental Monitoring and
  Assessment 195 (2023) 96.
\newblock \href {https://doi.org/https://doi.org/10.1007/s10661-023-11463-8}
  {\path{doi:https://doi.org/10.1007/s10661-023-11463-8}}.

\bibitem{Bellamoli-23}
F.~Bellamoli, M.~{Di Iorio}, M.~Vian, F.~Melgani, Machine learning methods for
  anomaly classification in wastewater treatment plants, Journal of
  Environmental Management 344 (2023) 118594.
\newblock \href {https://doi.org/https://doi.org/10.1016/j.jenvman.2023.118594}
  {\path{doi:https://doi.org/10.1016/j.jenvman.2023.118594}}.

\bibitem{Fan2022}
S.~Fan, X.~Zhang, Z.~Song, Imbalanced sample selection with deep reinforcement
  learning for fault diagnosis, IEEE Transactions on Industrial Informatics
  18~(4) (2022) 2518--2527.
\newblock \href {https://doi.org/10.1109/TII.2021.3100284}
  {\path{doi:10.1109/TII.2021.3100284}}.

\bibitem{Yang2023}
D.~Yang, H.~R. Karimi, M.~Pawelczyk,
  \href{https://www.sciencedirect.com/science/article/pii/S0967066123000448}{A
  new intelligent fault diagnosis framework for rotating machinery based on
  deep transfer reinforcement learning}, Control Engineering Practice 134
  (2023) 105475.
\newblock \href
  {https://doi.org/https://doi.org/10.1016/j.conengprac.2023.105475}
  {\path{doi:https://doi.org/10.1016/j.conengprac.2023.105475}}.
\newline\urlprefix\url{https://www.sciencedirect.com/science/article/pii/S0967066123000448}

\bibitem{Newhart-19}
K.~B. Newhart, R.~W. Holloway, A.~S. Hering, T.~Y. Cath, Data-driven
  performance analyses of wastewater treatment plants: A review, Water Research
  157 (2019) 498--513.
\newblock \href {https://doi.org/https://doi.org/10.1016/j.watres.2019.03.030}
  {\path{doi:https://doi.org/10.1016/j.watres.2019.03.030}}.

\bibitem{Alvi-23}
M.~Alvi, D.~Batstone, C.~K. Mbamba, P.~Keymer, T.~French, A.~Ward, J.~Dwyer,
  R.~Cardell-Oliver,
  \href{https://www.sciencedirect.com/science/article/pii/S0043135423009582}{Deep
  learning in wastewater treatment: a critical review}, Water Research 245
  (2023) 120518.
\newblock \href {https://doi.org/https://doi.org/10.1016/j.watres.2023.120518}
  {\path{doi:https://doi.org/10.1016/j.watres.2023.120518}}.
\newline\urlprefix\url{https://www.sciencedirect.com/science/article/pii/S0043135423009582}

\bibitem{Ismail-23}
W.~Ismail, N.~Niknejad, M.~Bahari, R.~Hendradi, N.~J.~M. {Zaizi}, M.~Z.
  {Zulkifli}, Water treatment and artificial intelligence techniques: a
  systematic literature review research, Environmental Science and Pollution
  Research 30 (2023) 71794–71812.
\newblock \href {https://doi.org/https://doi.org/10.1007/s11356-021-16471-0}
  {\path{doi:https://doi.org/10.1007/s11356-021-16471-0}}.

\bibitem{Harrou2018}
F.~Harrou, A.~Dairi, Y.~Sun, M.~Senouci,
  \href{https://www.sciencedirect.com/science/article/pii/S0301479718307394}{Statistical
  monitoring of a wastewater treatment plant: A case study}, Journal of
  Environmental Management 223 (2018) 807--814.
\newblock \href {https://doi.org/https://doi.org/10.1016/j.jenvman.2018.06.087}
  {\path{doi:https://doi.org/10.1016/j.jenvman.2018.06.087}}.
\newline\urlprefix\url{https://www.sciencedirect.com/science/article/pii/S0301479718307394}

\bibitem{Yu-19}
J.~Yu, X.~Yan,
  \href{https://www.sciencedirect.com/science/article/pii/S0019057818303902}{Active
  features extracted by deep belief network for process monitoring}, ISA
  Transactions 84 (2019) 247--261.
\newblock \href {https://doi.org/https://doi.org/10.1016/j.isatra.2018.10.011}
  {\path{doi:https://doi.org/10.1016/j.isatra.2018.10.011}}.
\newline\urlprefix\url{https://www.sciencedirect.com/science/article/pii/S0019057818303902}

\bibitem{Yu-21}
J.~Yu, X.~Yan,
  \href{https://www.sciencedirect.com/science/article/pii/S0957582021003438}{A
  new deep model based on the stacked autoencoder with intensified iterative
  learning style for industrial fault detection}, Process Safety and
  Environmental Protection 153 (2021) 47--59.
\newblock \href {https://doi.org/https://doi.org/10.1016/j.psep.2021.07.002}
  {\path{doi:https://doi.org/10.1016/j.psep.2021.07.002}}.
\newline\urlprefix\url{https://www.sciencedirect.com/science/article/pii/S0957582021003438}

\bibitem{Ba-Alawi-21}
A.~H. Ba-Alawi, P.~Vilela, J.~Loy-Benitez, S.~Heo, C.~Yoo,
  \href{https://www.sciencedirect.com/science/article/pii/S2214714421002932}{Intelligent
  sensor validation for sustainable influent quality monitoring in wastewater
  treatment plants using stacked denoising autoencoders}, Journal of Water
  Process Engineering 43 (2021) 102206.
\newblock \href {https://doi.org/https://doi.org/10.1016/j.jwpe.2021.102206}
  {\path{doi:https://doi.org/10.1016/j.jwpe.2021.102206}}.
\newline\urlprefix\url{https://www.sciencedirect.com/science/article/pii/S2214714421002932}

\bibitem{Ba-Alawi-22}
A.~H. Ba-Alawi, J.~Loy-Benitez, S.~Kim, C.~Yoo,
  \href{https://www.sciencedirect.com/science/article/pii/S0045653521031192}{Missing
  data imputation and sensor self-validation towards a sustainable operation of
  wastewater treatment plants via deep variational residual autoencoders},
  Chemosphere 288 (2022) 132647.
\newblock \href
  {https://doi.org/https://doi.org/10.1016/j.chemosphere.2021.132647}
  {\path{doi:https://doi.org/10.1016/j.chemosphere.2021.132647}}.
\newline\urlprefix\url{https://www.sciencedirect.com/science/article/pii/S0045653521031192}

\bibitem{Liu-23}
J.~Liu, L.~Xu, Y.~Xie, T.~Ma, J.~Wang, Z.~Tang, W.~Gui, H.~Yin, H.~Jahanshahi,
  Toward robust fault identification of complex industrial processes using
  stacked sparse-denoising autoencoder with softmax classifier, IEEE
  Transactions on Cybernetics 53~(1) (2023) 428--442.
\newblock \href {https://doi.org/10.1109/TCYB.2021.3109618}
  {\path{doi:10.1109/TCYB.2021.3109618}}.

\bibitem{Qian2022}
J.~Qian, Z.~Song, Y.~Yao, Z.~Zhu, X.~Zhang,
  \href{https://www.sciencedirect.com/science/article/pii/S0169743922002222}{A
  review on autoencoder based representation learning for fault detection and
  diagnosis in industrial processes}, Chemometrics and Intelligent Laboratory
  Systems 231 (2022) 104711.
\newblock \href
  {https://doi.org/https://doi.org/10.1016/j.chemolab.2022.104711}
  {\path{doi:https://doi.org/10.1016/j.chemolab.2022.104711}}.
\newline\urlprefix\url{https://www.sciencedirect.com/science/article/pii/S0169743922002222}

\bibitem{Mamandipoor-20}
B.~Mamandipoor, M.~Majd, S.~Sheikhalishahi, C.~M. ans Venet~Osmani, Monitoring
  and detecting faults in wastewater treatment plants using deep learning,
  Environmental Monitoring and Assessment 192 (2020) 148.
\newblock \href {https://doi.org/https://doi.org/10.1007/s10661-020-8064-1}
  {\path{doi:https://doi.org/10.1007/s10661-020-8064-1}}.

\bibitem{Wu2018}
H.~Wu, J.~Zhao,
  \href{https://www.sciencedirect.com/science/article/pii/S0098135418302990}{Deep
  convolutional neural network model based chemical process fault diagnosis},
  Computers \& Chemical Engineering 115 (2018) 185--197.
\newblock \href
  {https://doi.org/https://doi.org/10.1016/j.compchemeng.2018.04.009}
  {\path{doi:https://doi.org/10.1016/j.compchemeng.2018.04.009}}.
\newline\urlprefix\url{https://www.sciencedirect.com/science/article/pii/S0098135418302990}

\bibitem{Song-22}
Q.~Song, P.~Jiang,
  \href{https://www.sciencedirect.com/science/article/pii/S0957582021006170}{A
  multi-scale convolutional neural network based fault diagnosis model for
  complex chemical processes}, Process Safety and Environmental Protection 159
  (2022) 575--584.
\newblock \href {https://doi.org/https://doi.org/10.1016/j.psep.2021.11.020}
  {\path{doi:https://doi.org/10.1016/j.psep.2021.11.020}}.
\newline\urlprefix\url{https://www.sciencedirect.com/science/article/pii/S0957582021006170}

\bibitem{Chen-22}
H.~Chen, J.~Cen, Z.~Yang, W.~Si, H.~Cheng, Fault diagnosis of the dynamic
  chemical process based on the optimized cnn-lstm network, ACS Omega 7~(38)
  (2022) 34389–34400.
\newblock \href {https://doi.org/https://doi.org/10.1021/acsomega.2c04017}
  {\path{doi:https://doi.org/10.1021/acsomega.2c04017}}.

\bibitem{Dairi-19}
A.~Dairi, T.~Cheng, F.~Harrou, Y.~Sun, T.~Leiknes,
  \href{https://www.sciencedirect.com/science/article/pii/S2210670719304160}{Deep
  learning approach for sustainable wwtp operation: A case study on data-driven
  influent conditions monitoring}, Sustainable Cities and Society 50 (2019)
  101670.
\newblock \href {https://doi.org/https://doi.org/10.1016/j.scs.2019.101670}
  {\path{doi:https://doi.org/10.1016/j.scs.2019.101670}}.
\newline\urlprefix\url{https://www.sciencedirect.com/science/article/pii/S2210670719304160}

\bibitem{Peng-22}
C.~Peng, M.~Fanchao, Fault detection of urban wastewater treatment process
  based on combination of deep information and transformer network, IEEE
  Transactions on Neural Networks and Learning Systems (2022) 1--10\href
  {https://doi.org/10.1109/TNNLS.2022.3224804}
  {\path{doi:10.1109/TNNLS.2022.3224804}}.

\bibitem{Zhao2024}
L.~Zhao, Z.~Zhang, \href{https://doi.org/10.1038/s41598-024-51258-6}{A improved
  pooling method for convolutional neural networks}, Scientific Reports 14~(1)
  (2024) 1589.
\newblock \href {https://doi.org/10.1038/s41598-024-51258-6}
  {\path{doi:10.1038/s41598-024-51258-6}}.
\newline\urlprefix\url{https://doi.org/10.1038/s41598-024-51258-6}

\bibitem{Diamantis2021}
D.~E. Diamantis, D.~K. Iakovidis, Fuzzy pooling, IEEE Transactions on Fuzzy
  Systems 29~(11) (2021) 3481--3488.
\newblock \href {https://doi.org/10.1109/TFUZZ.2020.3024023}
  {\path{doi:10.1109/TFUZZ.2020.3024023}}.

\bibitem{Yager1988}
R.~Yager, On ordered weighted averaging aggregation operators in multicriteria
  decisionmaking, IEEE Transactions on Systems, Man, and Cybernetics 18~(1)
  (1988) 183--190.
\newblock \href {https://doi.org/10.1109/21.87068}
  {\path{doi:10.1109/21.87068}}.

\bibitem{Forcen2020}
J.~Forcén, M.~Pagola, E.~Barrenechea, H.~Bustince,
  \href{https://www.sciencedirect.com/science/article/pii/S0925231220309991}{Learning
  ordered pooling weights in image classification}, Neurocomputing 411 (2020)
  45--53.
\newblock \href {https://doi.org/https://doi.org/10.1016/j.neucom.2020.06.028}
  {\path{doi:https://doi.org/10.1016/j.neucom.2020.06.028}}.
\newline\urlprefix\url{https://www.sciencedirect.com/science/article/pii/S0925231220309991}

\bibitem{DominguezCatena2021}
I.~Dominguez-Catena, D.~Paternain, M.~Galar,
  \href{https://www.mdpi.com/2076-3417/11/16/7195}{A study of owa operators
  learned in convolutional neural networks}, Applied Sciences 11~(16) (2021).
\newblock \href {https://doi.org/10.3390/app11167195}
  {\path{doi:10.3390/app11167195}}.
\newline\urlprefix\url{https://www.mdpi.com/2076-3417/11/16/7195}

\bibitem{Catena2020}
I.~Dominguez-Catena, D.~Paternain, M.~Galar, Additional feature layers from
  ordered aggregations for deep neural networks, in: 2020 IEEE International
  Conference on Fuzzy Systems (FUZZ-IEEE), 2020, pp. 1--8.
\newblock \href {https://doi.org/10.1109/FUZZ48607.2020.9177555}
  {\path{doi:10.1109/FUZZ48607.2020.9177555}}.

\bibitem{Hussain2022a}
W.~Hussain, M.~R. Raza, M.~A. Jan, J.~M. Merigó, H.~Gao, Cloud risk management
  with owa-lstm and fuzzy linguistic decision making, IEEE Transactions on
  Fuzzy Systems 30~(11) (2022) 4657--4666.
\newblock \href {https://doi.org/10.1109/TFUZZ.2022.3157951}
  {\path{doi:10.1109/TFUZZ.2022.3157951}}.

\bibitem{Hussain2022}
W.~Hussain, H.~Gao, M.~R. Raza, F.~A. Rabhi, J.~M. Merig{\'o},
  \href{https://doi.org/10.1007/s00521-022-07297-z}{Assessing cloud qos
  predictions using owa in neural network methods}, Neural Computing and
  Applications 34~(17) (2022) 14895--14912.
\newblock \href {https://doi.org/10.1007/s00521-022-07297-z}
  {\path{doi:10.1007/s00521-022-07297-z}}.
\newline\urlprefix\url{https://doi.org/10.1007/s00521-022-07297-z}

\bibitem{Ghosal2021}
S.~Ghosal, A.~Jain, S.~Sharma, D.~K. Tayal,
  \href{https://doi.org/10.1007/s13748-021-00252-4}{Armlowa: aspect rating
  analysis with multi-layer approach}, Progress in Artificial Intelligence
  10~(4) (2021) 505--516.
\newblock \href {https://doi.org/10.1007/s13748-021-00252-4}
  {\path{doi:10.1007/s13748-021-00252-4}}.
\newline\urlprefix\url{https://doi.org/10.1007/s13748-021-00252-4}

\bibitem{Alex2017}
A.~Krizhevsky, I.~Sutskever, G.~E. Hinton,
  \href{https://doi.org/10.1145/3065386}{Imagenet classification with deep
  convolutional neural networks}, Commun. ACM 60~(6) (2017) 84–90.
\newblock \href {https://doi.org/10.1145/3065386} {\path{doi:10.1145/3065386}}.
\newline\urlprefix\url{https://doi.org/10.1145/3065386}

\bibitem{Szegedy_2016_CVPR}
C.~Szegedy, V.~Vanhoucke, S.~Ioffe, J.~Shlens, Z.~Wojna, Rethinking the
  inception architecture for computer vision, in: Proceedings of the IEEE
  Conference on Computer Vision and Pattern Recognition (CVPR), 2016.
\newblock \href {https://doi.org/https://doi.org/10.1016/j.neucom.2015.09.081}
  {\path{doi:https://doi.org/10.1016/j.neucom.2015.09.081}}.

\bibitem{Lecun2015}
Y.~LeCun, Y.~Bengio, G.~Hinton, \href{https://doi.org/10.1038/nature14539}{Deep
  learning}, Nature 521~(7553) (2015) 436--444.
\newblock \href {https://doi.org/10.1038/nature14539}
  {\path{doi:10.1038/nature14539}}.
\newline\urlprefix\url{https://doi.org/10.1038/nature14539}

\bibitem{Samira2018}
S.~Pouyanfar, S.~Sadiq, Y.~Yan, H.~Tian, Y.~Tao, M.~P. Reyes, M.-L. Shyu, S.-C.
  Chen, S.~S. Iyengar, \href{https://doi.org/10.1145/3234150}{A survey on deep
  learning: Algorithms, techniques, and applications}, ACM Comput. Surv. 51~(5)
  (sep 2018).
\newblock \href {https://doi.org/10.1145/3234150} {\path{doi:10.1145/3234150}}.
\newline\urlprefix\url{https://doi.org/10.1145/3234150}

\bibitem{Ausif2019}
A.~Shrestha, A.~Mahmood, Review of deep learning algorithms and architectures,
  IEEE Access 7 (2019) 53040--53065.
\newblock \href {https://doi.org/10.1109/ACCESS.2019.2912200}
  {\path{doi:10.1109/ACCESS.2019.2912200}}.

\bibitem{Jun2022}
Z.~Li, F.~Liu, W.~Yang, S.~Peng, J.~Zhou, A survey of convolutional neural
  networks: Analysis, applications, and prospects, IEEE Transactions on Neural
  Networks and Learning Systems 33~(12) (2022) 6999--7019.
\newblock \href {https://doi.org/10.1109/TNNLS.2021.3084827}
  {\path{doi:10.1109/TNNLS.2021.3084827}}.

\bibitem{Liu2008}
X.~Liu, S.~Han,
  \href{https://www.sciencedirect.com/science/article/pii/S0888613X07000709}{Orness
  and parameterized rim quantifier aggregation with owa operators: A summary},
  International Journal of Approximate Reasoning 48~(1) (2008) 77--97, special
  Section: Perception Based Data Mining and Decision Support Systems.
\newblock \href {https://doi.org/https://doi.org/10.1016/j.ijar.2007.05.006}
  {\path{doi:https://doi.org/10.1016/j.ijar.2007.05.006}}.
\newline\urlprefix\url{https://www.sciencedirect.com/science/article/pii/S0888613X07000709}

\bibitem{Yager2007}
R.~R. Yager, \href{https://doi.org/10.1007/s00500-006-0125-z}{Centered owa
  operators}, Soft Computing 11~(7) (2007) 631--639.
\newblock \href {https://doi.org/10.1007/s00500-006-0125-z}
  {\path{doi:10.1007/s00500-006-0125-z}}.
\newline\urlprefix\url{https://doi.org/10.1007/s00500-006-0125-z}

\bibitem{flores2024}
M.~Flores-Sosa, E.~Avilés-Ochoa, J.~M. Merigó, J.~Kacprzyk,
  \href{https://www.sciencedirect.com/science/article/pii/S1568494622003167}{The
  owa operator in multiple linear regression}, Applied Soft Computing 124
  (2022) 108985.
\newblock \href {https://doi.org/https://doi.org/10.1016/j.asoc.2022.108985}
  {\path{doi:https://doi.org/10.1016/j.asoc.2022.108985}}.
\newline\urlprefix\url{https://www.sciencedirect.com/science/article/pii/S1568494622003167}

\bibitem{Shang2021}
S.-M. Zhou, F.~Chiclana, R.~I. John, J.~M. Garibaldi, L.~Huo, Type-1 owa
  operators in aggregating multiple sources of uncertain information:
  Properties and real-world applications in integrated diagnosis, IEEE
  Transactions on Fuzzy Systems 29~(8) (2021) 2112--2121.
\newblock \href {https://doi.org/10.1109/TFUZZ.2020.2992909}
  {\path{doi:10.1109/TFUZZ.2020.2992909}}.

\bibitem{Serrano2020}
J.~Serrano-Guerrero, F.~Chiclana, J.~A. Olivas, F.~P. Romero, E.~Homapour,
  \href{https://www.sciencedirect.com/science/article/pii/S0950705119304927}{A
  t1owa fuzzy linguistic aggregation methodology for searching feature-based
  opinions}, Knowledge-Based Systems 189 (2020) 105131.
\newblock \href {https://doi.org/https://doi.org/10.1016/j.knosys.2019.105131}
  {\path{doi:https://doi.org/10.1016/j.knosys.2019.105131}}.
\newline\urlprefix\url{https://www.sciencedirect.com/science/article/pii/S0950705119304927}

\bibitem{Yager1993}
R.~R. Yager,
  \href{https://www.sciencedirect.com/science/article/pii/016501149390194M}{Families
  of owa operators}, Fuzzy Sets and Systems 59~(2) (1993) 125--148.
\newblock \href {https://doi.org/https://doi.org/10.1016/0165-0114(93)90194-M}
  {\path{doi:https://doi.org/10.1016/0165-0114(93)90194-M}}.
\newline\urlprefix\url{https://www.sciencedirect.com/science/article/pii/016501149390194M}

\bibitem{Yager1996}
R.~R. Yager, Quantifier guided aggregation using owa operators, International
  Journal of Intelligent Systems 11~(1) (1996) 49--73.
\newblock \href
  {https://doi.org/https://doi.org/10.1002/(SICI)1098-111X(199601)11:1<49::AID-INT3>3.0.CO;2-Z}
  {\path{doi:https://doi.org/10.1002/(SICI)1098-111X(199601)11:1<49::AID-INT3>3.0.CO;2-Z}}.

\bibitem{Li2022}
Z.~Li, F.~Liu, W.~Yang, S.~Peng, J.~Zhou, A survey of convolutional neural
  networks: Analysis, applications, and prospects, IEEE Transactions on Neural
  Networks and Learning Systems 33~(12) (2022) 6999--7019.
\newblock \href {https://doi.org/10.1109/TNNLS.2021.3084827}
  {\path{doi:10.1109/TNNLS.2021.3084827}}.

\bibitem{Lecun1998}
Y.~Lecun, L.~Bottou, Y.~Bengio, P.~Haffner, Gradient-based learning applied to
  document recognition, Proceedings of the IEEE 86~(11) (1998) 2278--2324.
\newblock \href {https://doi.org/10.1109/5.726791}
  {\path{doi:10.1109/5.726791}}.

\bibitem{Vrecko2006}
D.~Vrecko, K.~Gernaey, C.~Rosen, U.~Jeppsson,
  \href{https://doi.org/10.2166/wst.2006.773}{Benchmark simulation model no 2
  in matlab-simulink: towards plant-wide wwtp control strategy evaluation},
  Water Science and Technology 54~(8) (2006) 65--72.
\newblock \href
  {http://arxiv.org/abs/https://iwaponline.com/wst/article-pdf/54/8/65/431513/65.pdf}
  {\path{arXiv:https://iwaponline.com/wst/article-pdf/54/8/65/431513/65.pdf}},
  \href {https://doi.org/10.2166/wst.2006.773}
  {\path{doi:10.2166/wst.2006.773}}.
\newline\urlprefix\url{https://doi.org/10.2166/wst.2006.773}

\bibitem{Elakkiya2024}
M.~K. Elakkiya, Dejey,
  \href{https://www.sciencedirect.com/science/article/pii/S0957417423026040}{Novel
  deep learning models with novel integrated activation functions for autism
  screening: Autinet and minautinet}, Expert Systems with Applications 238
  (2024) 122102.
\newblock \href {https://doi.org/https://doi.org/10.1016/j.eswa.2023.122102}
  {\path{doi:https://doi.org/10.1016/j.eswa.2023.122102}}.
\newline\urlprefix\url{https://www.sciencedirect.com/science/article/pii/S0957417423026040}

\bibitem{Wang2024}
H.~Wang, S.~Zhang, G.~Yang, Z.~Zhao, F.~Ji,
  \href{https://api.semanticscholar.org/CorpusID:269017017}{Research on
  lithology recognition method for borehole camera based on deep learning},
  2024 IEEE 3rd International Conference on Electrical Engineering, Big Data
  and Algorithms (EEBDA) (2024) 1469--1471.
\newline\urlprefix\url{https://api.semanticscholar.org/CorpusID:269017017}

\bibitem{Rai2020}
H.~M. Rai, K.~Chatterjee,
  \href{https://www.sciencedirect.com/science/article/pii/S2666827020300049}{Detection
  of brain abnormality by a novel lu-net deep neural cnn model from mr images},
  Machine Learning with Applications 2 (2020) 100004.
\newblock \href {https://doi.org/https://doi.org/10.1016/j.mlwa.2020.100004}
  {\path{doi:https://doi.org/10.1016/j.mlwa.2020.100004}}.
\newline\urlprefix\url{https://www.sciencedirect.com/science/article/pii/S2666827020300049}

\bibitem{powers2020}
D.~Powers, Evaluation: From precision, recall and f-factor to roc,
  informedness, markedness \& correlation, Mach. Learn. Technol. 2 (01 2008).

\end{thebibliography}

\end{document}